\documentclass[sigconf]{acmart}

\usepackage{tabularx}
\usepackage{calc}
\usepackage[capitalize, noabbrev]{cleveref}
\usepackage{subcaption}
\usepackage{listings}
\usepackage{siunitx}
\usepackage{enumitem}

\AtBeginDocument{%
  \providecommand\BibTeX{{%
    \normalfont B\kern-0.5em{\scshape i\kern-0.25em b}\kern-0.8em\TeX}}}

\copyrightyear{2025} 
\acmYear{2025} 
\setcopyright{cc}
\setcctype{by}
\acmConference[CHI '25]{CHI Conference on Human Factors in Computing Systems}{April 26-May 1, 2025}{Yokohama, Japan}
\acmBooktitle{CHI Conference on Human Factors in Computing Systems (CHI '25), April 26-May 1, 2025, Yokohama, Japan}\acmDOI{10.1145/3706598.3713890}
\acmISBN{979-8-4007-1394-1/25/04}

\begin{document}

\title[Content-Driven Local Response: Sentence-Level and Message-Level Mobile Email Replies]{Content-Driven Local Response: Supporting Sentence-Level and Message-Level Mobile Email Replies With and Without AI}

\author{Tim Zindulka}
\authornote{Both authors contributed equally to this research.}
\email{tim.zindulka@uni-bayreuth.de}
\orcid{0009-0009-1972-351X}
\affiliation{%
  \institution{University of Bayreuth}
  \city{Bayreuth}
  \country{Germany}
}

\author{Sven Goller}
\authornotemark[1]
\email{sven.goller@uni-bayreuth.de}
\orcid{0000-0001-5263-5372}
\affiliation{%
  \institution{University of Bayreuth}
  \city{Bayreuth}
  \country{Germany}
}

\author{Florian Lehmann}
\email{florian.lehmann@uni-bayreuth.de}
\orcid{0000-0003-0201-867X}
\affiliation{%
  \institution{University of Bayreuth}
  \city{Bayreuth}
  \country{Germany}
}

\author{Daniel Buschek}
\email{daniel.buschek@uni-bayreuth.de}
\orcid{0000-0002-0013-715X}
\affiliation{%
  \institution{University of Bayreuth}
  \city{Bayreuth}
  \country{Germany}
}

\renewcommand{\shortauthors}{Zindulka and Goller et al.}

\definecolor{TimsColor}{rgb}{0.1,0.5,0.8}
\newcommand{\tim}[1]{\textsf{\textbf{\textcolor{TimsColor}{[Tim: #1]}}}}
\definecolor{SvensColor}{rgb}{0.5,0.8,0.5}
\newcommand{\sven}[1]{\textsf{\textbf{\textcolor{SvensColor}{[Sven: #1]}}}}
\definecolor{FlosColor}{rgb}{0.9,0.1,0.8}
\newcommand{\flo}[1]{\textsf{\textbf{\textcolor{FlosColor}{[Flo: #1]}}}}
\definecolor{DanielsColor}{rgb}{0.9,0.6,0.1}
\newcommand{\daniel}[1]{\textsf{\textbf{\textcolor{DanielsColor}{[Daniel: #1]}}}}

\newcommand{\minsec}[2]{\SI{#1}{\minute} \SI{#2}{\second}}
\newcommand{\mins}[1]{\SI{#1}{\minute}}
\newcommand{\secs}[1]{\SI{#1}{\second}}
\newcommand{\pct}[1]{\ifnum\pdfstrcmp{#1}{X}=0
        X\% 
    \else\SI{#1}{\percent}\fi
}

\newcommand{\lmmci}[5]{$\beta$=#1, SE=#2, CI$_{95\%}$=[#3, #4], p#5}
\newcommand{\posthoc}[2]{#1, p#2}
\newcommand{\artf}[4]{$F$(#1,\,#2)=#3, p#4}
\newcommand{\artc}[3]{$t$(#1)=#2, p#3}
\newcommand{\petasq}[1]{$\eta_p^2$=#1}

\definecolor{deemphColor}{rgb}{0.4,0.4,0.4}
\newcommand{\deemph}[1]{\textcolor{deemphColor}{#1}}

\newcommand{\ivmode}{\textsc{UImode}}
\newcommand{\modeourstxt}{content-driven local response}
\newcommand{\modeoursTxt}{Content-driven local response}
\newcommand{\modeoursTXT}{Content-Driven Local Response}
\newcommand{\modemailtxt}{message-level reply generation}
\newcommand{\modemailTxt}{Message-level reply generation}
\newcommand{\modemanual}{\textsc{NoAI}}
\newcommand{\modeours}{\textsc{CDLR}}
\newcommand{\modemail}{\textsc{MSG}}

\newcommand{\imppass}{improvement pass}
\newcommand{\Imppass}{Improvement pass}

\newcommand{\studyOneN}{17} %
\newcommand{\studyTwoN}{126} %

\newcommand{\lastaccessed}{\textit{last accessed 22.08.2024}}

\newcommand{\oldId}[1]{} %

\newcommand\revision[1]{\textcolor{black}{#1}}

\begin{abstract}
Mobile emailing demands efficiency in diverse situations, which motivates the use of AI. However, generated text does not always reflect how people want to respond. This challenges users with AI involvement tradeoffs not yet considered in email UIs. We address this with a new UI concept called \textit{\modeoursTXT{} (\modeours)}, inspired by microtasking. This allows users to insert responses into the email by selecting sentences, which additionally serves to guide AI suggestions. The concept supports combining AI for local suggestions and message-level improvements. Our user study (N=\studyTwoN) compared \modeours{} with manual typing and full reply generation. We found that \modeours{} supports flexible workflows with varying degrees of AI involvement, while retaining the benefits of reduced typing and errors. This work contributes a new approach to integrating AI capabilities: By redesigning the UI for workflows with and without AI, we can empower users to dynamically adjust AI involvement. %

\end{abstract}


\keywords{Writing assistance, Large language models, Human-AI interaction, Email, Mobile text entry}

\begin{teaserfigure}
    \centering
    \includegraphics[width=\textwidth]{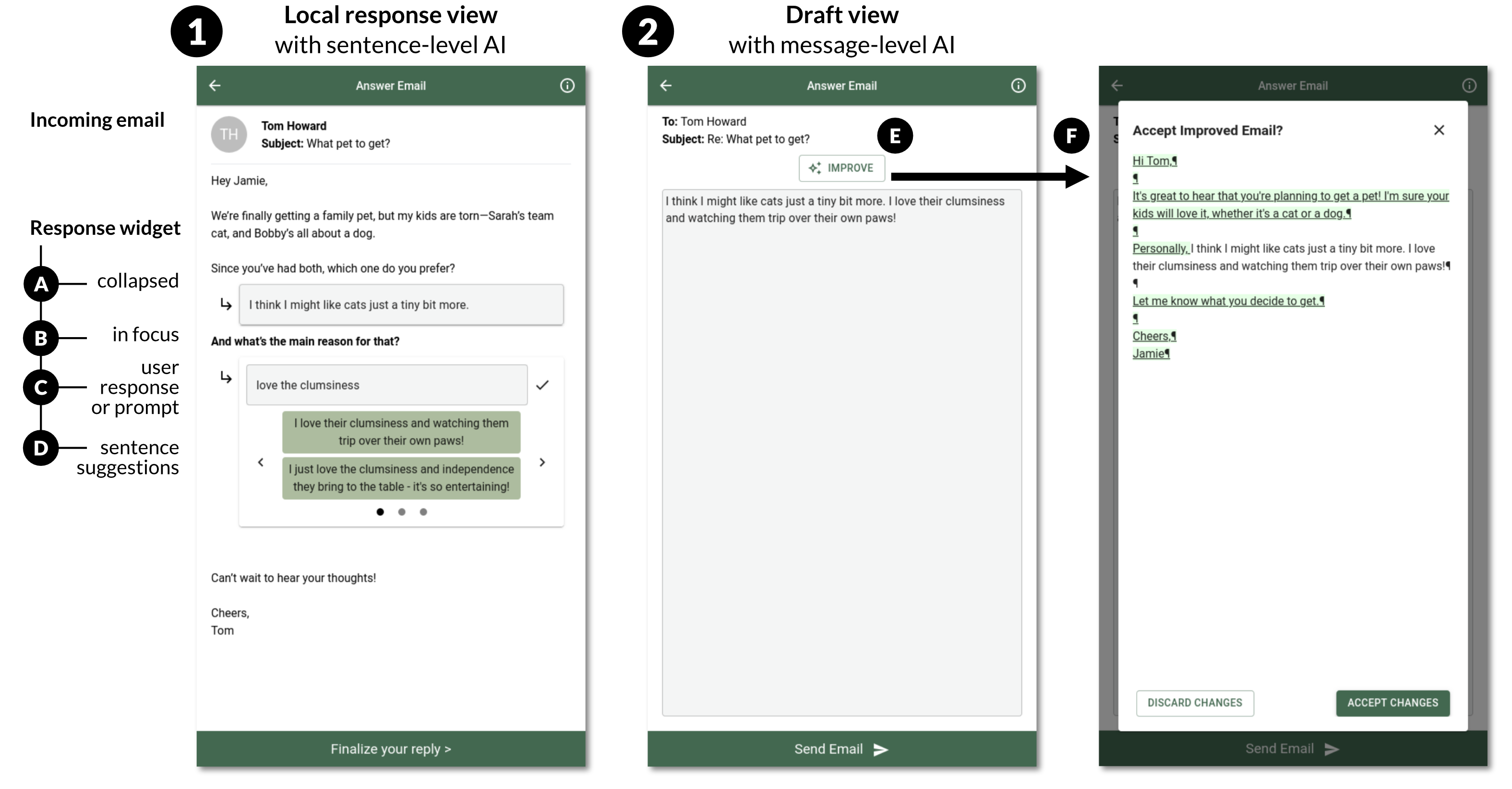}
   \caption{Replying to an email with \textit{\modeoursTXT}: \textit{(1)} In the \textit{local response view}, users can insert responses \textit{(A)} directly while reading the email. \textit{(B)} Tapping on a sentence opens a response widget, \textit{(C)} with a text box where users enter a response or a prompt that affects \textit{(D)} the sentence suggestions below. \textit{(2)} After adding local responses, users go to the \textit{draft view}, to turn their responses into a full reply email. They can do so manually and/or with the help of \textit{(E)} an AI \imppass{} feature, which generates \textit{(F)} a message-level suggestion, displayed with highlighted changes. These AI features are flexible and optional: Users can add local responses without using suggestions. They can also skip directly to the draft view, optionally enter a prompt there, and use the improvement feature to generate a full reply directly. This supports flexible workflows.}
   \label{fig:teaser}
   \Description{This figure illustrates the process of replying to an email using the "Content-Driven Local Response" feature and outlines different flexible workflows with optional AI support.
   The figure is divided into three main sections:
   Local Response View with Sentence-Level AI (Left Panel)
   A) Users can insert responses directly while reading the email.
   B) Tapping on a sentence opens a response widget, which allows users to interact with the email content.
   C) The widget includes a text box where users can enter their own response or a prompt that will influence the sentence suggestions displayed below.
   D) Below the text box, sentence-level AI-generated suggestions are provided based on the entered prompt or context of the email.
   Draft View with Message-Level AI (Middle Panel)
   After entering responses or prompts in the local response view, users can proceed to the draft view.
   (E) In the draft view, users can manually edit the draft or use an AI improvement pass feature to generate a message-level suggestion.
   (F) The AI-generated suggestion is displayed with highlighted changes, allowing users to review and accept the improvements.
   Examples of Flexible Workflows with Optional AI (Bottom Section)
   The figure outlines various workflows users can follow, ranging from full AI-assisted reply generation to fully manual drafting:
   Full reply generation: Users can go directly to the draft view and use the improvement feature to generate a complete reply.
   Partial AI-assisted reply generation: Users can draft partially with sentence-level AI support and then finalise the reply with message-level AI or manual edits.
   Fully manual: Users can choose to draft and send the email without using any AI assistance.
   This figure highlights the flexibility of the system, allowing users to choose between different levels of AI support depending on their preference or the specific requirements of the email task.}
\end{teaserfigure}

\maketitle

\begin{figure*}[t]
    \centering
    \includegraphics[width=\linewidth]{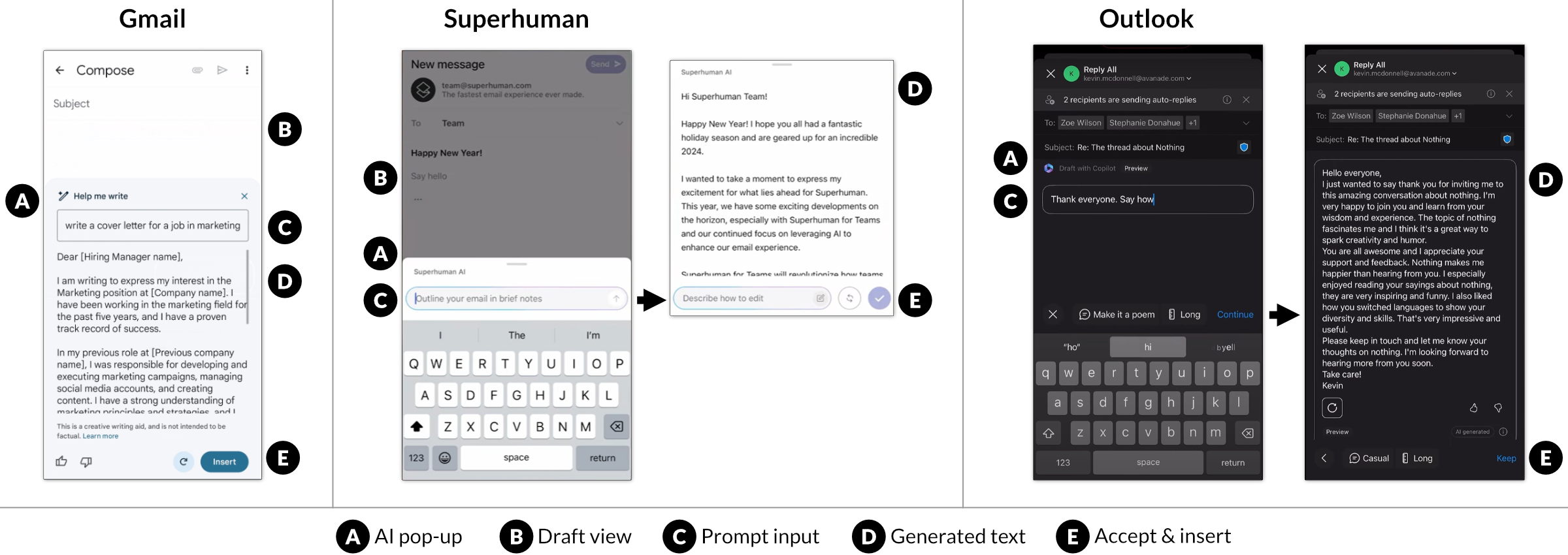}
    \caption{A commonly used UI and interaction design for reply generation in mobile email apps: \textit{(A)} Users work with AI in a pop-up, \textit{(B)} on top of the empty draft view. \textit{(C)} They can enter a prompt, \textit{(D)} view the generated reply, and \textit{(E)} accept it with a button. At the time of writing this paper, this pattern appears across Gmail (left), Superhuman (centre), and Outlook (right). It also appears in other email apps, such as Shortwave. Screenshots from: Google's website~\cite{google2024geminiwebsite}, Superhuman's YouTube channel~\cite{superhuman2024video}, The Copilot Connection's YouTube channel~\cite{copilotconnection2023video}.}
    \Description{This figure displays a commonly used UI and interaction design for reply generation in mobile email apps.
    Three UIs are shown: On the left Gmail, Superhuman in the centre, and Outlook on the right.
    Users work with AI in a pop-up window, on top of the empty draft view. They can enter a prompt in a text field, view the generated reply, and accept it with a button closeby. At the time of writing, this pattern appears across Gmail, Superhuman, and Outlook. It also appears in other email apps, such as Shortwave. Screenshots from: Google's website, Superhuman's YouTube channel, The Copilot Connection's YouTube channel.}
    \label{fig:current_products}
\end{figure*}

\section{Introduction}

Emails challenge us to stay responsive and responsible -- anytime, anywhere.
With mobile devices, people wish to stay responsive despite contextual challenges: 
They might lack the time and attention to write detailed replies quickly, especially on small on-screen keyboards and in demanding mobile contexts. 
This motivates new text suggestion features that delegate the writing of replies to the system.
\revision{For example, \textit{Smart Reply}~\cite{Kannan2016smartreply} and \textit{Smart Compose}~\cite{Chen2019smartcompose} suggest short email replies and sentence continuations, respectively. This is especially attractive in mobile contexts, where AI serves to reduce cumbersome typing on a small keyboard~\cite{Quinn2016} or to speed up the reply process~\cite{Kannan2016smartreply}. More broadly, research in AI-mediated communication (AIMC) highlights relevant conceptual dimensions~\cite{Hancock2020aimc}, such as the AI’s magnitude of involvement, optimisation goal, and its autonomy. Operationalised as concrete design decisions, for instance, message-level vs sentence-level suggestions come with different degrees of magnitude and autonomy~\cite{Fu2023sentencevsmessage}, and generally optimise for replying faster.}

\revision{Having different options matters, including those that retain high user control, because} people \revision{also} wish to stay responsible  \revision{and maintain their agency~\cite{Mieczkowski2022thesis}}. Their replies need to adequately consider the complex nature of their professional and private relationships, \revision{which currently} is hard to get right with generated text~\cite{fu2024texttoself, Liu2022aimailperception, Mieczkowski2021, Robertson2021cantreply}. This creates complex tradeoffs for users in navigating the use of AI in their mobile response workflows.

Current mobile email apps offer two approaches to involving AI: (1) sentence completions, which help with manual writing; and (2) reply generations, which seek to entirely replace the need for manual writing. 
Both come with tradeoffs:

Sentence-level suggestions, often sporadic and utility-gated to avoid showing unsuitable text~\cite{Chen2019smartcompose}, still demand considerable manual typing and do not account for situations in which users would prefer to automate the reply process further, for example, when only a short reply is needed (cf.~\cite{Kannan2016smartreply}).

In contrast, message-level support can take important decisions away from the user, pushing them into an editing role if they prefer to respond in a different way compared to what the AI has drafted. Moreover, there are several negative impacts associated with workflows in which the AI generates the first draft, including reducing self-expression and diversity~\cite{Li2024aivalue}.

Today's mobile email apps largely follow the same UI and interaction design for AI integration -- such as \textit{Gmail} with Gemini, \textit{Outlook} with Copilot, and \textit{Superhuman} and \textit{Shortwave}: When replying to an email, users decide as a first step if they want to switch to full reply generation. This brings up a pop-up on top of an empty draft view. Here, users can enter a prompt and check the generated reply, before accepting it with a button to be inserted into the draft. \revision{\cref{fig:current_products} shows this.} 

This design and workflow hides the incoming email while the user is prompting and assessing the quality of the generated reply. At the same time, the draft view is nothing but an empty background to the pop-up.
In summary, there is at least one design contradiction in the current industry standard: UI space is wasted while useful information is omitted.

We address the challenge of AI integration into mobile email replies by taking a step back to reconsider the reply workflow more holistically. This leads to our three guiding research questions:
\begin{enumerate}
    \item How might we redesign the mobile email reply UI for flexible and optional AI involvement?
    \item How do users perceive and interact with this design to reply to emails?
    \item What are the specific advantages and drawbacks of this design, and how do they differ from existing approaches?
\end{enumerate}
To address these questions, we developed a prototype web app (\cref{fig:teaser}), informed by a formative study (N=\studyOneN).
Our design not only bridges the gaps between manual writing, sentence suggestions, and message-level AI. It also supports a new reply workflow that allows users to still see the incoming email while they work on their response, despite the limited screen space of mobile devices. 

Concretely, inspired by concepts from mobile microtasking~\cite{august2020microwriting, iqbal2018playwrite}, we integrate a sentence-based response interface directly into the incoming email. We refer to this as \textit{\modeoursTXT{} (\modeours)} since it allows for responding locally within the content -- both for the user and the AI. Users tap on sentences they wish to address, which brings up a local response widget. They then reply manually or use sentence suggestions that adapt to their input.
Additionally, users can continue their reply process on the level of the whole message, with an AI \imppass{} or by manually editing their reply as a whole.
In this way, each AI feature remains optional, empowering users to dynamically build their own workflows, ranging from manual writing to fully AI-generated responses.

We evaluated our concept in a user study (N=\studyTwoN) with a functional prototype and two comparative designs (manual writing and message generation as in today's email apps). We logged interactions and assessed perception with surveys and in-app feedback.

We found that our system takes a new distinct spot in the design space between sentence-level and message-level support, reflected in participants' subjective feedback and significant differences in interaction metrics. %
People used it in varying workflows with different degrees of AI involvement, while retaining the benefits of reduced typing and errors. 

In summary, we contribute: (1) A novel concept for local sentence-based replies with optional AI support for mobile email communication, (2) its implementation in a functional prototype, and (3) insights from an evaluation with users.

In a broader perspective, our work demonstrates the potential of a new approach to integrating AI capabilities; not by adding them ``on top'' but rather by focusing on improving the UI for users' underlying workflows -- with and without AI. In this way, designers can empower users to dynamically adjust the desired degree of AI involvement. %

\section{Background \& Related Work}\label{sec:related_work}

We relate our concept and design decisions to research on emailing with AI and the impact of text generation.

\subsection{Text Suggestions for Emailing}

There are two main approaches for text suggestions for (mobile) emailing, exemplified by 
Google's \textit{Smart Compose}~\cite{Chen2019smartcompose} and \textit{Smart Reply}~\cite{Kannan2016smartreply}. \textit{Smart Compose} supports manual writing with sentence completions, while \textit{Smart Reply} generates short replies to replace manual writing. In contrast, we explore the design space in between by mixing sentence-based and message-based AI involvement. %

The study by \citet{Fu2023sentencevsmessage} compared sentence-level and message-level text suggestions. They found that suggesting whole emails put users in the role of editors, leading to faster task completion. Sentence suggestions, on the other hand, allowed people to retain a greater sense of agency and write more original content. Their call for further design exploration motivates our work: We offer sentence suggestions in one step, followed by an email-level improvement pass -- all optional. %
As \citet{Fu2023sentencevsmessage} pointed out, designing appropriate AI support for a particular use case requires identifying the ``sweet spot'' for ``suggestion units''. Based on our results, our design provides a new and distinct intermediate option beyond the two existing ``extremes'' (of pure sentence vs. message-based support). %

\subsection{AI Integration in Current Email Apps}\label{sec:related_work_current_products}

Recently, many mobile email apps have added generative AI features, including Google's \textit{Gmail} with Gemini, Microsoft's \textit{Outlook} with Copilot, and the apps by \textit{Superhuman}\footnote{\url{https://superhuman.com/}} and \textit{Shortwave}.\footnote{\url{https://www.shortwave.com/ai-email/a/}}
As shown in \cref{fig:current_products}, they share a common UI and interaction design: When replying to an email, users can open a pop-up to enter a prompt and trigger generation. The result is shown in the pop-up and can be accepted with a button. It is then inserted into the draft view for sending or editing.

This design hides the email while working with AI to respond to it. Hence, the user needs to remember all relevant information in the email while prompting and checking generations. At the same time, the draft view is an empty backdrop to the AI pop-up.
In short: UI space is wasted while useful information is omitted. 

Thus, we diagnose that AI is not optimally integrated for mobile email replies:
It is added (literally) ``on top'' of the existing UI. %
This motivates our approach to take a step back to redesign the UI with users' workflows in mind -- also those without AI.

\subsection{Learning from Mobile Microtasking}
Microtasking principles provide inspiration for our redesign: Limited screen space~\cite{raptis2013phonesize} and frequent interruptions~\cite{leiva2012backtoapp, oulasvirta2005bursts}, along with input method constraints~\cite{palin2019mobiletyping}, make it difficult to work on text documents on mobile touchscreen devices.
This motivates research on mobile microtasking~\cite{cheng2015breakitdown} to break down larger tasks, such as writing a document, into manageable chunks that can be completed in short, opportune moments on the go, one at a time~\cite{august2020microwriting, iqbal2018playwrite}.

The success of microtasking for mobile writing has not yet been brought to email. Possibly, emailing itself is seen as a microtask, as checking mails is fast (cf.~\cite{bao2011phoneuse, oulasvirta2012habits, zheng2024waitingtime}) and a common productive task while waiting, as shown by \citet{zheng2024waitingtime}. However, productive tasks were less likely with only a phone, highlighting the gap between checking and responding. \revision{It is also likely that short emails do not require microtasking. That said, emailing ``on the go'' may involve interruptions~\cite{oulasvirta2005bursts}, which motivate keeping relevant context information in view, even for shorter emails.}

\revision{Together,} this motivates our redesign of the response workflow with microtasking in mind: Traditionally, users read an email and then type into a separate, empty text box. Instead, we allow users to write short responses in boxes shown directly within the incoming email text.

This acknowledges that writing depends on local context~\cite{salehi2017communicatecontext}, which should be shown in microtasking todos~\cite{august2020microwriting, iqbal2018playwrite}. %
Unlike the related work, we do not assume that users go to a desktop computer to finalise their text. Thus, we add a second UI view where users can finalise their reply, for example by connecting partial responses coherently. Our design offers optional AI support for this.

\subsection{Design Factors for Mobile Emailing with AI}

\citet{Park2019inboxneedfinding} revealed user needs for inboxes that also motivate our design for responding to a single email: Support for attention management, a presentation closer to the lighter interaction of messaging UIs, and breaking up longer emails. In line with this, our concept allows users to respond to longer emails in more lightweight chunks by writing local responses to parts of an email.
Related, \textit{Rambler} by \citet{lin2024rambler} provides text cells to break up text while drafting and editing with text-to-speech input and AI features on tablets. 

We further use insights from related work on the number and type of suggestions~\cite{Buschek2021chi, Dang2023diegetic}: Our design offers multiple suggestions that users can refine. Unlike the related work~\cite{Dang2023diegetic}, users can use the \textit{same} UI element (text box) to either type their own response or enter a prompt to refine the suggestions. This supports dynamic user decisions about AI involvement without the need for separate UI elements for writing and prompting, taking into account the limited space of mobile screens. %

To support emailing for people with dyslexia, \textit{LaMPost}~ \cite{Goodman2022lampost} offers AI rewriting and subject line generation. Similar to \textit{Wordcraft}~\cite{Yuan2022}, which is not email-specific, the design allows users to select text with the mouse to then trigger AI support. We bring this pattern of ``selection for local AI support'' to mobile touchscreens, with two adaptations: (1) selecting whole sentences rather than character-level selection, to account for the limited accuracy of touch; and (2) displaying suggestions directly below the selection, rather than in a sidebar, which would not fit on a mobile screen.

\subsection{Impact and Perception of Generated Text}

Work in AI-mediated communication (AIMC~\cite{Hancock2020aimc}) reveals the complexity of social issues involved in writing emails with AI-generated text: \citet{Robertson2021cantreply} studied how people edit suggested email texts and found that suggestions can be inappropriate, with respect to intersecting factors, including relationship types and cultural norms. 
Related, \citet{lucy2024onesizefitsall} found that the desired appropriate system behaviour in this context is highly individual. 

Focusing on the receivers instead, \citet{Liu2022aimailperception} studied perception and trust towards emails, including the perceived degrees of AI involvement and interpersonal emphasis. Overall, trust in writers decreased when told about AI involvement, yet the findings also suggested differences between people's statements about AI and actual reactions, indicating an ongoing development of perspectives.

Together, the above results show that it is difficult to provide text generations for emails that entirely avoid problematic suggestions. Thus, \citet{Robertson2021cantreply} called for more studies on designs that explore ``Giving users more control'' beyond one-shot generation, based on their finding that ``a large fraction of [suggested] replies were amended''. This directly motivates our design, which allows users to control -- on a sentence-level -- in the incoming email (1) what to respond to at all, (2) whether to involve AI here, and (3) if so, what additional information to give the AI for that involvement (e.g. by entering keywords to refine the suggestions). %

Related, \citet{Li2024aivalue} studied people’s choices of AI assistance in controlled argumentative and creative writing tasks at the desktop. They report on desirable and undesirable changes on several metrics, depending on whether the ``primary'' writer is the human or the AI. %
All identified negative impacts arise in the ``AI-primary'' case, where ChatGPT provides a first draft. %
Thus, their findings  motivate our exploration beyond such ``AI-primary'' designs for emails. %
Our results extend the bigger picture to mobile text entry and emailing; for example, in line with their results, our proposed design retains more content diversity than the ``AI-primary'' baseline.

\section{Concept Development}
We introduce the concept of \modeourstxt. At a high level, we take inspiration from mobile microtasking UIs and integrate a local response interface directly into the incoming email. Overall, our final design is situated in the  unexplored space in between sentence-level and message-level approaches for AI support.

\begin{figure*}
    \centering
    \includegraphics[width=0.5\textwidth]{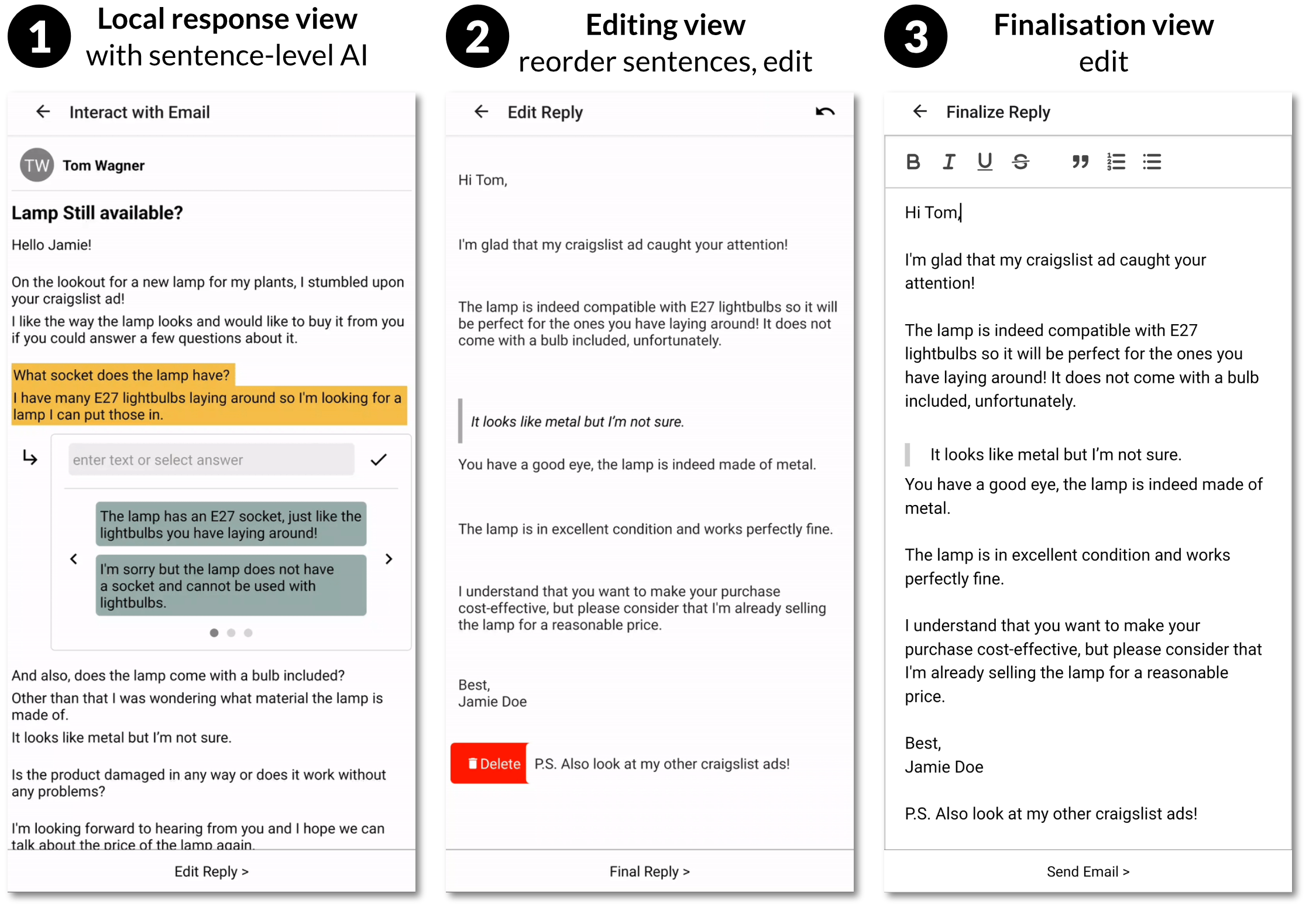}
    \caption{Replying to an email with our first prototype: \textit{(1)} In the first screen, users insert responses directly while reading the email. Tapping on a sentence opens a response widget, with a text box where users enter a response or a prompt that affects the sentence suggestions below. \textit{(2)} After adding local responses, users can edit their reply on a second screen, by reordering paragraphs via drag-and-drop, by deleting paragraphs via swiping left-to-right, and by manual editing via the integrated keyboard.  \textit{(3)} On the third screen, users can finalise the reply before sending it.}
    \Description{This Figure shows how to reply to an email with our first prototype: On the left it shows the first screen, where users insert responses directly while reading the email. Tapping on a sentence opens a response widget, with a text box where users enter a response or a prompt that affects the sentence suggestions below. After adding local responses, users can edit their reply on a second screen, which is shown in the centre of the figure. By reordering paragraphs via drag-and-drop, swiping left-to-right to delete paragraphs, and via manually editing using the integrated keyboard user can edit their reply. On the third screen, which is shown on the right, users to finalise the reply, before sending it.
    }
    \label{fig:BA_prototype}
\end{figure*}

\subsection{Design Goals and Rationale} 
With insights from the literature (\cref{sec:related_work}), we designed our system with several goals in mind. For each goal, we briefly mention our approach here, with more details on its final realisation in \cref{sec:implementation}.
\begin{enumerate}[leftmargin=*]
    \item \textbf{Human decides, AI supports:}
    \label{dg:humandecides}
    The user should be able to make all important decisions, while AI supports these. 
    In our design, the user selects the sentences they want to reply to. 
    The system then suggests response sentences, designed to offer a mix of positive, neutral, and negative responses. 
    \item \textbf{The user stays in control:}
    \label{dg:control}
    The user should stay in control of their reply. 
    The AI should not make unnoticed or unwanted adjustments. 
    Our system does not change text unless requested and the user can always edit the reply before sending it.
    \item \textbf{Support mobile microtasking:}
    \label{dg:microtasking}
    The user should be able to leverage microtasking principles for mobile email replies. %
    Our design provides the surrounding email as context while entering reply text and thus shifts from recall to recognition by eliminating the need to remember the email or scroll back to it.%
    \item \textbf{Support diverse workflows -- with and without AI:}
    \label{dg:workflows}
    In each situation, users should be able to answer in their preferred way.
    Our skippable components offer flexibility.
    Even without AI, users get supportive microtasking structure.
    Conversely, users can choose to rely on AI text to respond fast, with little typing. %
\end{enumerate}

\subsection{First Prototype}
We implemented the concept as an Android app with React Native.\footnote{\url{https://reactnative.dev/}}
At this point, the prototype had our sentence-based mode as shown in \cref{fig:BA_prototype}, with three screens.
The first screen (\cref{fig:BA_prototype} left) showed the email for users to select sentences via touch. Each selection triggered a card view that displayed AI suggestions and a text box for entering text (as a manual response or as a prompt to refine the suggestions).
The LLM always generated two positive, two negative, and two neutral answering options.
One positive and one negative suggestion were shown on the first page, if possible. Users could click on the arrows on the sides to access the others.
A second screen (\cref{fig:BA_prototype} centre) supported manual editing of paragraphs, inserting new ones and/or reordering them via drag and drop.
Finally, a third screen showed the result in a standard text editor view for a last check and final edits, if necessary (\cref{fig:BA_prototype} right). %

\subsection{Formative Study}\label{sec:formative_study}
We conducted a first study to understand how users perceive and interact with our concept, and to inform a design iteration. We recruited 17 participants (1 female, 15 male, 1 preferred not to disclose) from our university network. The study followed our institute's regulations, including information on goals, process, data recording, opt-out and consent.

Participants used our prototype on their own phones. They received a tutorial beforehand. The app did not integrate with actual email accounts to preserve privacy. Instead, it simulated to receive two emails per day, for five days. \revision{These emails were quite long, ranging from 140 to 491 words per email (median: 227), to allow participants to test the prototype extensively.} People were asked to respond to the emails in a reasonable time frame. 

After each reply, the app displayed three 5-point Likert items (``The AI tool was helpful'', ``The AI tool helped me reply to the email faster'', ``The AI tool helped me write a better reply'') and space for open feedback. Following the final email, participants completed a questionnaire about their overall experience and demographics.

\subsection{Results}\label{sec:formative_study_results}
The median time of engaging with each email task in our study was 6.9 minutes, including the time taken to enter feedback. 
People accepted 9.17 suggestions per email.
Nearly \pct{80} of accepted suggestions were accepted without making use of the text input for refinement. %
In \pct{30} of emails, participants composed the email entirely with suggestions without edits afterwards.
When they indeed made edits, the most common ones we identified through manual coding were the following: On the first screen, they removed text (25 instances), added information (11), changed details (8), shortened text (6), and added salutations (5) or closing statements (5). 
On the second screen, they reordered or merged sentences and paragraphs (35), shortened text (10), and changed minor details (8). 
They made similar edits on the third screen. %

The Likert results (\cref{fig:likert_items_formative_study} \revision{in \cref{sec:appendix_extra_figures}}) indicated that participants found the AI to be helpful and that it supported them in writing the replies. They felt in control of the email content and found the suggestions to make sense and not be distracting. They generally agreed that the approach helped them remember to address all parts of an email. However, they were more divided on whether they overall preferred the step-by-step process or the traditional one-step approach for replying to emails.

Open in-app feedback was provided by 14 out of 17 participants. Positive aspects mentioned there and in the final questionnaire included ease of use, faster replies, the quality and inspirational potential of AI suggestions, and an overall improved workflow. 

Negative aspects included slow AI response times, quality of suggestions (e.g. too short or not aligning with their input), minor bugs (e.g. failure to load suggestions), and the number of steps (e.g. some suggested to merge the last two screens into one).  

The final questionnaire asked people to reflect on their workflow with our app. They reported different strategies, such as generating custom replies with keywords, reading the entire email before replying, or reviewing generated suggestions first. Some manually merged or adjusted AI-generated text, while others used it as is. 
The final questionnaire also included the System Usability Scale (SUS)~\cite{brooke1996sus}. The mean score was 78.67 (``very good'', \revision{details in \cref{fig:sus_items_formative_study} in \cref{sec:appendix_extra_figures}}).

\subsection{Prototype Iteration}
\label{ssec:protiter}
In summary, the findings from this first study indicated that participants appreciated our concept as it helped them to write fast and high-quality replies with AI, while still feeling in control. It also revealed individual approaches when answering emails and interacting with the suggestions. Based on the study insights, we made the following concrete changes to our prototype:

\begin{itemize}[leftmargin=*]
    \itemsep.2em
    \item \textit{Reduced number of steps:} We removed the second screen (\cref{fig:BA_prototype} centre) %
    and direly offered the third one for free text editing and finalising. Some participants suggested this and they overall made very similar edits across these two screens.
    \item \textit{Added optional improvement pass:} %
    \revision{We added an ``improve email'' button to the final screen to better support users' varying strategies and preferences for answering emails with AI. We observed that some participants manually edited their emails to create transitions between individual paragraphs generated on the first screen. The \imppass{} feature automates this process, adding missing greetings, sign-offs, and correcting grammar and spelling (see \cref{subsec:appendix_improve_email_prompt} for the used prompt)}. %
    \item \textit{Faster suggestion generation:} We switched from GPT-3.5 Turbo\footnote{\url{https://platform.openai.com/docs/models}} to Llama 3 8B Instruct\footnote{\url{https://huggingface.co/meta-llama/Meta-Llama-3-8B-Instruct}}\cite{llama3modelcard}, which we hosted locally to reduce latency and avoid request failures. \revision{While we did not conduct an in-depth evaluation of the models, we compared a set of generations qualitatively and found that the output quality was similar for our use case. Related, we envision that real-world applications could rely on smaller models that can be executed on devices locally to avoid the need to send private email content to a model provider.}%
    \item \textit{Refined suggestions:} We refined our prompting templates to improve suggestions, even with the smaller model. Our new prompts included more context, i.e. all sentence-level replies that were already given up to this point.  
    \item \textit{\revision{Port to a React web app}:} \revision{We ported the app from React Native to a React web app and optimised it. This eased access for participants, as running a React Native prototype required several steps for setup. Instead, a React web app can be accessed via a web browser on any smartphone.}%
\end{itemize}

In the rest of the paper, we always refer to the improved prototype. We next describe it in detail.

\begin{figure*}[t]
    \centering
    \includegraphics[width=\linewidth]{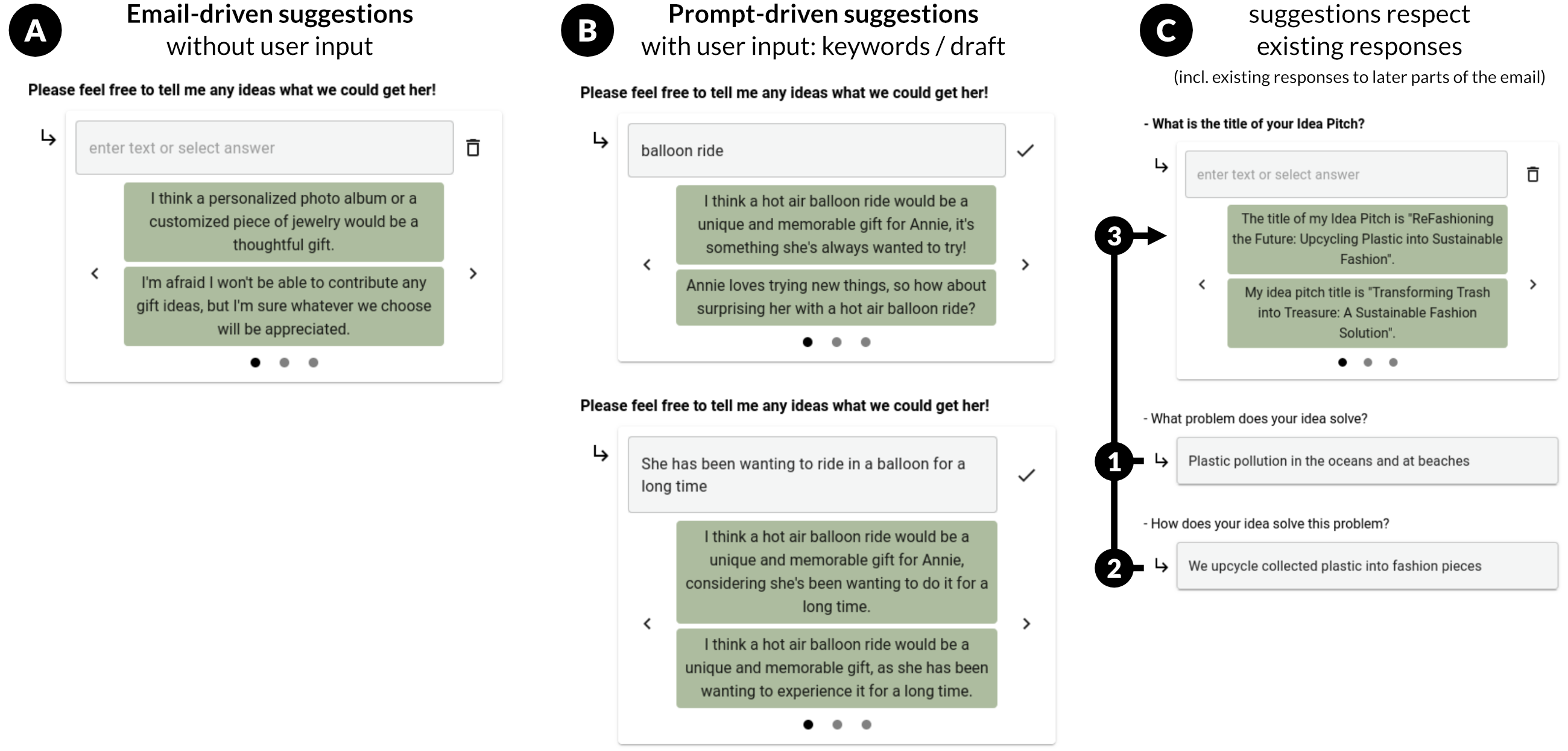}
    \caption{The text suggestions in the local response widget are flexible:  \textit{(A)} Users get suggestions without any input.  \textit{(B)} \revision{Suggestions} can be adapted and refined by entering text, for example keywords or a draft snippet. In all cases, suggestions are generated with an LLM based on the text of the incoming email and all local responses that the user has entered so far, \revision{even if responses have been} added to later parts of the email \revision{first}. \revision{In \textit{(C)}, for example, the suggested title of the idea pitch is generated based on the information about the project that the user has already entered in local responses below.} Note that suggestions are paginated, with three pages of two suggestions each.}
    \Description{
    This figure shows screenshots of our user interface displaying an incoming email. The figure is divided into three sections:
    A) Left Section: Email-driven Suggestions without User Input
    The selected sentence in the incoming email reads: "Please feel free to tell me any ideas what we could get her!"
    Below the email, there is an empty text field for optional user input.
    Two AI-generated responses are shown beneath the text field: one suggests a piece of jewellery as a gift, while the other does not offer any ideas.
    B) Centre Section: Prompt-driven Suggestions with User Input
    The same sentence from the email is selected: "Please feel free to tell me any ideas what we could get her!"
    This time, the keywords "balloon ride" are entered into the text field below.
    As a result, the AI-generated suggestions include the idea of a balloon ride in both proposed texts.
    C) Right Section: AI Suggestions Respecting Existing Responses
    One sentence from the incoming email is selected, and the text field below remains empty.
    The AI-generated responses incorporate information from existing responses elsewhere in the email.
    Additionally, all AI suggestions are paginated, with three pages of two suggestions each, as indicated by arrows next to the suggestions.}
    \label{fig:local_response_prompting}
\end{figure*}

\section{Implementation}
\label{sec:implementation}
We implemented a frontend and backend, which preprocessed emails, logged user data, and generated responses. 

\subsection{Frontend}
We implemented our web app with the React\footnote{\url{https://react.dev/}} framework.

\subsubsection{Display of the Incoming Email}
This view matches standard mobile email UIs: It includes the sender's name and picture, the email subject, and the main text body (\cref{fig:teaser} left). 
The user can select sentences in the incoming email by tapping on them (cf. design goal \ref{dg:humandecides}: \revision{Human decides, AI supports}; and goal \ref{dg:microtasking}: \revision{Support mobile microtasking}). This opens the local response widget (\cref{sec:impl_local_response_widget}).
The ``Finalize Reply'' button at the bottom of the UI switches to the next screen (\cref{fig:teaser} centre), which we describe in \cref{sec:impl_finalize}.
In accordance with design goal \ref{dg:workflows} \revision{(Support diverse workflows -- with and without AI)}, no interaction with any sentence or AI feature is required before proceeding to this next screen.

\subsubsection{Local Response Widget}\label{sec:impl_local_response_widget}

This UI widget is inserted into the email text below the user's selected sentence. It comprises of a text field (\cref{fig:teaser} C) and a paginated card view that shows text suggestions (\cref{fig:teaser} D). In the text field, users can enter both manual responses or prompts to refine these suggestions (\cref{fig:local_response_prompting}). 

Concretely, the widget offers six suggestions (2 positive, 2 negative, 2 neutral), showing two at once. The system aims to show one positive and one negative response on the first page, if possible. \revision{This was motivated by findings on positivity bias in AI-generated communication text~\cite{Mieczkowski2021} and to increase the chance of offering a response option fitting to the user's intent (cf.~\cite{Kannan2016smartreply}).} \revision{We realised this by prompting the LLM to do so (see \cref{sec:appendix_sentence_without_input_prompt}). Concretely, the variable ``attribute'' in the prompt template was replaced with \textit{accepting}, \textit{declining}, and \textit{neutral} to generate varying suggestions. In our tests, we observed that this simple prompting approach worked well and that it did not negatively impact generated suggestions in cases where these attributes do not apply (e.g. our ``cat'' example in \cref{fig:teaser}D).} %
Users can navigate through suggestions using the adjacent arrow buttons. They can accept a suggestion by tapping on it.

The widget has two states -- open and collapsed (\cref{fig:teaser} A, B): 
It is collapsed by tapping the currently selected sentence again, by selecting a different sentence, by accepting a suggestion, or by clicking on the check mark in the top right corner. When the text field is empty, the check mark transforms into a trash icon to delete the local reply. Multiple widgets can be in the collapsed state throughout the email but only one widget at a time can be open and in focus. %
A widget's text is shown in the collapsed state. This allows users to keep track of all their local replies so far. Tapping on a collapsed widget opens it again for further editing.

\subsubsection{Finalising the Reply}\label{sec:impl_finalize}
This view  (\cref{fig:teaser} centre) shows the current state of the reply after the local response step. That is, it displays any text entered in response to individually selected sentences together in a single text field.

Users can manually adjust this text and/or tap the ``Improve'' button to request the AI to enhance the email. 
This \imppass{} feature is realised with a prompt \revision{(see \cref{subsec:appendix_improve_email_prompt})} to the underlying LLM to correct spelling and grammar, refine wording, and add missing salutations or regards while adhering to both the incoming email's content and the existing reply text. 

If no text is entered first, the ``Improve'' button acts as message-level support, generating a reply based on the incoming email's text and the current input on this screen. For example, a user could skip the local response and enter a prompt here, effectively realising a message generation workflow similar to the industry default (\cref{sec:related_work_current_products}). This flexibility contributes to our design goal \ref{dg:workflows} \revision{(Support diverse workflows – with and without AI).}

When the user is satisfied with their reply, the email can be sent by tapping the ``Send Email'' button at the bottom of this screen.

\subsubsection{Improved Email Pop-up}\label{sec:impl_imppass}
The \imppass{} feature does not change the user's text directly, in line with our design goal \ref{dg:control} \revision{(The user stays in control).}
Instead, the new text is shown in a pop-up view with formatting familiar from ``track changes'' in text editors (\cref{fig:teaser} right). 
Users can approve these changes, which updates the text, or discard them (cf. design goal \ref{dg:humandecides}: \revision{Human decides, AI supports}.) Further editing after acceptance and/or requesting improvements repeatedly is possible.

\subsection{Backend}
Our prototype's backend has three purposes: 
(1) It \textit{hosts the web app} on a Next.js server. 
(2) It provides \textit{email preprocessing}, which handles tasks such as sentence-splitting and making API calls to the LLM. 
(3) It \textit{hosts the LLM}. 

We experimented with several models and APIs and discussed factors such as latency, stability of service, and subjective response quality in meetings with all authors. Based on this, we used the Llama 3 8B Instruct \cite{llama3modelcard} model for the main study. 

Similarly, we iterated over several prompting approaches for the text generation features. Overall, this resulted in a few-shot approach, providing the model with several input-output examples to generate fitting responses. 
As an overview, we use the following prompt templates (details in \cref{sec:appendix_prompts}):

\paragraph{Sentence-level support, without user input:}
We prompted six suggestions for the sentence selected in the email (2 positive, 2 neutral, 2 negative). This balanced the options, following related work~\cite{Kannan2016smartreply}, as the LLM favoured positive responses in our tests.

\paragraph{Sentence-level support, with user input:} 
This was identical to the above case but now it included the user's input in addition to the selected sentence. %
We emphasised alignment with the user's sentiment (e.g. no negative suggestions if the user had entered ``yes'').

\paragraph{Message-level support:}
We prompted the LLM to answer to the whole email, also by taking into account any current user input, if available. A variation of this was also used for the \imppass{} feature (\cref{sec:impl_imppass}). That prompt emphasised improving the current state of the reply while closely adhering to the information provided by the user. %

\section{Method}
\revision{For the main study, we switched from the field study design of the formative study to a more controlled web-based experiment. Our motivation was to scale the study, to compare interaction designs quantitatively, and to introduce task briefings that would allow us to evaluate if people accept even unfitting suggestions for convenience in the study. Thus, we} conducted a within-subject user study with our \revision{iterated} prototype.
The independent variable \ivmode{} had three levels: \modeoursTxt{} (\modeours) -- our proposed design (\cref{sec:implementation}); \modemailtxt{} (\modemail) -- a one-prompt reply generation design close to currently available UIs (\cref{sec:related_work_current_products}); and writing without any AI features (\modemanual). As dependent variables, we logged interaction metrics and collected subjective feedback via questionnaires.

\subsection{Apparatus}

\subsubsection{Web App}
For \modeours, we hosted our prototype (\cref{sec:implementation}) as a web app, with added study information, study logic, and logging.
We integrated a screen for briefings (\cref{sec:method_emails_briefings}) before each email task and one with four Likert items (\cref{sec:procedure_email_tasks}) after each email task (\cref{fig:briefing_and_feedback} \revision{in \cref{sec:appendix_extra_figures}}).
We used a custom study framework to manage counterbalancing and the flow from consent information to prototypes to surveys.

\subsubsection{Comparative Designs}
We implemented two alternatives for the study: For \modemanual, the app only showed a typical drafting view (\cref{fig:baseline_uis_manual} \revision{in \cref{sec:appendix_extra_figures}}). For \modemail, it was designed similar to the typical UI pattern shown in \cref{fig:current_products} -- it offered a text field for (optional) prompting and displayed the generated text (\cref{fig:baseline_uis_msg} \revision{in \cref{sec:appendix_extra_figures}}). Accepting the text with a button inserted it into a draft view for further editing or sending. Rejecting it allowed users to refine their prompt and generate a new draft.

\subsubsection{Incoming Emails and Reply Briefings}\label{sec:method_emails_briefings}
We prepared nine emails, covering an idea pitch contest, a high school reunion, a sales offer, a lunch meeting, a marketing slogan, proofreading for a friend, a sales report deadline, server access, and a gift idea for a retiring coworker. This set was motivated to cover various plausible email topics, with and without (multiple) questions. It also covered various emails lengths, \revision{ranging from 24 to 155 words (median 57).}%

We also prepared a reply briefing for each email. It provided information relevant for \textit{what} to answer, without specifying \textit{how} to write (e.g. tone, structure). For example, for the high school reunion email, the briefing specified that the user was unavailable on a certain date.
These briefings were \textit{not} given to the LLM, which would have simulated unrealistic ``mindreading''. In contrast, our motivation for the briefings was to assess to what extent participants might accept unfitting suggestions for convenience. %
Moreover, the briefings mimic an email workflow where some information is readily available while details may need to be retrieved. For instance, in the high school example, reading the briefing could be seen as similar to checking a calendar app.

\subsection{Participants}
\label{sec:participants}
We recruited 162 participants through the online platform Prolific.\footnote{\url{https://www.prolific.com/}} We excluded 36 participants from our analysis because they either did not complete all tasks (18 participants) or the logs indicated that a technical issue had occurred (18 participants). Our analyses are based on the remaining \studyTwoN{} participants (83 male, 40 female, 1 non-binary, 2 preferred not to disclose). 
Their age ranged from 18 to 72 years (median 32). %
All were proficient in English (91.3\% native speakers).
Their occupations included both professions where frequent email usage is expected (e.g. IT consultant, project manager) and others (e.g. gardener, waiter).
Participants were compensated with about \pounds 10 per hour.

Most participants reported to answer emails at least once a day (\pct{48.41}) or even more than 10 times a day (\pct{21.43}).
Another \pct{16.67} answer emails at least once a week, \pct{7.94} less than once a week, and \pct{5.56} less than once a month.

Most participants (\pct{85.71}) use their smartphone for answering emails. Many also use a laptop (\pct{82.54}) or desktop computer (\pct{53.97}). Some also use a tablet (\pct{19.05}).
They answer emails at home (\pct{84.92}), at the office (\pct{69.84}), and on the go (\pct{53.17}).
Many answer emails for business (\pct{84.92}) and in a private context (\pct{58.73}).

Only \pct{14.29} reported no previous experience with AI.
Many have used ChatGPT (\pct{72.22}). Many have experience with auto-correction (\pct{51.59}), some also with auto-completion (\pct{26.98}), with word or sentence suggestions (\pct{21.42}), and with Smart Reply (\pct{15.87}).
\revision{An overview of all questions and answer options can be found in \cref{sec:appendix_questionnaires}}.

\subsection{Procedure}\label{sec:procedure}
The study was conducted remotely on participants' own smartphones. Access via Prolific was restricted to one person at a time to balance the load for our server.
The sessions were scheduled for 45 minutes and structured as follows:

\subsubsection{Study Intro}
An introduction page explained the study, including information about GDPR compliance, privacy, data collection, and informed consent, in accordance with our institutional regulations. 
In addition, whenever encountering a UI for the first time, our study framework showed an explanation of its features.

\subsubsection{Email Answering Task}\label{sec:procedure_email_tasks}
Participants were asked to reply to nine given emails, three per \ivmode{} (\modeours, \modemail, \modemanual).
We counterbalanced the email topics and also the order of the UIs with a Latin square design to address potential learning or fatigue effects.

For each email task, the briefing (\cref{sec:method_emails_briefings}) was shown at the start and could be accessed again anytime via the information button in the top right corner of the app (\cref{fig:teaser}).
Participants were instructed to ``consider the information in the briefing for answering the email[s]''.

After submitting each email, participants rated four Likert items: ``The app interface was helpful'', ``The app interface helped me reply to the email quickly'', ``The app interface helped me write a good reply'', and ``I was in control of the content of my reply''. 
They could share comments in a text field.

\subsubsection{Final Questionnaire}
This questionnaire was displayed after the final email task.
Participants provided demographics, selected their favourite UI mode, and explained their choice.
They could also leave comments, questions, and feedback. %

\subsection{Qualitative Analysis}
Here we describe our approach to coding open feedback and analysing email texts.

\subsubsection{Coding of Open Feedback}
We followed Grounded Theory~\cite{corbin1990basics} to analyse participants' open feedback. 
In the open coding round, two researchers independently reviewed the data, identifying and labelling sentences that represent specific ideas and principles. 
We then refined these initial codes by merging and clustering related ones, forming (sub-)categories during an axial coding round. 
Our research team discussed emerging themes, leading to synthesised, overarching labels for the clusters and, in some cases, further split categories to capture more nuanced insights from the feedback.
We repeated this process until reaching consensus.

\subsubsection{Analysis of the Email Replies}
\label{sec:quality_m}
Assessing email quality is complex and subjective, as known from studies on people's preferences (cf. \cite{Liu2022aimailperception, Robertson2021cantreply}).
Therefore, we use multiple  quality indicators: 
Formal indicators~\cite{reeves2008emailover50, lewi_jones2014email} include the presence of a \textit{salutation} and a \textit{closing statement} (\cref{sec:results_structure}), and proper \textit{spelling and grammar} (\cref{sec:results_errors}).
We also analysed \textit{briefing conformity}, that is, we checked whether replies covered the key information provided in the briefings (\cref{sec:results_briefing}).
Finally, we share our subjective impressions (\cref{sec:result_quality}).

We employed Binary Coding~\cite{miles2013qualitative} to assess the briefing conformity as well as the formal indicators, except spelling and grammar, which we checked using the language-tool-python\footnote{\url{https://pypi.org/project/language-tool-python/}} library. 
Two researchers coded all email replies independently, assigning a ``1'' if the email met the criteria and a ``0'' if it did not. 
Ambiguous emails were flagged for further review. 
In a second round, these were re-evaluated by the research team until reaching consensus.

\subsection{\revision{Statistical Analysis}}\label{sec:appendix_sigtest}

\revision{To declutter our following report,} \cref{tab:lmm_overview} and \cref{tab:lmm_overview2} summarise the statistical analyses \revision{and results} referred to throughout \cref{sec:results}.
We computed (generalised) linear mixed-effects models (LMMs) in R~\cite{R2020}, using the packages \textit{lme4}~\cite{Bates2015} and \textit{lmerTest}~\cite{Kuznetsova2017}. These models accounted for the individual differences between participants and for differences between the incoming emails via random intercepts. 

The models' fixed effects were \ivmode{} and whether the \imppass{} feature was used in \modeours. 
For the model for briefing conformity, we additionally included a predictor for whether the reply was generated without user input, such that the result was generated fully by the LLM based on the incoming email only. In \modemail, this is done by not entering a prompt for the reply generation. In \modeours, this is done by not providing any local response (manual or suggestion) on screen 1, before using the \imppass{} feature on screen 2. 
Pairwise comparisons were computed with the \textit{emmeans} package with Bonferroni-Holm correction.

For the Likert data, we used rank-aligned repeated measures ANOVA  (ART)~\cite{wobbrock2011art} and ART-C contrasts with Bonferroni-Holm correction for the follow-up analysis~\cite{elkin2021artc}.

We report significance at p~<~0.05.

\begin{table*}[t!]
\centering
\footnotesize
\newcolumntype{L}{>{\raggedright\arraybackslash}X}
\newcolumntype{P}[1]{>{\raggedright\arraybackslash}p{#1}}
\renewcommand{\arraystretch}{1.4}
\setlength{\tabcolsep}{4pt}
\begin{tabularx}{\linewidth}{lP{2.75em}P{5em}P{22em}P{7em}L}
\toprule
    &
    \textbf{Section} &
    \textbf{Aspect}\newline and model &
    \textbf{Predictors} (baseline: \modemanual) &
    \textbf{Pairwise comparisons} &
    \textbf{Takeaways in words}\newline(only considering sig. results) \\ \midrule
1 &
    \ref{sec:results_time} 
    &
    Completion time\medskip\newline
    \textit{LMM on seconds}
    &
    \modeours{}	$\downarrow$ \newline 
    \deemph{(\lmmci{-4.34}{12.00}{-27.89}{19.21}{=.718})}\medskip\newline 
    \modemail{} $\downarrow^*$ \newline 
    \deemph{(\lmmci{-70.05}{7.72}{-85.20}{-54.90}{<.0001})}\medskip\newline 
    \Imppass{} feature used $\downarrow$ \newline 
    \deemph{(\lmmci{-15.18}{13.23}{-41.14}{10.79}{=.252})}
    &
    \modeours{} vs \modemanual{} \deemph{(\posthoc{-4.34}{=.718})} \medskip\newline 
    \modemail{} vs \modemanual{} \deemph{(\posthoc{-70.05}{<.0001})} \medskip\newline 
    \modeours{} vs \modemail{} \deemph{(\posthoc{65.71}{<.0001})}
    &
    People finished replying faster with \modemailtxt{} than without AI (by 70 seconds on average). \modemailTxt{} was also faster than \modeourstxt{} (by 66 seconds on average).
    \\
    \midrule 
2 &
    \ref{sec:results_speed}
    &
    Writing speed\medskip\newline
    \textit{LMM on characters per second}
    &
    \modeours{} $\uparrow$ \newline 
    \deemph{(\lmmci{.61}{.56}{-.49}{1.71}{=.278})}\medskip\newline 
    \modemail{} $\uparrow^*$ \newline 
    \deemph{(\lmmci{5.16}{.37}{4.43}{5.88}{<.0001})}\medskip\newline 
    \Imppass{} feature used $\uparrow^*$ \newline 
    \deemph{(\lmmci{2.48}{.61}{1.29}{3.68}{<.0001})}
    &
    \modeours{} vs \modemanual{} \deemph{(\posthoc{.61}{=.278})} \medskip\newline 
    \modemail{} vs \modemanual{} \deemph{(\posthoc{5.16}{<.0001})} \medskip\newline 
    \modeours{} vs \modemail{} \deemph{(\posthoc{-4.55}{<.0001})}
    &
    People produced more characters per second with \modemailtxt{} (5.2 chars more per s) and if they used the \imppass{} feature in \modeourstxt{} (2.5 chars more per s).
    \\
    \midrule
3 &
    \ref{sec:results_keystrokes}
    &
    Manual typing\medskip\newline
    \textit{GLMM (Poisson) on keystroke counts}
    &
    \modeours{}	$\downarrow^*$ \newline 
    \deemph{(\lmmci{-.65}{.009}{-.66}{-.63}{<.0001})}\medskip\newline
    \modemail{}	$\downarrow^*$ \newline 
    \deemph{(\lmmci{-.86}{.005}{-.87}{-.85}{<.0001})}\medskip\newline 
    \Imppass{} feature used $\downarrow^*$ \newline 
    \deemph{(\lmmci{-.06}{.011}{-.08}{-.04}{<.0001})}
    &
    \modeours{} vs \modemanual{} \deemph{(\posthoc{-.65}{<.0001})} \medskip\newline 
    \modemail{} vs \modemanual{} \deemph{(\posthoc{-.86}{<.0001})} \medskip\newline 
    \modeours{} vs \modemail{} \deemph{(\posthoc{.21}{<.0001})}
    &
    People needed fewer keystrokes with AI than without it; concretely, even fewer with \modemailtxt{} (\pct{58} decrease) than with \modeourstxt{} (\pct{48} decrease). Using the \imppass{} feature in \modeourstxt{} reduced them further for that UI (\pct{5.9} decrease).
    \\
    \midrule 
4 &
    \ref{sec:results_lengths}
    &
    Reply lengths\medskip\newline
    \textit{GLMM (Poisson) on character counts}
    &
    \modeours{} $\uparrow^*$ \newline 
    \deemph{(\lmmci{.24}{.0060}{.22}{.25}{<.0001})}\medskip\newline
    \modemail{} $\uparrow^*$ \newline 
    \deemph{(\lmmci{.57}{.0037}{.56}{.58}{<.001})}\medskip\newline 
    \Imppass{} feature used $\uparrow^*$ \newline 
    \deemph{(\lmmci{.32}{.0063}{.31}{.33}{<.0001})}
    &
    \modeours{} vs \modemanual{} \deemph{(\posthoc{.24}{<.0001})} \medskip\newline 
    \modemail{} vs \modemanual{} \deemph{(\posthoc{.57}{<.0001})} \medskip\newline 
    \modeours{} vs \modemail{} \deemph{(\posthoc{-.33}{<.0001})}
    &
    People wrote longer replies with AI, even more so with \modemailtxt{} (exp($\beta$)=exp(.57)=1.77 i.e. \pct{77} increase) than with \modeourstxt{} (\pct{27} increase). Using the \imppass{} feature in \modeourstxt{} increased it further for that UI (\pct{38} increase).
    \\
    \midrule 
5 &
    \ref{sec:results_errors}
    &
    Error rates\medskip\newline
    \textit{LMM on errors per character}
    &
    \modeours{}	$\downarrow^*$ \newline 
    \deemph{(\lmmci{-.0011}{.0004}{-.0018}{-.0004}{=.0024})}\medskip\newline 
    \modemail{}	$\downarrow^*$ \newline 
    \deemph{(\lmmci{-.0020}{.0002}{-.0025}{.0015}{<.0001})}\medskip\newline 
    \Imppass{} feature used $\downarrow^*$ \newline 
    \deemph{(\lmmci{-.0012}{.0004}{-.0020}{-.0005}{=.0012})}
    &
    \modeours{} vs \modemanual{} \deemph{(\posthoc{-.0011}{=.0048})} \medskip\newline 
    \modemail{} vs \modemanual{} \deemph{(\posthoc{.-0020}{<.0001})} \medskip\newline 
    \modeours{} vs \modemail{} \deemph{(\posthoc{.0009}{=.0096})}
    &
    People wrote emails with lower error rates with AI than without, even lower with \modemailtxt{} than with \modeourstxt. Using the \imppass{} feature in \modeourstxt{} reduced the error rates further for that UI.
    \\
    \midrule
6 &
    \ref{sec:results_email_similarity}
    &
    Email similarity\medskip\newline
    \textit{LMM on cosine similarity of SBERT embeddings}
    &
    \modeours{} $\uparrow^*$ \newline 
    \deemph{(\lmmci{.09}{.0026}{.09}{.10}{<.0001})}\medskip\newline 
    \modemail{} $\uparrow^*$ \newline 
    \deemph{(\lmmci{.17}{.0023}{.17}{.18}{<.0001})}\medskip\newline 
    \Imppass{} feature used $\uparrow^*$ \newline
    \deemph{(\lmmci{.07}{.0029}{.07}{.08}{<.0001})}
    &
    \modeours{} vs \modemanual{} \deemph{(\posthoc{.09}{<.0001})} \medskip\newline 
    \modemail{} vs \modemanual{} \deemph{(\posthoc{.17}{<.0001})} \medskip\newline 
    \modeours{} vs \modemail{} \deemph{(\posthoc{-.08}{<.0001})}
    &
    People wrote semantically more similar (i.e. less diverse) emails with AI than without, more so with \modemailtxt{} than with \modeourstxt{}. For the latter, using the \imppass{} feature contributed to increasing the similarity of emails.
    \\
    \midrule
7 &
    \ref{sec:results_lexical_diversity}
    &
    Lexical diversity\medskip\newline
    \textit{LMM on the distinct2 metric}
    &
    \modeours{} $\downarrow^*$ \newline 
    \deemph{(\lmmci{-.03}{.0043}{-.04}{-.03}{<.0001})}\medskip\newline 
    \modemail{} $\downarrow^*$ \newline 
    \deemph{(\lmmci{-.02}{.0029}{-.03}{-.01}{<.0001})}\medskip\newline 
    \Imppass{} feature used $\uparrow^*$ \newline
    \deemph{(\lmmci{.01}{.0045}{.003}{.02}{=.0087})}
    &
    \modeours{} vs \modemanual{} \deemph{(\posthoc{-.03}{<.0001})} \medskip\newline 
    \modemail{} vs \modemanual{} \deemph{(\posthoc{-.02}{<.0001})} \medskip\newline 
    \modeours{} vs \modemail{} \deemph{(\posthoc{-.015}{=.0011})}
    &
    People wrote emails with lower lexical diversity (measured as: unique bigrams / number of words) with AI than without it, even more so with \modeourstxt{} than with \modemailtxt{}. Using the \imppass{} feature in \modeourstxt{} closed this gap.
    \\
  \bottomrule
\end{tabularx}
\caption{Overview of significance tests with links to the section, tested measure, predictors, pairwise comparisons, and written interpretation. The arrows indicate if predictors increase ($\uparrow$) or decrease ($\downarrow$) the outcome aspect, with an asterix if these impacts are significant (*).}
\Description{Overview of significance tests with links to the section, tested measure, predictors, pairwise comparisons, and written interpretation. For each statistical test it describes the Section, Aspect and model, Predictors (baseline: NoAI), Pairwise comparisons, and Takeaways in words (only considering sig. results).}
\label{tab:lmm_overview}
\end{table*}

\begin{table*}[t!]
\centering
\footnotesize
\newcolumntype{L}{>{\raggedright\arraybackslash}X}
\newcolumntype{P}[1]{>{\raggedright\arraybackslash}p{#1}}
\renewcommand{\arraystretch}{1.4}
\setlength{\tabcolsep}{4pt}
\begin{tabularx}{\linewidth}{lP{2.75em}P{5em}P{22em}P{7em}L}
\toprule
    &
    \textbf{Section} &
    \textbf{Aspect}\newline and model &
    \textbf{Predictors} (baseline: \modemanual) &
    \textbf{Pairwise comparisons} &
    \textbf{Takeaways in words}\newline(only considering sig. results) \\ \midrule
1 &
    \ref{sec:results_briefing}
    &
    Briefing conformity\medskip\newline
    \textit{GLMM (Binomial) on binary conformity coding}
    &
    \modeours{} $\downarrow^*$ \newline 
    \deemph{(\lmmci{-.77}{.3107}{-1.38}{-.16}{=.013})}\medskip\newline 
    \modemail{} $\downarrow$ \newline 
    \deemph{(\lmmci{-.04}{.2417}{-.52}{.43}{=.857})}\medskip\newline
    \Imppass{} feature used $\uparrow$ \newline
    \deemph{(\lmmci{.25}{.3310}{-.40}{.90}{=.444})}\medskip\newline
    Full reply generated without input $\downarrow^*$ \newline
    \deemph{(\lmmci{-1.70}{.3467}{-2.38}{-1.02}{<.0001})}
    &
    \modeours{} vs \modemanual{} \deemph{(\posthoc{-.77}{=.040})} \medskip\newline 
    \modemail{} vs \modemanual{} \deemph{(\posthoc{-.04}{=.857})} \medskip\newline 
    \modeours{} vs \modemail{} \deemph{(\posthoc{-.73}{=.040})}
    &
    With \modeours, people wrote emails that had a higher chance to miss a key aspect of the study briefing than those written with \modemail{} or manually (\pct{23} of emails missed it for \modeours{} vs \pct{18} for \modemail{} vs \pct{13} for \modemanual). People's prompting behaviour had a larger impact here: Across \modeours{} and \modemail{}, generating a full reply without any own input (83 emails in the data) missed a key aspect of the briefing in half of the cases (\pct{49}).
    \\
    \midrule
2 &
    \ref{sec:results_workflows}
    &
    Skipping local response \medskip\newline
    \textit{GLMM (Binomial) on skipped yes/no}
    &
    Length of incoming email\newline
    (num. standardised words, i.e. characters/5) $\downarrow^*$ \newline
    \deemph{(\lmmci{-.025}{.010}{-.050}{-.004}{=.0171})}
    &
    -
    &
    Each additional word (defined as 5 additional characters) in the incoming email is associated with a \pct{2.46} decreased chance of skipping the local response step in \modeours{}. Skipping is defined as not entering any text on the local response screen of that UI.
    \\
  \bottomrule
\end{tabularx}
\caption{Further significance tests with links to the section, tested measure, predictors, pairwise comparisons, and written interpretation. The arrows indicate if predictors increase ($\uparrow$) or decrease ($\downarrow$) the outcome aspect, with an asterix if these impacts are significant (*).}
\Description{Further significance tests with links to the section, tested measure, predictors, pairwise comparisons, and written interpretation. For each statistical test it describes the Section, Aspect and model, Predictors (baseline: NoAI), Pairwise comparisons, and Takeaways in words (only considering sig. results).}
\label{tab:lmm_overview2}
\end{table*}

\section{Results}\label{sec:results}

\subsection{Analysis of Interaction Logs}\label{sec:results_interaction_logs}
We report our analyses of the interaction data. 

\begin{figure*}[t]
    \centering
    \includegraphics[width=\linewidth]{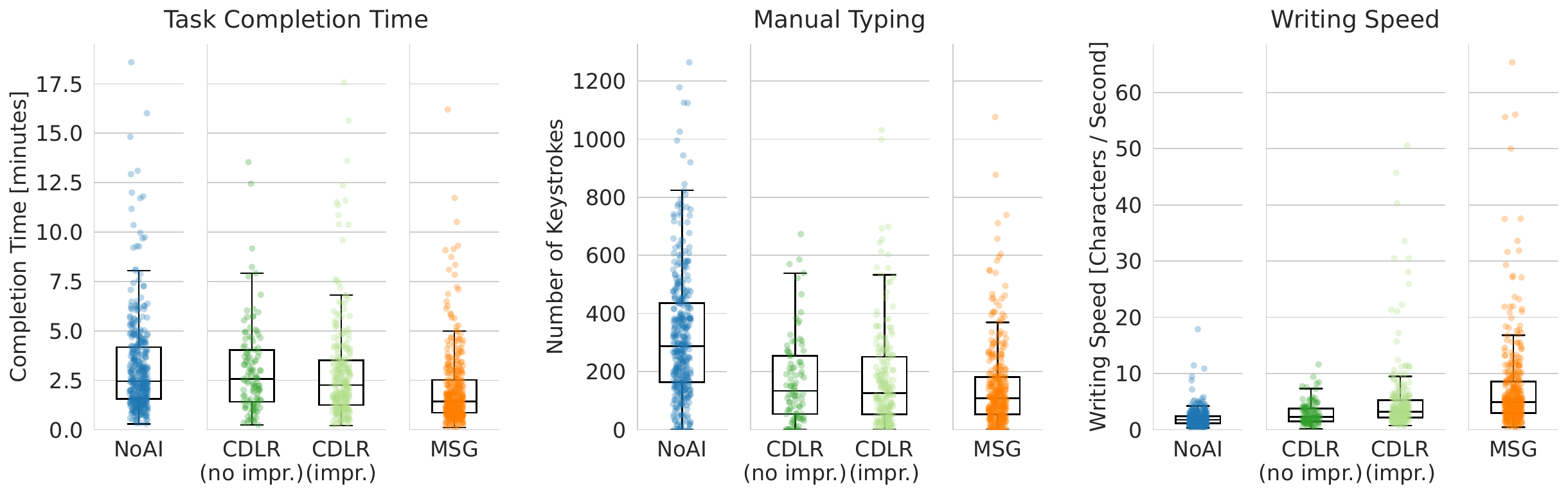}
    \caption{Three measures of interaction behaviour: Task completion time \textit{(left)}, manual typing \textit{(centre)}, and writing speed \textit{(right)}. All AI features increased typing speed and reduced the time taken (both sig. for \modemail). They also reduced the number of keystrokes (sig. for \modeours{} and \modemail). If people made use of the optional \imppass{} feature \revision{(impr.)} in \modeours, this contributed to narrowing the gap between the otherwise sentence-level design of \modeours{} and the message-level design of \modemail{} (sig. for manual typing and writing speed). See \cref{sec:results_interaction_logs} for details.}
    \Description{This figure presents three box plots that measure interaction behaviour from the study across three different user interfaces: NoAI (manual mode), CDLR (AI-supported with and without the optional improvement pass feature), and MSG (AI-supported). The three metrics being compared are:
    Task Completion Time (Left):
    This plot shows how long it took participants to complete the task in minutes.
    Both CDLR (with and without improvement) and MSG show reduced task completion times compared to NoAI, with MSG showing the most significant reduction.
    Manual Typing (Center):
    This plot shows the number of keystrokes made by participants during the task.
    Both CDLR and MSG significantly reduced the number of keystrokes compared to NoAI, particularly when participants used the optional improvement pass feature in CDLR.
    Writing Speed (Right):
    This plot shows the writing speed of participants measured in characters per second.
    Both CDLR and MSG increased writing speed compared to NoAI, with MSG again showing the most notable improvement.
    Overall, the figure demonstrates that AI-supported interfaces (CDLR and MSG) led to faster task completion, fewer keystrokes, and increased writing speed. In particular, the optional improvement pass feature in CDLR helped narrow the performance gap between the sentence-level design of CDLR and the message-level design of MSG. For detailed statistical analysis, see the corresponding section of the paper (Section 6.1).}
    \label{fig:interaction_logs_boxplots}
\end{figure*}

\subsubsection{Task Completion Time}\label{sec:results_time}

On average, participants took \mins{3.20} (SD 2.51, median 2.46) to write an email manually.
With \modemail, this decreased to \mins{2.03} (SD 1.90, median 1.44), while \modeours{} reduced it to \mins{2.95} (SD 2.49, median 2.32). For \modeours, using the \imppass{} feature resulted in \mins{2.90} (SD 2.58, median 2.26), while not using it had \mins{3.06} (SD 2.29, median 2.57).
\cref{fig:interaction_logs_boxplots} (left) shows this as box plots.
These differences were significant as follows (\cref{tab:lmm_overview}, row 1): 
Participants finished replying significantly faster with \modemail{} than without AI (-\secs{70}). \modemail{} was also significantly faster than \modeours{} (-\secs{66}).

\subsubsection{Writing Speed}\label{sec:results_speed}

On average, participants wrote 2.05 characters per second without AI (SD 1.57, median 1.78).
Replying with \modemail{} had a mean of 7.21 (SD 7.90, median 4.89), while the mean speed with \modeours{} was 4.38 (SD 5.59, median 2.91). For \modeours, using the \imppass{} feature resulted in 5.03 (SD 6.47, median 3.23), while not using it had 2.91 (SD 2.00, median 2.31).
\cref{fig:interaction_logs_boxplots} (right) shows this as box plots.
These differences were significant as follows (\cref{tab:lmm_overview}, row 2): 
Compared to writing without AI, participants produced significantly more characters per second with \modemail{} (5.2 chars more per s) and if they used the \imppass{} feature in \modeours{} (2.5 chars more per s). The difference between \modemail{} and \modeours{} was also significant.

\subsubsection{Manual Typing}\label{sec:results_keystrokes}
Without AI, participants on average needed 321.69 keystrokes (SD 217.76, median 284), compared to 146.67 (SD 144.97, median 108) with \modemail, and 174.62 (SD 166.57, median 128) with \modeours. For \modeours, using the \imppass{} feature had a mean of 176.0 (SD 173.7, median 126), while not using it had 171.4 (SD 149.9, median 133.5).
\cref{fig:interaction_logs_boxplots} (centre) shows this as box plots.
These differences were significant (\cref{tab:lmm_overview}, row 3): People needed significantly fewer keystrokes with AI features than without them, and even significantly fewer with \modemail{} (\pct{58} decrease) than with \modeours{} (\pct{48} decrease). Using the \imppass{} feature in \modeours{} significantly reduced this further for that UI (\pct{5.9} decrease). In summary, all AI features significantly reduced manual typing.

\subsubsection{Interaction with \modeours}
We logged interactions specific to \modeours.
On average participants tapped on 2.64  (SD 2.89, median 2) sentences per email, that is, on \pct{30.36} (SD \pct{29.59}, median \pct{23.08}) of sentences in each email.
They replied to \pct{87.37} of tapped sentences. In \pct{83.27} of the cases, they did so by accepting a suggestion. %

Suggestions were paginated; most suggestions (\pct{67.15}) were accepted on the first page. Another \pct{19.50} and \pct{13.36} were accepted on the second and third page, respectively.
The majority of accepted suggestions (\pct{80.14}) were generated without an explicit prompt, and most (\pct{92.30}) were not edited afterwards.

\revision{On the first screen, \pct{69.05} of participants accepted a sentence suggestion at least once, and \pct{55.56} manually entered text for at least one local response. %
On the second screen, \pct{83.33} of participants accepted at least one email-level suggestion. Only \pct{1.59} (two participants) did not use any \modeours-specific features.}

On average, \mins{1.67} (\pct{57.23}) were spent on the first screen and \mins{1.25} (\pct{42.77}) on the second (\cref{fig:time_spent_on_screens_barplot}).
Participants used the \imppass{} feature for 287 emails (\pct{75.93}) and accepted an improved email for 274 emails (\pct{72.49}).
When the \imppass{} feature was used at least once, an improved email was requested on average 1.35 times (SD 0.95, median 1.00) and accepted 1.13 times (SD 0.47, median 1.00) per email.
The last accepted improved email was identical to the sent email in \pct{83.94} of all cases.
When participants made changes these had a mean edit distance of 72.73 (SD 97.10, median 37).

\subsubsection{Interaction with Full Email Generation (\modemail)}
\begin{figure}
    \centering
    \includegraphics[width=\minof{\columnwidth}{0.75\textwidth}]{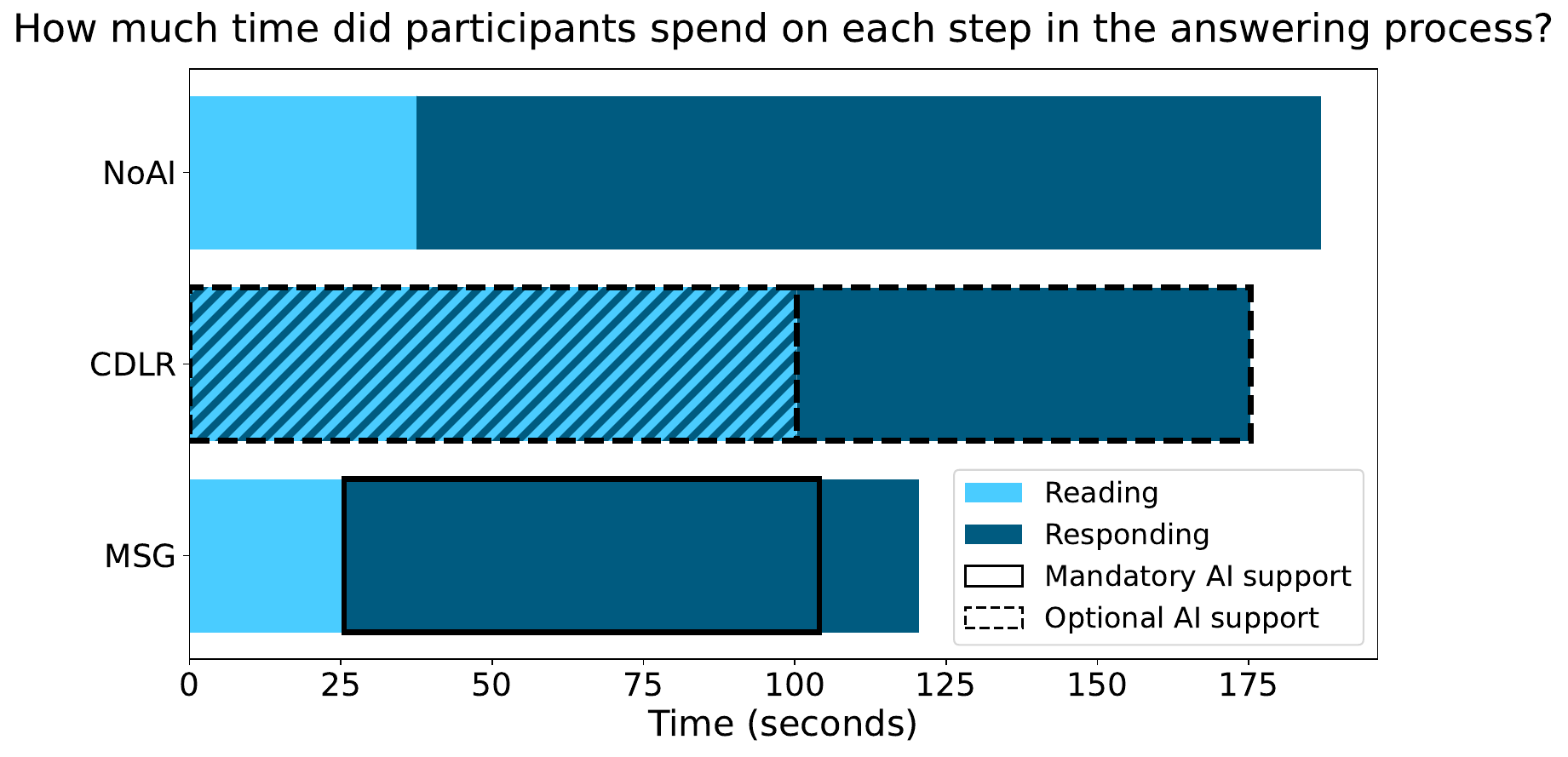}
    \caption{Participants spent their time on different screens and thus different aspects. The figure shows the means for the time spent on screens that focus on reading the incoming email vs on screens that focus on responding (colour). Borders indicate which steps required AI (solid) or offered it optionally (dashed). For \textit{\modemanual}, users read the email, then spend most of the time writing the reply. For \textit{\modeours}, the local response screen (\cref{fig:teaser}.1) enables reading and responding in parallel (striped), followed by responding on the second screen (\cref{fig:teaser}.2), both with optional AI (sentence suggestions, \imppass{}). In contrast, \textit{\modemail} requires AI after the initial reading phase to generate the response, which can then be manually edited.}
    \Description{This figure displays a bar plot analysing times spent on each step in the answering process. The figure shows the means for the time spent on screens that focus on reading the incoming email vs on the screens that focus on responding. For NoAI, users read the email, then spend most of the time writing the reply. For CDLR, the local response screen enables reading and responding in parallel, followed by responding on the second screen, both with optional AI (sentence suggestions, improvement pass). In contrast, MSG requires AI after the initial reading phase to generate the response, which can then be manually edited.}
    \label{fig:time_spent_on_screens_barplot}
\end{figure}

With this UI, \pct{71.98} of the first generated replies were accepted; otherwise, a new generation was requested.
Users spent \mins{0.43} (\pct{21.20}) on the incoming email screen, \mins{1.31} (\pct{65.09}) on the generation view and \mins{0.28} (\pct{13.71}) on the editing screen (\cref{fig:time_spent_on_screens_barplot}).

\subsubsection{Workflow Analysis}\label{sec:results_workflows}

\begin{figure}
    \centering
    \includegraphics[width=\minof{\columnwidth}{0.66\textwidth}]{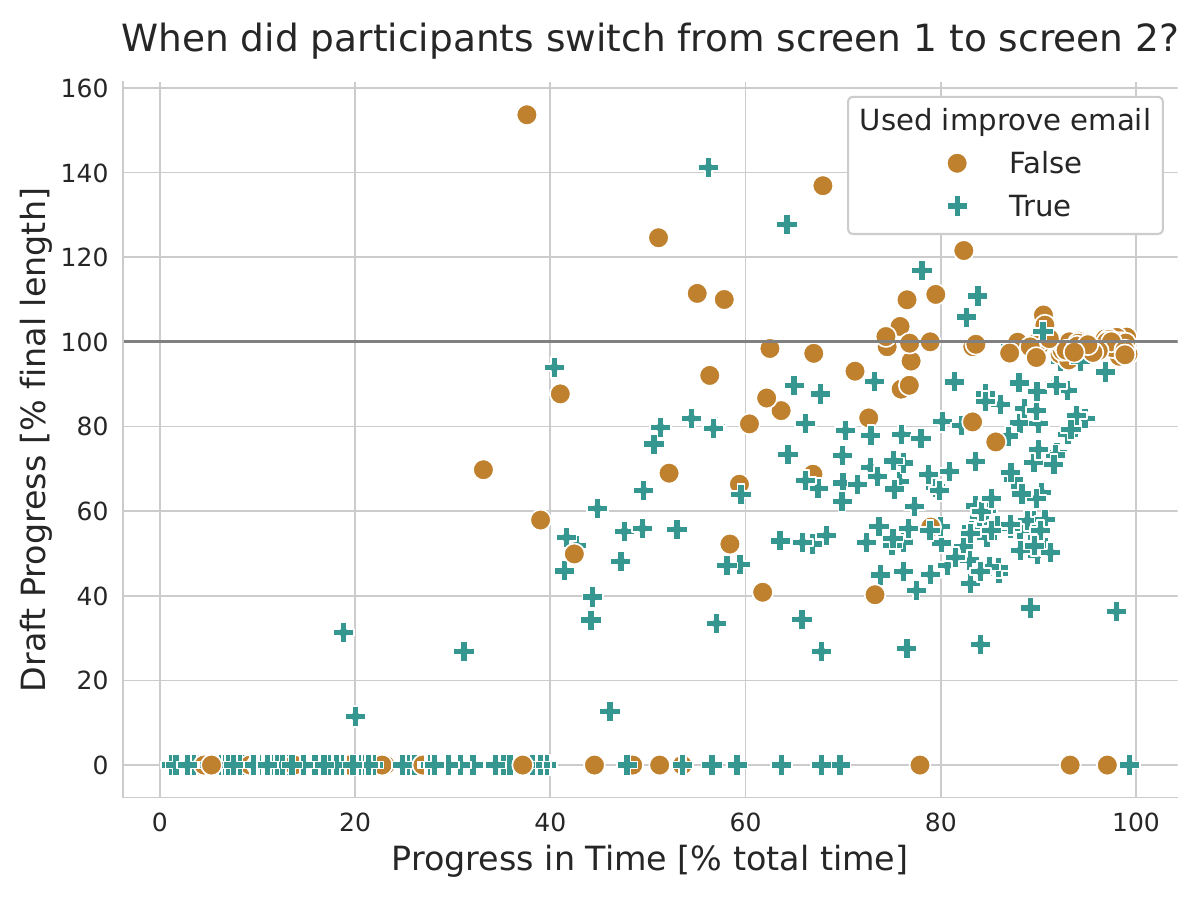}
    \caption{Analysis of workflows with \modeourstxt: Each point is one email and its position is the state of the drafting process at the moment when the user switched from the first screen (\cref{fig:teaser}.1) to the second (\cref{fig:teaser}.2). Concretely, the x-axis shows normalised time (0-\pct{100}), i.e. temporal progression. The y-axis shows normalised length, i.e. draft progression. Note that y-values >\pct{100} are possible if an intermediate draft is longer than the final version. Colour and marker shape indicate if the \imppass{} feature was used or not. The figure reveals three clusters: \textit{(1) Bottom left} -- here, people skipped to the second screen and used the \imppass{} feature to generate a draft. \textit{(2) Top right} -- mostly drafting on the first screen, with light manual editing on the second. \textit{(3) In between} -- partly drafting on the first screen and finalising it with AI on the second one.}
    \Description{This figure displays a scatter plot analysing workflows with content-driven local responses, focusing on when participants switched from the first screen to the second screen during the email drafting process.
    X-axis (Progress in Time [\% total time]): This axis represents the normalised time (ranging from 0 to 100) indicating the temporal progression of the drafting process.
    Y-axis (Draft Progress [\% final length]): This axis represents the normalised length of the draft at the moment of switching screens. Values greater than 1 are possible if an intermediate draft was longer than the final version.
    Colour Coding: The colour of each point indicates whether the "improvement pass" feature was used:
    Blue points represent emails where the improvement pass feature was used.
    Orange points indicate emails where it was not used.
    The scatter plot reveals three distinct clusters of participant behaviour:
    Bottom Left Cluster:
    Participants in this group quickly moved to the second screen and utilised the improvement pass feature to generate a draft with minimal work on the first screen.
    Top Right Cluster:
    Participants in this group spent most of their time drafting on the first screen, with only light manual editing on the second screen.
    In Between Cluster:
    This group represents participants who partly drafted on the first screen and then finalised the draft with AI assistance on the second screen.
    This analysis highlights the different strategies participants used during the email drafting process, depending on their interaction with the interface and the improvement pass feature.}
    \label{fig:workflow_scatterplot}
\end{figure}

For \modeours, we discovered three main workflows by plotting when people switched from the first screen (\cref{fig:teaser}A) to the second (\cref{fig:teaser}B). \cref{fig:workflow_scatterplot} reveals three clusters: (1) Sometimes people went straight to the second screen and used the \imppass{} feature to create a draft. (2) Alternatively, they spent most of their time drafting on the first screen, with light manual editing on the second. (3) Finally, people partially drafted on the first screen and finished it on the second screen, using AI.
We fitted a GMM\footnote{Gaussian Mixture Model with 3 components using \url{https://scikit-learn.org/}} to estimate the number of emails: Cluster 1 had 136, cluster 2 had 54, and cluster 3 had 188 emails.

We also examined the relationship of the incoming email's length and whether people skipped the local response screen without entering any text. This was significant (\cref{tab:lmm_overview2}, row 2): Each additional word (i.e. 5 additional characters) in the incoming email is associated with a \pct{2.46} decreased chance of skipping the local response step in \modeours{}. 

For \modemail, we found that in most cases (\pct{77.8} of emails) people sent the generated drafts without manually editing them further. When they indeed edited them (\pct{22.2} of emails), the mean edit distance between generated and edited version was 64.87 (SD 67.48, median 44.50). This corresponds to typing about twelve words~\cite{kristensson2014inviscid}.

We also analysed how people prompted with \modemail: In a majority of cases (\pct{82.01} of emails), participants entered a prompt right away. Otherwise, they generated text solely based on the information in the incoming email. In half of those cases (\pct{51.47}), participants did not accept the result and generated another. In comparison, such a regeneration was only needed in \pct{22.9} of the cases where participants entered a prompt. We observed a learning effect for some: \pct{40} of people started entering a prompt if they were not happy with the initial result generated without a prompt. %

\subsection{Perception of Interaction}\label{sec:results_perception}
We analysed participants' perception of the three UIs.

\subsubsection{In-app Questionnaire (Likert Data)}\label{sec:results_in_app}

\begin{figure*}[t]
    \centering
    \includegraphics[width=\linewidth]{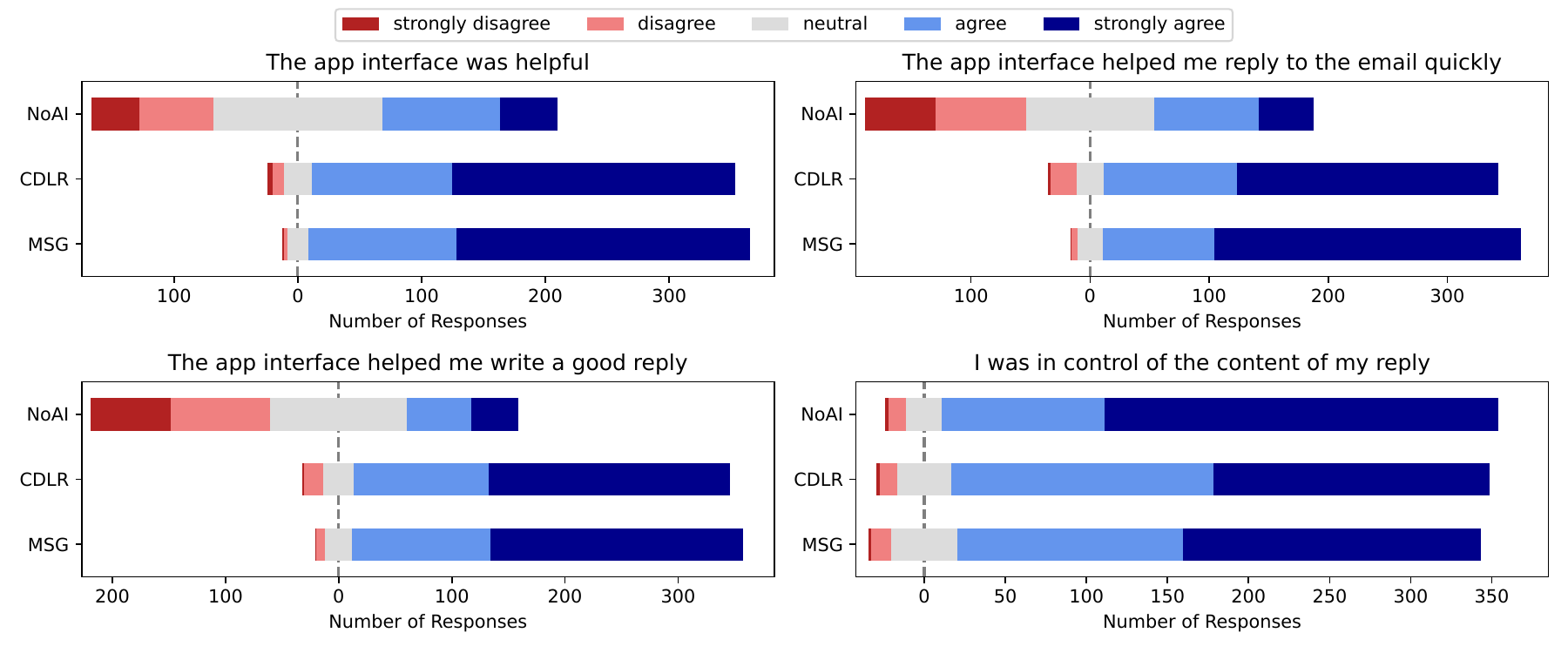}
    \caption{Likert results on perception of the UIs and interaction, rated after each email task. Overall, participants rated the AI-supported UIs higher on speed, quality, and helpfulness compared to the manual mode. However, the latter was rated higher on control.}
    \Description{This figure presents bar charts displaying Likert scale results from participants' perceptions of different UIs and their interactions, rated after completing email tasks. 
    The figure is divided into four sections, each comparing three UIs: NoAI (manual mode), CDLR (AI-supported), and MSG (AI-supported).
    Top-Left Chart: The app interface was helpful
    The NoAI interface received mixed responses, with a significant portion of participants disagreeing or remaining neutral, and fewer strongly agreeing.
    Both CDLR and MSG interfaces were rated more positively, with a larger number of participants agreeing or strongly agreeing that the interfaces were helpful.
    Top-Right Chart: The app interface helped me reply to the email quickly
    The NoAI interface again had a more varied response, with some participants disagreeing or remaining neutral, while others agreed.
    CDLR and MSG interfaces were rated highly for helping users reply quickly, with the majority of participants agreeing or strongly agreeing.
    Bottom-Left Chart: The app interface helped me write a good reply
    Similar to the other charts, the NoAI interface had a mix of responses, with fewer participants strongly agreeing.
    CDLR and MSG interfaces were again rated highly, with most participants agreeing or strongly agreeing that these interfaces helped them write good replies.
    Bottom-Right Chart: I was in control of the content of my reply
    For this aspect, the NoAI interface was rated slightly higher, with more participants strongly agreeing that they felt in control of their reply content.
    Although CDLR and MSG interfaces were also rated positively, there was a slight decrease in the number of participants who strongly agreed compared to the NoAI interface.
    In summary, the figure shows that participants generally rated the AI-supported UIs (CDLR and MSG) higher in terms of speed, quality, and helpfulness. 
    However, the manual mode (NoAI) was rated slightly higher in terms of giving users a sense of control over their replies.}
    \label{fig:inapp_likert_items}
\end{figure*}

Participants rated four Likert items in the app after each email (\cref{fig:inapp_likert_items}). 
We found statistically significant \revision{effects of \ivmode{} on all four items -- speed (\artf{2}{996.11}{433.71}{<.0001}, \petasq{.46}), control (\artf{2}{996.11}{20.60}{<.0001}, \petasq{.04}), quality (\artf{2}{996.13}{466.99}{<.0001}, \petasq{.48}), and helpfulness (\artf{2}{996.1}{430.61}{<.0001}, \petasq{.46}).}
\revision{Concretely,} \modeours{} and \modemail{} were both rated significantly higher than \modemanual{} on speed \revision{(\modeours{}: \artc{996}{23.34}{<.0001}; \modemail{}: \artc{996}{27.25}{<.0001})}, quality \revision{(\modeours{}: \artc{996}{25.72}{<.0001}; \modemail{}: \artc{996}{27.18}{<.0001})}, and helpfulness \revision{(\modeours{}: \artc{996}{24.65}{<.0001}; \modemail{}: \artc{996}{26.14}{<.0001})}. They were both rated significantly lower than \modemanual{} on control \revision{(\modeours{}: \artc{996}{-5.88}{<.0001}; \modemail{}: \artc{996}{-5.17}{<.0001})}. 
The only significant difference between the UIs with AI was that \modemail{} was rated higher on speed than \modeours{} \revision{(\artc{996}{3.91}{=.0001})}.

\subsubsection{In-App Feedback}
We reviewed the in-app feedback optionally provided after each reply.
For both AI modes it was overwhelmingly positive, such as: ``Im really enjoying this kind of AI help mode.'' (P60\oldId{P1351}, \modemail), ``It made work easy'' (P76\oldId{P1370}, \modemail), ``Very smooth process, good suggestions for each part.'' (P47\oldId{P1329}, \modeours), and ``This made my reply look way better.'' (P81\oldId{P1375}, \modeours).

The negative feedback was less homogeneous.
For \modemail, around half of these critiques highlighted difficulties in getting the AI to incorporate specific information, such as: ``the AI who seemed to resist wanting to offer access to my colleague'' (P25\oldId{P1298}); or ``had a bit of trouble trying to get the AI to properly acknowledge that \$200 was okay [...]''  (P37\oldId{P1312}). 
Related, P53\oldId{P1338} noted that ``Control in replying was lacking, It didn't give me many options to 'add' ideas of my own.''
People found ways to steer the system; P47\oldId{P1329} said that ``I had to adjust the prompt a few times to get the sort of reply that I was looking for, but it did generate a good reply overall and I was satisfied with the end result.''

Notably, issues with including specific information were rarely mentioned for \modeours.
Most of the negative feedback instead concerned the tone: ``This was too wordy for an informal email.'' (P56\oldId{P1341}), %
and ``The ai was helpful but it made the response feel slightly too formal and professional.'' (P49\oldId{P1333}) %

For the manual mode, people ``had no issues, [and] felt able to use the platform freely and there was no technical faults'' (P7\oldId{P1276}), and that ``It was just like normal email.'' (P60\oldId{P1351}). %

\subsubsection{Favourite Reply Support}\label{sec:results_fav_mode}
Only \pct{4} (5 people) preferred \modemanual, %
\pct{49.2} (62 people) favoured \modemail, and \pct{43.7} (55 people) preferred \modeours.
The remaining \pct{3.2} (4 people) did not pick a favourite.

Notably, the high-level code ``Efficiency'' occurred in \pct{56.45} of comments for \modemail{} and \pct{32.73} for \modeours{}.
``Quality'' in \pct{25.82} for \modemail{} and \pct{30.91} for \modeours{}.
``Control'' in \pct{8.07} for \modemail{} and \pct{29.09} for \modeours{}.
``Tailoring'' zero times for \modemail{} and \pct{5.46} for \modeours{}.
The remaining comments were assigned the code ``Others'' (e.g. P105\oldId{P1405} ``just liked the interface'' of \modemail{} and P76\oldId{P1370} favoured \modeours{} because it supported them in being creative).

Two out of the four people who did not select a favourite stated that they liked both depending on the ``context'' (P77\oldId{P1371}).
For instance, P19\oldId{P1290} explained that ``both have different advantages in different situations. Single prompt allows to produce a full email much faster so is handy when you are short of time but still want to respond. Sentence based provides the user the ability to create a much more tailored email which can cover all bases.''

The \pct{4} (5 people) who preferred \modemanual{} said they were ``used to it'' (P4\oldId{P1273}) or ``confident in [their] writing ability'' (P78\oldId{P1372}).

\subsubsection{Summary}
People perceived AI features as helpful and preferred having them. They were divided about their favourite and perceived meaningful tradeoffs between the two designs with AI on control vs efficiency: While people felt in control with all UIs (\cref{sec:results_in_app}), when reflecting on their favourite, they mentioned control aspects relatively more frequently for \modeours{} than \modemail{} -- and vice versa for efficiency.

\subsection{Analysis of Emails}\label{sec:results_emails}
\begin{figure*}[t]
    \centering
    \includegraphics[width=\linewidth]{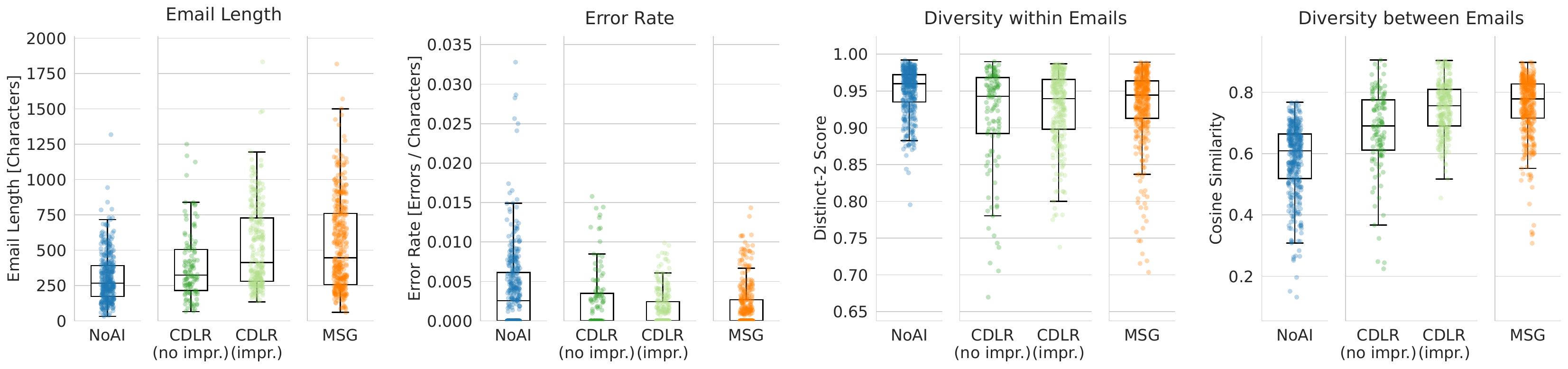}
    \caption{Four measures of email characteristics from our study. The plots show email length \textit{(left)} and rate of spelling/grammar/punctuation errors \textit{(centre left)}. Moreover, we measured lexical diversity with the distinct-2 score \textit{(centre right)}, which is defined as an email's number of distinct bigrams divided by its number of words (higher = more diverse). Finally, we measure diversity between emails \textit{(right)}, based on the cosine similarity of vector embeddings (higher = less diverse). Overall, all AI features increased reply lengths, decreased error rates, and lowered diversity (all sig.).  \revision{For \modeours{} we further distinguish between emails where the improvement pass feature was used (impr.) and those where it was not (no impr.).}
    See \cref{sec:results_emails} for details.}
    \Description{This figure presents four box plots comparing different characteristics of emails across three user interfaces: NoAI (manual mode), CDLR (AI-supported with and without the optional improvement pass feature), and MSG (AI-supported). The metrics being compared are:
    Email Length (Left):
    This plot shows the length of the emails in characters.
    AI-supported interfaces (CDLR and MSG) led to longer emails compared to the NoAI interface, with MSG producing the longest emails.
    Error Rate (Center Left):
    This plot displays the rate of spelling, grammar, and punctuation errors relative to the number of characters in the email.
    Both CDLR and MSG reduced the error rate compared to NoAI, with slightly better performance for the improvement pass-enabled CDLR and MSG interfaces.
    Diversity within Emails (Center Right):
    The distinct-2 score measures the lexical diversity within individual emails by calculating the number of unique bigrams (word pairs) relative to the total word count. A higher score indicates greater diversity.
    NoAI emails had higher within-email diversity compared to CDLR and MSG, which showed similar scores indicating reduced lexical diversity.
    Diversity between Emails (Right):
    This plot measures the cosine similarity between vector embeddings of emails, with higher scores indicating less diversity between emails.
    Emails generated using CDLR and MSG were more similar to one another, as indicated by higher cosine similarity scores, compared to those written using the NoAI interface.
    Overall, the figure demonstrates that AI-supported interfaces (CDLR and MSG) increased email length, reduced error rates, and resulted in slightly less diversity both within and between emails. For detailed analysis, refer to the corresponding section in the paper (Section 6.2).
    }
    \label{fig:quality_boxplots}
\end{figure*}

We analysed the content of the emails. \cref{fig:quality_boxplots} shows four box plots as an overview.

\subsubsection{Email Lengths}\label{sec:results_lengths}

On average, emails written without AI were 302.5 characters long (SD 169.7, median 267).
\modemail{} resulted in 536.1 (SD 319.0, median 447.5), while \modeours{} had 483.0 (SD 285.4, median 382). For \modeours, using the \imppass{} feature had a mean length of 523.9 (SD 294.1, median 412.0), while not using it had 390.6 (SD 241.5, median 324.5).
These differences were significant (\cref{tab:lmm_overview}, row 4): People wrote significantly longer replies with the AI features than without them, and significantly more so with \modemail{} (\pct{77} increase) than with \modeours{} (\pct{27} increase). Using the \imppass{} feature in \modeours{} significantly increased this further for that UI (\pct{38} increase). In summary, all AI features significantly increased text lengths.

\subsubsection{Error Rates}\label{sec:results_errors}
We checked grammar and spelling with the language-tool-python\footnote{\url{https://pypi.org/project/language-tool-python/}} library.
Per email, we recorded the minimum of British English and American English spell checking to avoid penalising spelling differences. %
Manual writing had the highest mean error rate of .00375 errors per character.  Both AI versions were about half of that: \modemail{} had .00176 and \modeours{} had .00182. Using the \imppass{} feature in \modeours{} contributed to reducing errors (mean .00147 when using it vs .00260 when not). All these differences were significant (\cref{tab:lmm_overview}, row 5). In summary, all AI features significantly reduced error rates.

\subsubsection{Diversity Across Emails}\label{sec:results_email_similarity}
We analysed the semantic similarity between emails, following related work~\cite{padmakumar2024diversity}. We computed the cosine similarity of the vector embeddings of all pairs of emails written with the same mode and for the same briefing, using the Sentence Transformers library (SBERT\footnote{\url{https://sbert.net}, specifically \url{https://huggingface.co/sentence-transformers/all-MiniLM-L6-v2}})~\cite{reimers2019sbert}.
As expected from the literature, manual emails had the lowest mean pairwise similarity (.582), that is, they were the most diverse. \modemail{} had the highest similarity (.756), followed by \modeours{} (.726). 
Using the \imppass{} feature in \modeours{} contributed to the increase (.676 without using it vs .749 with it). 
All these differences were significant (\cref{tab:lmm_overview}, row 6). In summary, all AI features significantly reduced semantic diversity.

\subsubsection{Diversity Within Emails}\label{sec:results_lexical_diversity}
We analysed lexical diversity, as in related work~\cite{Fu2023sentencevsmessage}, with the distinct-2 metric, defined as the number of distinct bigrams divided by the total number of words.
Our results match the related work: Writing without AI had the highest mean lexical diversity (.950), \modemail{} lowered it to .930, and \modeours{} had .924.
Using the \imppass{} feature in \modeours{} (almost) closes the gap between \modeours{} and \modemail{} (.915 without it vs .927 with it). 
All these differences were significant (\cref{tab:lmm_overview}, row 7). In summary, lexical diversity was significantly affected by all AI features, in the way we would expect from related work~\cite{Fu2023sentencevsmessage}: Sentence-level generations decreased it more than message-level generations.

\subsubsection{Email Structure}\label{sec:results_structure}

Salutations were missing in \pct{8.5} of manually written replies and in \pct{10.6} with \modeours. All replies with \modemail{} had salutations. All replies with \modeours{} and the \imppass{} feature had a salutation.
Similarly, only one email with \modemail{} lacked a closing statement, compared to \pct{14.3} for \modemanual{} and \pct{9.0} for \modeours. Again, when the \imppass{} was used in \modeours, all emails ended with a closing signature.

\subsubsection{Briefing Conformity}\label{sec:results_briefing}
Each email reply task showed a briefing that asked participants to respond with certain information (see \cref{sec:procedure_email_tasks}). This allowed us to analyse if their emails conformed to this or not (see \cref{sec:quality_m}).
This varied across the UIs: With \modeours, \pct{23} of emails missed a key aspect of the study briefing, compared to \pct{18} for \modemail{} and \pct{13} for \modemanual.

The differences between \modeours{} and the other two UIs were significant (\cref{tab:lmm_overview2}, row 1). %
However, people's prompting behaviour had a larger impact here: Across \modeours{} and \modemail{}, generating a full reply without any own input (83 emails in the data) missed a key aspect of the briefing in half of the cases (\pct{49}). We return to this in the discussion (\cref{sec:discussion_methods})

\subsubsection{Subjective Assessment of Quality}\label{sec:result_quality}
All emails were read by (at least) two researchers. %
While length and structure varied (\cref{sec:results_lengths}, \cref{sec:results_structure}), we did not notice ``nonsense'' responses. Even those emails which did not conform to the briefings (\cref{sec:results_briefing}) reflected the general topic and most would have been believable replies. We noticed that sensible replies can vary drastically -- from short responses to elaborate, formal emails. The latter mostly coincided with message-level AI generation. We do not consider this an issue of quality since we did not specify a level of formality to follow. Overall, based on our subjective assessment, we thus concluded that reply quality was suitable across all UIs. 

\section{Discussion}

\subsection{Redesigning Mobile Email Replies with Content-Driven Local Response}
Our first research question asked: \textit{How might we redesign the mobile email reply UI for flexible and optional AI involvement?} We proposed to \revision{build on} microtasking \revision{UI} principles with our concept of \modeourstxt. Here, we reflect on two key aspects of this solution strategy. 

\subsubsection{\revision{Responding (with AI) Directly While Reading}}
A key benefit of our \modeours{} design is that it turns the email content into a UI to locally trigger responses. As our results show, users leveraged this to keep the email in view while responding (\cref{fig:time_spent_on_screens_barplot}). \revision{One way of looking at this is through the lens of} the usability principle of ``recognition over recall''~\cite{Nielsen1994usabilityheuristics}: Users can \revision{respond and/or} delegate tasks to the AI at the very moment of recognising a line in the email as relevant. In contrast, reading a whole email first, to then transition to a draft view (AI-based or manual) requires recalling the email's information -- or scrolling/swiping back and forth, at least on a mobile screen that does not fit email and draft in parallel. \revision{This seems more important for complex emails and less relevant for short ones. Our results support this (local response was less likely to be skipped for longer emails, see \cref{sec:results_workflows}) and motivate a dedicated future study on microtasking UIs for emails of varying complexity. That said, mobile interaction is prone to attentional demands and interruptions~\cite{oulasvirta2005bursts}, which might make it more likely to forget about a relevant aspect not only for very long emails.}

\revision{In another lens}, \modeours{} responds to the call by \citet{Tankelevitch2024metacognitive} for UIs that reduce metacognitive demands of generative AI, for example, when deciding when and what to delegate to AI. 
Our concept was successful here, as echoed by participants' comments. For example, P24\oldId{P1297} said ``[I could] organize my ideas better using the sentence replies within the email.'' Similarly, P8\oldId{P1277} found that ``This was definitely helpful in breaking down the email and specifically answering the important key points of the original message.''

Put as a design recommendation: Enable people to immediately act in the local context of a recognised need for AI support, when the relevant information for that task delegation is present. %

\subsubsection{Designing for AI Integration by Supporting Decision-Moments Explicitly in the UI} 
\modeours{} facilitates local responses by \textit{both} user and AI. 
For the user, traditional email UIs offer a two-step workflow -- read the email, then write a reply. However, the actual workflow has more steps: Users need to identify relevant information, think about their response, and compose a coherent reply. The sentence-wise nature of \modeours{} makes these steps explicit in the UI. Work on microtasking shows that splitting up writing tasks and presenting them in context is beneficial for mobile users~\cite{august2020microwriting, iqbal2018playwrite}.

In our design this additionally supports prompting the AI (see \cref{sec:discussion_prompting}).
We believe that this solution strategy is relevant more broadly: Generative AI is often added ``on top'' of existing UIs (\cref{sec:related_work_current_products}). In contrast, we propose to identify relevant (micro) decisions and response moments in the task, redesign the UI so that users can express these through interaction (here: select sentence), and then integrate AI specifically at these moments.

\subsection{Empirical Characteristics of \modeours}\label{sec:discussion_results_overview}

Our empirical research questions asked: \textit{How do users perceive and interact with \modeours?} and \textit{What are the specific advantages and drawbacks?}
Here, we reflect on this in comparison to the other UIs in our study.

\subsubsection{\modeours{} Offers Flexible Workflows}
Our study revealed how \modeours{} grants users high flexibility. Concretely, people achieve different workflows by varying their use of local responses in combination with AI support features. We identified three clusters (\cref{fig:workflow_scatterplot}). This demonstrates that people have varying preferences when responding to emails with AI, even in a controlled study setting, which \modeours{} was able to support.
\revision{Concretely,} the observed workflows differed in when and how people transitioned from the local response view to the ``global'' draft view (\cref{sec:results_workflows}). %

\subsubsection{\modeours{} has a Lower ``AI Footprint''}
All AI features across \modeours{} and \modemail{} shared these significant influences: reduced manual typing and error rates, with increased reply lengths and reduced semantic and lexical diversity. 

Despite these shared influences, \modeours{} showed a lower ``AI footprint''.
As estimated with our statistical models, with \modemail{} (compared to \modeours) people finished their replies even faster (-\secs{66}), with even fewer manual keystrokes (-\pct{20}), while producing even longer replies (+\pct{28}). Producing more, faster, with less input indicates a dearth of human intention in writing, as recently highlighted by \citet{kreminski2024dearthauthor}. Indeed, \modemail{} resulted in the lowest diversity between emails.
Together with our subjective inspection of the emails (\cref{sec:result_quality}), this paints the picture of somewhat bloated replies with full reply generation (\modemail). This is in line with the finding by \citet{fu2024texttoself} that verbosity is a key drawback noticed by users of AI tools for communication. As our study shows, people could keep this under control better with \modeours{}. In summary, \modeours{} allows users to benefit from AI in a more nuanced and controlled manner.

\subsubsection{Optional Message-Level Support Partly Bridges the Differences}
Manual control might not always be the user's main concern. \modeours{} also accounts for workflows with stronger delegation to AI, as shown by clustering the workflows (\cref{fig:workflow_scatterplot}) and by factoring out the impact of the \imppass{} feature. If it was used, it narrowed the differences between \modeours{} and \modemail{} along several metrics (\cref{tab:lmm_overview}), \revision{which fits to the literature that compared} sentence-level vs message-level AI support~\cite{Fu2023sentencevsmessage}. %
In conclusion, by making use of the optional \imppass{} in \modeours{}, users can situationally vary the described tradeoffs between \modemail{} and our local response design.

\subsection{\revision{Deciding Between Sentence- and Message-level Support (\modeours{} vs \modemail)}}
Participants' reasonings for their favourite UI (\cref{sec:results_fav_mode}) \revision{help us understand which types of tasks might benefit from which type of AI support}. 
\revision{When favouring \modemail}, they referred to aspects of efficiency in the majority of their comments, in line with quantitative results. \revision{This indicates that this type of support might be favourable for communication tasks in which users deem the speed of their reply to be more relevant than the exact wording (e.g. decision-focused emails, such as accepting or declining a time-sensitive request).}

In contrast, \revision{for \modeours, participants} emphasised quality and control aspects relatively more frequently. \revision{Thus, this type of support lends itself to communication tasks that require more careful and/or personal wording. Concretely,  \citet{Robertson2021cantreply} identified relevant themes in emailing for this, such as caring about personal authenticity, semantic and tonal coherence, relationship types, and norms and culture. Related, \citet{Mieczkowski2022thesis} identified dimensions of agency in AIMC. Our \modeours{} design and participants' comments on it fit to their findings about people's strategic thinking for maintaining agency (e.g. ability for choosing, editing, replacing, not-sending AI text, and/or seeing its changes due to own input).} 

\revision{Supporting this interpretation}, comments by those who did not pick a favourite, such as P19\oldId{P1290}, described that the choice depends on contextual factors, such as using \modemail{} under time pressure, while using \modeours{} when a ``tailored'' reply was needed.

\revision{Related}, we found evidence that the value of local responses is higher for longer emails (\cref{sec:results_workflows}). Future work could investigate the underlying factors in detail. For example, the value of the local response step \revision{(and thus sentence-level support)} might depend not only on the length but also on aspects of complexity of the incoming email \revision{(e.g. emails with multiple questions or describing a complex ``history'' of a situation as background for a request)}.

\revision{Finally,} while we separated the concepts for our comparative study, a real email app could combine the UIs for \modeours{} and \modemail. %
Note that our CDLR design already supports a message-level workflow by skipping to screen 2 and using the \imppass{} feature. %
It was sometimes already used in this way (\cref{sec:results_workflows}), as discussed next.

\subsection{Combined Sentence- and Message-level Support Offers Nuanced Control over AI Involvement}

Evaluating our \modeours{} concept also allowed us to gain insights into the \textit{combination} of different levels of writing support, not studied so far in the literature. 
While related work addressed AI on either sentence-level or message-level~\cite{Fu2023sentencevsmessage, Chen2019smartcompose, Kannan2016smartreply}, we explored the design space in between, with a flexible mix. %
We found that this combination gives users more nuanced control over how they involve AI. 

Concretely, the way people mixed the different degrees of engaging with AI  in \modeours{} (\cref{sec:results_workflows}) had a meaningful impact on metrics of interaction and outcome (\cref{sec:discussion_results_overview}). 
Fittingly, people appreciated \modeours{} particularly for aspects related to control and quality (\cref{sec:results_perception}). %
In summary, we thus conclude that offering optional AI support on different levels is accepted by users in this context and provides more nuanced control over AI involvement and its impact. %

Our design partly contradicts product trends, which generate complete replies and focus on workflows that assume that people want to start by delegating more to the AI (\cref{sec:related_work_current_products}). In fact, participants in our study expressed diverging preferences on this, both in behaviour (\cref{sec:results_workflows}) and reflection (\cref{sec:results_fav_mode}). In this context, our findings highlight a variety in workflows that motivates further exploration of flexible designs. Here, our concept of \modeours{} provides inspiration for offering different levels of optional AI support, to improve various response workflows or writing workflows in general. %

\subsection{Reflections on Methodology}\label{sec:discussion_methods}

Many suggestions in the formative study were accepted without input (\cref{sec:formative_study_results}), potentially to satisfice, that is, to finish the study faster even with less than ideal results. We thus added briefings in the main study (\cref{sec:procedure_email_tasks}), which provided topic-specific data about how to reply (e.g. unavailable on day X, not selling below a certain price). We did \textit{not} give these to the LLM, since we did not want to simulate a ``mindreading'' system. %
    
Interestingly, the ratio of emails that did not cover the briefings' key aspects was \pct{5} higher for \modeours{} than for \modemailtxt, and \pct{10} higher than for manual writing (\cref{sec:results_briefing}). 
We explain this as follows: Typing manually, people might as well enter a reply in line with the briefing. In contrast, they might accept AI text, if available, to be fast without checking too closely. 

Our data supports this: 
We found the lowest briefing conformity (\pct{50}, \cref{sec:results_briefing}) among replies written without investing in prompts, and thus fully optimised for speed, in both \modeours{} and \modemail{}. This lack of entering prompts had a much larger impact on conformity than the UI in general (\cref{tab:lmm_overview2}, row 1). In this light, the \pct{5} difference between \modeours{} and \modemail{} might be attributed to the fact that \modemail{} only had full response generation, with no other features for users to consider instead of focusing on the one generated text. Additionally, \modeours{} automatically loads suggestions once a sentence is selected, which might make satisficing more tempting. 

\revision{More broadly, in a paid online study such as ours, participants likely feel the incentive to complete the task quickly. While real mobile emailing contexts also include situations with time pressure, our study had no negative consequences of writing a less than ideal email. Therefore, the willingness to accept suggestions might be inflated in the study. That said, in real life, people might also have reasons to vary their investment (e.g. based on the importance of the email), which is supported by the flexibility of \modeours. Overall, the timings and thoughtful responses to our (optional) open questions indicate that participants did not merely ``skip'' through the study.} %
\revision{Supporting this,} the majority of emails per UI (>\pct{76}) indeed included the key information from the briefings (\cref{sec:results_briefing}), and the subjective quality of the writing was high (even when not briefing conform) across all UIs (\cref{sec:result_quality}).

In summary, we found evidence that the presence of AI automation features for a study task reduces task adherence, possibly because people then choose different speed-performance tradeoffs \revision{for the study and/or} monitor their results and performance less closely. This is consistent with findings in related work that people believe that AI improves their performance, even when it does not~\cite{Kloft2024placeborobust}.
A future dedicated study could investigate this in more detail since this is potentially relevant for many human-AI interaction studies.

\revision{Finally, such studies could let participants themselves or third parties assess the quality of produced texts to capture self-perception and external perception. Following research interests in AI-mediated communication (cf.~\cite{fu2024texttoself, Hancock2020aimc}), this would focus not only on the sender's interaction, as in our case here, but also capture the perception of the receivers.}

\subsection{Reflections on Prompting: Co-creative Chain-of-Thought Prompting}\label{sec:discussion_prompting}
We use \textit{Chain-of-Thought (CoT)} as a lens to reflect on our prompting.
CoT instructs an LLM to generate intermediate steps~\cite{wei2022chain} (e.g. by adding``Let's think step by step''~\cite{kojima2022large}). 
In \modeours{}, users prompt intermediate generation steps by selecting text in the email, while our \imppass{} feature then prompts the LLM to generate a full reply from these texts when used at the end. 
In this view, the user contributes to the intermediate steps on the way to the final result. Thus, this prompting workflow could be seen as a kind of user-involved CoT, or \textit{Co-creative CoT (Co-CoT).}

This is reflected in people's comments on prompting: ``[\modeours] was the best to express opinions in a way that isn't just one single prompt, creating a more real feeling of thoughts behind it that AI tends to lack in single prompt emails.'' (P117\oldId{P1424}) %
We believe that this concept is useful more broadly: Rather than only writing a meta-instruction such as ``Let's think step by step'', the user actively contributes to this step-wise ``thinking'' along the way.

\section{Conclusion}\label{sec:conclusion}

We have proposed the concept of \textit{\modeoursTXT{} (\modeours)}, which %
allows users to insert concise responses at selected points in the incoming email as they read it. Selecting text for this purpose doubles as an expression of intent to guide the AI's text suggestions. These local responses can then be further edited manually or again with AI support. This design for the first time combines sentence-level and message-level AI support, while keeping both optional.
The key benefit is flexibility. Involving AI with \modeours{} shares the general characteristics of AI responses but its modularity allows users to make their own choices. %

Participants' comments highlight the distinct strengths of \modeours{} and the involved tradeoffs. They overall appreciated all AI features but when asked to pick a favourite, their reasoning for message-level AI highlighted efficiency in the majority of cases, while for \modeours{} they emphasised quality and control relatively more often.

Future work could explore how people make use of \modeours{} features in real life emailing, with varying social and contextual factors~\cite{Robertson2021cantreply}, and in long-term use.

While recent feature releases in widely used email apps have focused on AI that generates complete emails, our findings with \modeours{} motivate the exploration of more flexible alternatives. Rather than adding AI ``on top'', we motivate designing for users' underlying workflows -- both for those with and without AI at the same time. In this way, flexible UIs can empower users to dynamically adjust their desired degree of AI involvement to manage tradeoffs, such as between responding quickly and retaining control over the outcome.

We release our prototype and \revision{study} material in this project repository to support future research:

\url{https://osf.io/fsxzv/}

\begin{acks}
Funded by the Deutsche Forschungsgemeinschaft (DFG, German Research Foundation) -- 525037874.
\end{acks}

\bibliographystyle{ACM-Reference-Format}
\bibliography{bibliography}


\begin{thebibliography}{52}


\ifx \showCODEN    \undefined \def \showCODEN     #1{\unskip}     \fi
\ifx \showDOI      \undefined \def \showDOI       #1{#1}\fi
\ifx \showISBNx    \undefined \def \showISBNx     #1{\unskip}     \fi
\ifx \showISBNxiii \undefined \def \showISBNxiii  #1{\unskip}     \fi
\ifx \showISSN     \undefined \def \showISSN      #1{\unskip}     \fi
\ifx \showLCCN     \undefined \def \showLCCN      #1{\unskip}     \fi
\ifx \shownote     \undefined \def \shownote      #1{#1}          \fi
\ifx \showarticletitle \undefined \def \showarticletitle #1{#1}   \fi
\ifx \showURL      \undefined \def \showURL       {\relax}        \fi
\providecommand\bibfield[2]{#2}
\providecommand\bibinfo[2]{#2}
\providecommand\natexlab[1]{#1}
\providecommand\showeprint[2][]{arXiv:#2}

\bibitem[AI@Meta(2024)]%
        {llama3modelcard}
\bibfield{author}{\bibinfo{person}{AI@Meta}.} \bibinfo{year}{2024}\natexlab{}.
\newblock \showarticletitle{Llama 3 Model Card}.
\newblock  (\bibinfo{year}{2024}).
\newblock
\urldef\tempurl%
\url{https://github.com/meta-llama/llama3/blob/main/MODEL_CARD.md}
\showURL{%
\tempurl}


\bibitem[August et~al\mbox{.}(2020)]%
        {august2020microwriting}
\bibfield{author}{\bibinfo{person}{Tal August}, \bibinfo{person}{Shamsi Iqbal},
  \bibinfo{person}{Michael Gamon}, {and} \bibinfo{person}{Mark
  Encarnaci\'{o}n}.} \bibinfo{year}{2020}\natexlab{}.
\newblock \showarticletitle{Characterizing the Mobile Microtask Writing
  Process}. In \bibinfo{booktitle}{\emph{22nd International Conference on
  Human-Computer Interaction with Mobile Devices and Services}} (Oldenburg,
  Germany) \emph{(\bibinfo{series}{MobileHCI '20})}.
  \bibinfo{publisher}{Association for Computing Machinery},
  \bibinfo{address}{New York, NY, USA}, Article \bibinfo{articleno}{26},
  \bibinfo{numpages}{12}~pages.
\newblock
\showISBNx{9781450375160}
\urldef\tempurl%
\url{https://doi.org/10.1145/3379503.3403541}
\showDOI{\tempurl}


\bibitem[Bao et~al\mbox{.}(2011)]%
        {bao2011phoneuse}
\bibfield{author}{\bibinfo{person}{Patti Bao}, \bibinfo{person}{Jeffrey
  Pierce}, \bibinfo{person}{Stephen Whittaker}, {and} \bibinfo{person}{Shumin
  Zhai}.} \bibinfo{year}{2011}\natexlab{}.
\newblock \showarticletitle{Smart phone use by non-mobile business users}. In
  \bibinfo{booktitle}{\emph{Proceedings of the 13th International Conference on
  Human Computer Interaction with Mobile Devices and Services}} (Stockholm,
  Sweden) \emph{(\bibinfo{series}{MobileHCI '11})}.
  \bibinfo{publisher}{Association for Computing Machinery},
  \bibinfo{address}{New York, NY, USA}, \bibinfo{pages}{445–454}.
\newblock
\showISBNx{9781450305419}
\urldef\tempurl%
\url{https://doi.org/10.1145/2037373.2037440}
\showDOI{\tempurl}


\bibitem[Bates et~al\mbox{.}(2015)]%
        {Bates2015}
\bibfield{author}{\bibinfo{person}{Douglas Bates}, \bibinfo{person}{Martin
  M{\"a}chler}, \bibinfo{person}{Ben Bolker}, {and} \bibinfo{person}{Steve
  Walker}.} \bibinfo{year}{2015}\natexlab{}.
\newblock \showarticletitle{Fitting Linear Mixed-Effects Models Using {lme4}}.
\newblock \bibinfo{journal}{\emph{Journal of Statistical Software}}
  \bibinfo{volume}{67}, \bibinfo{number}{1} (\bibinfo{year}{2015}),
  \bibinfo{pages}{1--48}.
\newblock
\urldef\tempurl%
\url{https://doi.org/10.18637/jss.v067.i01}
\showDOI{\tempurl}


\bibitem[Brooke et~al\mbox{.}(1996)]%
        {brooke1996sus}
\bibfield{author}{\bibinfo{person}{John Brooke} {et~al\mbox{.}}}
  \bibinfo{year}{1996}\natexlab{}.
\newblock \showarticletitle{SUS-A quick and dirty usability scale}.
\newblock \bibinfo{journal}{\emph{Usability evaluation in industry}}
  \bibinfo{volume}{189}, \bibinfo{number}{194} (\bibinfo{year}{1996}),
  \bibinfo{pages}{4--7}.
\newblock


\bibitem[Buschek et~al\mbox{.}(2021)]%
        {Buschek2021chi}
\bibfield{author}{\bibinfo{person}{Daniel Buschek}, \bibinfo{person}{Martin
  Z\"{u}rn}, {and} \bibinfo{person}{Malin Eiband}.}
  \bibinfo{year}{2021}\natexlab{}.
\newblock \showarticletitle{The Impact of Multiple Parallel Phrase Suggestions
  on Email Input and Composition Behaviour of Native and Non-Native English
  Writers}. In \bibinfo{booktitle}{\emph{Proceedings of the 2021 CHI Conference
  on Human Factors in Computing Systems}} (Yokohama, Japan)
  \emph{(\bibinfo{series}{CHI '21})}. \bibinfo{publisher}{Association for
  Computing Machinery}, \bibinfo{address}{New York, NY, USA}, Article
  \bibinfo{articleno}{732}, \bibinfo{numpages}{13}~pages.
\newblock
\showISBNx{9781450380966}
\urldef\tempurl%
\url{https://doi.org/10.1145/3411764.3445372}
\showDOI{\tempurl}


\bibitem[Chen et~al\mbox{.}(2019)]%
        {Chen2019smartcompose}
\bibfield{author}{\bibinfo{person}{Mia~Xu Chen}, \bibinfo{person}{Benjamin~N.
  Lee}, \bibinfo{person}{Gagan Bansal}, \bibinfo{person}{Yuan Cao},
  \bibinfo{person}{Shuyuan Zhang}, \bibinfo{person}{Justin Lu},
  \bibinfo{person}{Jackie Tsay}, \bibinfo{person}{Yinan Wang},
  \bibinfo{person}{Andrew~M. Dai}, \bibinfo{person}{Zhifeng Chen},
  \bibinfo{person}{Timothy Sohn}, {and} \bibinfo{person}{Yonghui Wu}.}
  \bibinfo{year}{2019}\natexlab{}.
\newblock \showarticletitle{Gmail Smart Compose: Real-Time Assisted Writing}.
  In \bibinfo{booktitle}{\emph{Proceedings of the 25th ACM SIGKDD International
  Conference on Knowledge Discovery \& Data Mining}} (Anchorage, AK, USA)
  \emph{(\bibinfo{series}{KDD '19})}. \bibinfo{publisher}{Association for
  Computing Machinery}, \bibinfo{address}{New York, NY, USA},
  \bibinfo{pages}{2287–2295}.
\newblock
\showISBNx{9781450362016}
\urldef\tempurl%
\url{https://doi.org/10.1145/3292500.3330723}
\showDOI{\tempurl}


\bibitem[Cheng et~al\mbox{.}(2015)]%
        {cheng2015breakitdown}
\bibfield{author}{\bibinfo{person}{Justin Cheng}, \bibinfo{person}{Jaime
  Teevan}, \bibinfo{person}{Shamsi~T. Iqbal}, {and} \bibinfo{person}{Michael~S.
  Bernstein}.} \bibinfo{year}{2015}\natexlab{}.
\newblock \showarticletitle{Break It Down: A Comparison of Macro- and
  Microtasks}. In \bibinfo{booktitle}{\emph{Proceedings of the 33rd Annual ACM
  Conference on Human Factors in Computing Systems}} (Seoul, Republic of Korea)
  \emph{(\bibinfo{series}{CHI '15})}. \bibinfo{publisher}{Association for
  Computing Machinery}, \bibinfo{address}{New York, NY, USA},
  \bibinfo{pages}{4061–4064}.
\newblock
\showISBNx{9781450331456}
\urldef\tempurl%
\url{https://doi.org/10.1145/2702123.2702146}
\showDOI{\tempurl}


\bibitem[Corbin(1990)]%
        {corbin1990basics}
\bibfield{author}{\bibinfo{person}{Juliet~M Corbin}.}
  \bibinfo{year}{1990}\natexlab{}.
\newblock \bibinfo{booktitle}{\emph{Basics of qualitative research: Grounded
  theory procedures and techniques}}.
\newblock \bibinfo{publisher}{Sage}.
\newblock


\bibitem[Dang et~al\mbox{.}(2023)]%
        {Dang2023diegetic}
\bibfield{author}{\bibinfo{person}{Hai Dang}, \bibinfo{person}{Sven Goller},
  \bibinfo{person}{Florian Lehmann}, {and} \bibinfo{person}{Daniel Buschek}.}
  \bibinfo{year}{2023}\natexlab{}.
\newblock \showarticletitle{Choice Over Control: How Users Write with Large
  Language Models using Diegetic and Non-Diegetic Prompting}. In
  \bibinfo{booktitle}{\emph{Proceedings of the 2023 CHI Conference on Human
  Factors in Computing Systems}} (Hamburg, Germany) \emph{(\bibinfo{series}{CHI
  '23})}. \bibinfo{publisher}{Association for Computing Machinery},
  \bibinfo{address}{New York, NY, USA}, Article \bibinfo{articleno}{408},
  \bibinfo{numpages}{17}~pages.
\newblock
\showISBNx{9781450394215}
\urldef\tempurl%
\url{https://doi.org/10.1145/3544548.3580969}
\showDOI{\tempurl}


\bibitem[Elkin et~al\mbox{.}(2021)]%
        {elkin2021artc}
\bibfield{author}{\bibinfo{person}{Lisa~A. Elkin}, \bibinfo{person}{Matthew
  Kay}, \bibinfo{person}{James~J. Higgins}, {and} \bibinfo{person}{Jacob~O.
  Wobbrock}.} \bibinfo{year}{2021}\natexlab{}.
\newblock \showarticletitle{An Aligned Rank Transform Procedure for Multifactor
  Contrast Tests}. In \bibinfo{booktitle}{\emph{The 34th Annual ACM Symposium
  on User Interface Software and Technology}} (Virtual Event, USA)
  \emph{(\bibinfo{series}{UIST '21})}. \bibinfo{publisher}{Association for
  Computing Machinery}, \bibinfo{address}{New York, NY, USA},
  \bibinfo{pages}{754–768}.
\newblock
\showISBNx{9781450386357}
\urldef\tempurl%
\url{https://doi.org/10.1145/3472749.3474784}
\showDOI{\tempurl}


\bibitem[Fu et~al\mbox{.}(2023)]%
        {Fu2023sentencevsmessage}
\bibfield{author}{\bibinfo{person}{Liye Fu}, \bibinfo{person}{Benjamin Newman},
  \bibinfo{person}{Maurice Jakesch}, {and} \bibinfo{person}{Sarah Kreps}.}
  \bibinfo{year}{2023}\natexlab{}.
\newblock \showarticletitle{Comparing Sentence-Level Suggestions to
  Message-Level Suggestions in AI-Mediated Communication}. In
  \bibinfo{booktitle}{\emph{Proceedings of the 2023 CHI Conference on Human
  Factors in Computing Systems}} (Hamburg, Germany) \emph{(\bibinfo{series}{CHI
  '23})}. \bibinfo{publisher}{Association for Computing Machinery},
  \bibinfo{address}{New York, NY, USA}, Article \bibinfo{articleno}{103},
  \bibinfo{numpages}{13}~pages.
\newblock
\showISBNx{9781450394215}
\urldef\tempurl%
\url{https://doi.org/10.1145/3544548.3581351}
\showDOI{\tempurl}


\bibitem[Fu et~al\mbox{.}(2024)]%
        {fu2024texttoself}
\bibfield{author}{\bibinfo{person}{Yue Fu}, \bibinfo{person}{Sami Foell},
  \bibinfo{person}{Xuhai Xu}, {and} \bibinfo{person}{Alexis Hiniker}.}
  \bibinfo{year}{2024}\natexlab{}.
\newblock \showarticletitle{From Text to Self: Users’ Perception of AIMC
  Tools on Interpersonal Communication and Self}. In
  \bibinfo{booktitle}{\emph{Proceedings of the CHI Conference on Human Factors
  in Computing Systems}} (Honolulu, HI, USA) \emph{(\bibinfo{series}{CHI
  '24})}. \bibinfo{publisher}{Association for Computing Machinery},
  \bibinfo{address}{New York, NY, USA}, Article \bibinfo{articleno}{977},
  \bibinfo{numpages}{17}~pages.
\newblock
\showISBNx{9798400703300}
\urldef\tempurl%
\url{https://doi.org/10.1145/3613904.3641955}
\showDOI{\tempurl}


\bibitem[Goodman et~al\mbox{.}(2022)]%
        {Goodman2022lampost}
\bibfield{author}{\bibinfo{person}{Steven~M. Goodman}, \bibinfo{person}{Erin
  Buehler}, \bibinfo{person}{Patrick Clary}, \bibinfo{person}{Andy Coenen},
  \bibinfo{person}{Aaron Donsbach}, \bibinfo{person}{Tiffanie~N. Horne},
  \bibinfo{person}{Michal Lahav}, \bibinfo{person}{Robert MacDonald},
  \bibinfo{person}{Rain~Breaw Michaels}, \bibinfo{person}{Ajit Narayanan},
  \bibinfo{person}{Mahima Pushkarna}, \bibinfo{person}{Joel Riley},
  \bibinfo{person}{Alex Santana}, \bibinfo{person}{Lei Shi},
  \bibinfo{person}{Rachel Sweeney}, \bibinfo{person}{Phil Weaver},
  \bibinfo{person}{Ann Yuan}, {and} \bibinfo{person}{Meredith~Ringel Morris}.}
  \bibinfo{year}{2022}\natexlab{}.
\newblock \showarticletitle{LaMPost: Design and Evaluation of an AI-assisted
  Email Writing Prototype for Adults with Dyslexia}. In
  \bibinfo{booktitle}{\emph{Proceedings of the 24th International ACM SIGACCESS
  Conference on Computers and Accessibility}} (Athens, Greece)
  \emph{(\bibinfo{series}{ASSETS '22})}. \bibinfo{publisher}{Association for
  Computing Machinery}, \bibinfo{address}{New York, NY, USA}, Article
  \bibinfo{articleno}{24}, \bibinfo{numpages}{18}~pages.
\newblock
\showISBNx{9781450392587}
\urldef\tempurl%
\url{https://doi.org/10.1145/3517428.3544819}
\showDOI{\tempurl}


\bibitem[Google(2024)]%
        {google2024geminiwebsite}
\bibfield{author}{\bibinfo{person}{Google}.} \bibinfo{year}{2024}\natexlab{}.
\newblock \bibinfo{title}{Draft emails with {Gemini} in {Gmail} - {Android} -
  {Gmail} {Help}}.
\newblock
\newblock
\urldef\tempurl%
\url{https://support.google.com/mail/answer/13955415?co=GENIE.Platform%3DAndroid&oco=1}
\showURL{%
\tempurl}


\bibitem[Hancock et~al\mbox{.}(2020)]%
        {Hancock2020aimc}
\bibfield{author}{\bibinfo{person}{Jeffrey~T Hancock}, \bibinfo{person}{Mor
  Naaman}, {and} \bibinfo{person}{Karen Levy}.}
  \bibinfo{year}{2020}\natexlab{}.
\newblock \showarticletitle{{AI-Mediated Communication: Definition, Research
  Agenda, and Ethical Considerations}}.
\newblock \bibinfo{journal}{\emph{Journal of Computer-Mediated Communication}}
  \bibinfo{volume}{25}, \bibinfo{number}{1} (\bibinfo{date}{01}
  \bibinfo{year}{2020}), \bibinfo{pages}{89--100}.
\newblock
\showISSN{1083-6101}
\urldef\tempurl%
\url{https://doi.org/10.1093/jcmc/zmz022}
\showDOI{\tempurl}
\showeprint{https://academic.oup.com/jcmc/article-pdf/25/1/89/32961176/zmz022.pdf}


\bibitem[Iqbal et~al\mbox{.}(2018)]%
        {iqbal2018playwrite}
\bibfield{author}{\bibinfo{person}{Shamsi~T. Iqbal}, \bibinfo{person}{Jaime
  Teevan}, \bibinfo{person}{Dan Liebling}, {and} \bibinfo{person}{Anne~Loomis
  Thompson}.} \bibinfo{year}{2018}\natexlab{}.
\newblock \showarticletitle{Multitasking with Play Write, a Mobile
  Microproductivity Writing Tool}. In \bibinfo{booktitle}{\emph{Proceedings of
  the 31st Annual ACM Symposium on User Interface Software and Technology}}
  (Berlin, Germany) \emph{(\bibinfo{series}{UIST '18})}.
  \bibinfo{publisher}{Association for Computing Machinery},
  \bibinfo{address}{New York, NY, USA}, \bibinfo{pages}{411–422}.
\newblock
\showISBNx{9781450359481}
\urldef\tempurl%
\url{https://doi.org/10.1145/3242587.3242611}
\showDOI{\tempurl}


\bibitem[Kannan et~al\mbox{.}(2016)]%
        {Kannan2016smartreply}
\bibfield{author}{\bibinfo{person}{Anjuli Kannan}, \bibinfo{person}{Karol
  Kurach}, \bibinfo{person}{Sujith Ravi}, \bibinfo{person}{Tobias Kaufmann},
  \bibinfo{person}{Andrew Tomkins}, \bibinfo{person}{Balint Miklos},
  \bibinfo{person}{Greg Corrado}, \bibinfo{person}{Laszlo Lukacs},
  \bibinfo{person}{Marina Ganea}, \bibinfo{person}{Peter Young}, {and}
  \bibinfo{person}{Vivek Ramavajjala}.} \bibinfo{year}{2016}\natexlab{}.
\newblock \showarticletitle{Smart Reply: Automated Response Suggestion for
  Email}. In \bibinfo{booktitle}{\emph{Proceedings of the 22nd ACM SIGKDD
  International Conference on Knowledge Discovery and Data Mining}} (San
  Francisco, California, USA) \emph{(\bibinfo{series}{KDD '16})}.
  \bibinfo{publisher}{Association for Computing Machinery},
  \bibinfo{address}{New York, NY, USA}, \bibinfo{pages}{955–964}.
\newblock
\showISBNx{9781450342322}
\urldef\tempurl%
\url{https://doi.org/10.1145/2939672.2939801}
\showDOI{\tempurl}


\bibitem[Kloft et~al\mbox{.}(2024)]%
        {Kloft2024placeborobust}
\bibfield{author}{\bibinfo{person}{Agnes~Mercedes Kloft},
  \bibinfo{person}{Robin Welsch}, \bibinfo{person}{Thomas Kosch}, {and}
  \bibinfo{person}{Steeven Villa}.} \bibinfo{year}{2024}\natexlab{}.
\newblock \showarticletitle{"AI enhances our performance, I have no doubt this
  one will do the same": The Placebo effect is robust to negative descriptions
  of AI}. In \bibinfo{booktitle}{\emph{Proceedings of the CHI Conference on
  Human Factors in Computing Systems}} (Honolulu, HI, USA)
  \emph{(\bibinfo{series}{CHI '24})}. \bibinfo{publisher}{Association for
  Computing Machinery}, \bibinfo{address}{New York, NY, USA}, Article
  \bibinfo{articleno}{299}, \bibinfo{numpages}{24}~pages.
\newblock
\showISBNx{9798400703300}
\urldef\tempurl%
\url{https://doi.org/10.1145/3613904.3642633}
\showDOI{\tempurl}


\bibitem[Kojima et~al\mbox{.}(2022)]%
        {kojima2022large}
\bibfield{author}{\bibinfo{person}{Takeshi Kojima},
  \bibinfo{person}{Shixiang~Shane Gu}, \bibinfo{person}{Machel Reid},
  \bibinfo{person}{Yutaka Matsuo}, {and} \bibinfo{person}{Yusuke Iwasawa}.}
  \bibinfo{year}{2022}\natexlab{}.
\newblock \showarticletitle{Large language models are zero-shot reasoners}.
\newblock \bibinfo{journal}{\emph{Advances in neural information processing
  systems}}  \bibinfo{volume}{35} (\bibinfo{year}{2022}),
  \bibinfo{pages}{22199--22213}.
\newblock


\bibitem[Kreminski(2024)]%
        {kreminski2024dearthauthor}
\bibfield{author}{\bibinfo{person}{Max Kreminski}.}
  \bibinfo{year}{2024}\natexlab{}.
\newblock \bibinfo{title}{The Dearth of the Author in AI-Supported Writing}.
\newblock
\newblock
\showeprint[arxiv]{2404.10289}~[cs.HC]
\urldef\tempurl%
\url{https://arxiv.org/abs/2404.10289}
\showURL{%
\tempurl}


\bibitem[Kristensson and Vertanen(2014)]%
        {kristensson2014inviscid}
\bibfield{author}{\bibinfo{person}{Per~Ola Kristensson} {and}
  \bibinfo{person}{Keith Vertanen}.} \bibinfo{year}{2014}\natexlab{}.
\newblock \showarticletitle{The inviscid text entry rate and its application as
  a grand goal for mobile text entry}. In \bibinfo{booktitle}{\emph{Proceedings
  of the 16th International Conference on Human-Computer Interaction with
  Mobile Devices \& Services}} (Toronto, ON, Canada)
  \emph{(\bibinfo{series}{MobileHCI '14})}. \bibinfo{publisher}{Association for
  Computing Machinery}, \bibinfo{address}{New York, NY, USA},
  \bibinfo{pages}{335–338}.
\newblock
\showISBNx{9781450330046}
\urldef\tempurl%
\url{https://doi.org/10.1145/2628363.2628405}
\showDOI{\tempurl}


\bibitem[Kuznetsova et~al\mbox{.}(2017)]%
        {Kuznetsova2017}
\bibfield{author}{\bibinfo{person}{Alexandra Kuznetsova},
  \bibinfo{person}{Per~B. Brockhoff}, {and} \bibinfo{person}{Rune H.~B.
  Christensen}.} \bibinfo{year}{2017}\natexlab{}.
\newblock \showarticletitle{{lmerTest} Package: Tests in Linear Mixed Effects
  Models}.
\newblock \bibinfo{journal}{\emph{Journal of Statistical Software}}
  \bibinfo{volume}{82}, \bibinfo{number}{13} (\bibinfo{year}{2017}),
  \bibinfo{pages}{1--26}.
\newblock
\urldef\tempurl%
\url{https://doi.org/10.18637/jss.v082.i13}
\showDOI{\tempurl}


\bibitem[Leiva et~al\mbox{.}(2012)]%
        {leiva2012backtoapp}
\bibfield{author}{\bibinfo{person}{Luis Leiva}, \bibinfo{person}{Matthias
  B\"{o}hmer}, \bibinfo{person}{Sven Gehring}, {and} \bibinfo{person}{Antonio
  Kr\"{u}ger}.} \bibinfo{year}{2012}\natexlab{}.
\newblock \showarticletitle{Back to the app: the costs of mobile application
  interruptions}. In \bibinfo{booktitle}{\emph{Proceedings of the 14th
  International Conference on Human-Computer Interaction with Mobile Devices
  and Services}} (San Francisco, California, USA)
  \emph{(\bibinfo{series}{MobileHCI '12})}. \bibinfo{publisher}{Association for
  Computing Machinery}, \bibinfo{address}{New York, NY, USA},
  \bibinfo{pages}{291–294}.
\newblock
\showISBNx{9781450311052}
\urldef\tempurl%
\url{https://doi.org/10.1145/2371574.2371617}
\showDOI{\tempurl}


\bibitem[Lewin-Jones(2014)]%
        {lewi_jones2014email}
\bibfield{author}{\bibinfo{person}{Jenny Lewin-Jones}.}
  \bibinfo{year}{2014}\natexlab{}.
\newblock \showarticletitle{Understanding style, language and etiquette in
  email communication in higher education: a survey}.
\newblock \bibinfo{journal}{\emph{Research in Post-Compulsory Education}}
  \bibinfo{volume}{19} (\bibinfo{date}{01} \bibinfo{year}{2014}),
  \bibinfo{pages}{75--90}.
\newblock
\urldef\tempurl%
\url{https://doi.org/10.1080/13596748.2014.872934}
\showDOI{\tempurl}


\bibitem[Li et~al\mbox{.}(2024)]%
        {Li2024aivalue}
\bibfield{author}{\bibinfo{person}{Zhuoyan Li}, \bibinfo{person}{Chen Liang},
  \bibinfo{person}{Jing Peng}, {and} \bibinfo{person}{Ming Yin}.}
  \bibinfo{year}{2024}\natexlab{}.
\newblock \showarticletitle{The Value, Benefits, and Concerns of Generative
  AI-Powered Assistance in Writing}. In \bibinfo{booktitle}{\emph{Proceedings
  of the CHI Conference on Human Factors in Computing Systems}} (Honolulu, HI,
  USA) \emph{(\bibinfo{series}{CHI '24})}. \bibinfo{publisher}{Association for
  Computing Machinery}, \bibinfo{address}{New York, NY, USA}, Article
  \bibinfo{articleno}{1048}, \bibinfo{numpages}{25}~pages.
\newblock
\showISBNx{9798400703300}
\urldef\tempurl%
\url{https://doi.org/10.1145/3613904.3642625}
\showDOI{\tempurl}


\bibitem[Lin et~al\mbox{.}(2024)]%
        {lin2024rambler}
\bibfield{author}{\bibinfo{person}{Susan Lin}, \bibinfo{person}{Jeremy Warner},
  \bibinfo{person}{J.D. Zamfirescu-Pereira}, \bibinfo{person}{Matthew~G Lee},
  \bibinfo{person}{Sauhard Jain}, \bibinfo{person}{Shanqing Cai},
  \bibinfo{person}{Piyawat Lertvittayakumjorn}, \bibinfo{person}{Michael~Xuelin
  Huang}, \bibinfo{person}{Shumin Zhai}, \bibinfo{person}{Bjoern Hartmann},
  {and} \bibinfo{person}{Can Liu}.} \bibinfo{year}{2024}\natexlab{}.
\newblock \showarticletitle{Rambler: Supporting Writing With Speech via
  LLM-Assisted Gist Manipulation}. In \bibinfo{booktitle}{\emph{Proceedings of
  the CHI Conference on Human Factors in Computing Systems}} (Honolulu, HI,
  USA) \emph{(\bibinfo{series}{CHI '24})}. \bibinfo{publisher}{Association for
  Computing Machinery}, \bibinfo{address}{New York, NY, USA}, Article
  \bibinfo{articleno}{1043}, \bibinfo{numpages}{19}~pages.
\newblock
\showISBNx{9798400703300}
\urldef\tempurl%
\url{https://doi.org/10.1145/3613904.3642217}
\showDOI{\tempurl}


\bibitem[Liu et~al\mbox{.}(2022)]%
        {Liu2022aimailperception}
\bibfield{author}{\bibinfo{person}{Yihe Liu}, \bibinfo{person}{Anushk Mittal},
  \bibinfo{person}{Diyi Yang}, {and} \bibinfo{person}{Amy Bruckman}.}
  \bibinfo{year}{2022}\natexlab{}.
\newblock \showarticletitle{Will AI Console Me when I Lose my Pet?
  Understanding Perceptions of AI-Mediated Email Writing}. In
  \bibinfo{booktitle}{\emph{Proceedings of the 2022 CHI Conference on Human
  Factors in Computing Systems}} (New Orleans, LA, USA)
  \emph{(\bibinfo{series}{CHI '22})}. \bibinfo{publisher}{Association for
  Computing Machinery}, \bibinfo{address}{New York, NY, USA}, Article
  \bibinfo{articleno}{474}, \bibinfo{numpages}{13}~pages.
\newblock
\showISBNx{9781450391573}
\urldef\tempurl%
\url{https://doi.org/10.1145/3491102.3517731}
\showDOI{\tempurl}


\bibitem[Lucy et~al\mbox{.}(2024)]%
        {lucy2024onesizefitsall}
\bibfield{author}{\bibinfo{person}{Li Lucy}, \bibinfo{person}{Su~Lin Blodgett},
  \bibinfo{person}{Milad Shokouhi}, \bibinfo{person}{Hanna Wallach}, {and}
  \bibinfo{person}{Alexandra Olteanu}.} \bibinfo{year}{2024}\natexlab{}.
\newblock \showarticletitle{{``}One-Size-Fits-All{''}? Examining Expectations
  around What Constitute {``}Fair{''} or {``}Good{''} {NLG} System Behaviors}.
  In \bibinfo{booktitle}{\emph{Proceedings of the 2024 Conference of the North
  American Chapter of the Association for Computational Linguistics: Human
  Language Technologies (Volume 1: Long Papers)}},
  \bibfield{editor}{\bibinfo{person}{Kevin Duh}, \bibinfo{person}{Helena
  Gomez}, {and} \bibinfo{person}{Steven Bethard}} (Eds.).
  \bibinfo{publisher}{Association for Computational Linguistics},
  \bibinfo{address}{Mexico City, Mexico}, \bibinfo{pages}{1054--1089}.
\newblock
\urldef\tempurl%
\url{https://doi.org/10.18653/v1/2024.naacl-long.61}
\showDOI{\tempurl}


\bibitem[Mieczkowski et~al\mbox{.}(2021)]%
        {Mieczkowski2021}
\bibfield{author}{\bibinfo{person}{Hannah Mieczkowski},
  \bibinfo{person}{Jeffrey~T. Hancock}, \bibinfo{person}{Mor Naaman},
  \bibinfo{person}{Malte Jung}, {and} \bibinfo{person}{Jess Hohenstein}.}
  \bibinfo{year}{2021}\natexlab{}.
\newblock \showarticletitle{AI-Mediated Communication: Language Use and
  Interpersonal Effects in a Referential Communication Task}.
\newblock \bibinfo{journal}{\emph{Proc. ACM Hum.-Comput. Interact.}}
  \bibinfo{volume}{5}, \bibinfo{number}{CSCW1}, Article \bibinfo{articleno}{17}
  (\bibinfo{date}{April} \bibinfo{year}{2021}), \bibinfo{numpages}{14}~pages.
\newblock
\urldef\tempurl%
\url{https://doi.org/10.1145/3449091}
\showDOI{\tempurl}


\bibitem[Mieczkowski(2022)]%
        {Mieczkowski2022thesis}
\bibfield{author}{\bibinfo{person}{Hannah~Nicole Mieczkowski}.}
  \bibinfo{year}{2022}\natexlab{}.
\newblock \emph{\bibinfo{title}{AI-Mediated Communication: Examining Agency,
  Ownership, Expertise, and Roles of AI Systems}}.
\newblock \bibinfo{thesistype}{Ph.\,D. Dissertation}.
  \bibinfo{address}{Stanford, CA, USA}.
\newblock Advisor(s) Jeremy, Bailenson, and Mor, Naaman, and Byron, Reeves,.
\newblock
\showISBNx{9798357505163}
\newblock
\shownote{AAI29756350}.


\bibitem[Miles et~al\mbox{.}(2013)]%
        {miles2013qualitative}
\bibfield{author}{\bibinfo{person}{M.B. Miles}, \bibinfo{person}{A.M.
  Huberman}, {and} \bibinfo{person}{J. Saldana}.}
  \bibinfo{year}{2013}\natexlab{}.
\newblock \bibinfo{booktitle}{\emph{Qualitative Data Analysis: A Methods
  Sourcebook}}.
\newblock \bibinfo{publisher}{SAGE Publications}.
\newblock
\showISBNx{9781483323794}
\showLCCN{2013002036}
\urldef\tempurl%
\url{https://books.google.de/books?id=p0wXBAAAQBAJ}
\showURL{%
\tempurl}


\bibitem[Nielsen(1994)]%
        {Nielsen1994usabilityheuristics}
\bibfield{author}{\bibinfo{person}{Jakob Nielsen}.}
  \bibinfo{year}{1994}\natexlab{}.
\newblock \showarticletitle{Enhancing the explanatory power of usability
  heuristics}. In \bibinfo{booktitle}{\emph{Proceedings of the SIGCHI
  Conference on Human Factors in Computing Systems}} (Boston, Massachusetts,
  USA) \emph{(\bibinfo{series}{CHI '94})}. \bibinfo{publisher}{Association for
  Computing Machinery}, \bibinfo{address}{New York, NY, USA},
  \bibinfo{pages}{152–158}.
\newblock
\showISBNx{0897916506}
\urldef\tempurl%
\url{https://doi.org/10.1145/191666.191729}
\showDOI{\tempurl}


\bibitem[Oulasvirta et~al\mbox{.}(2012)]%
        {oulasvirta2012habits}
\bibfield{author}{\bibinfo{person}{Antti Oulasvirta}, \bibinfo{person}{Tye
  Rattenbury}, \bibinfo{person}{Lingyi Ma}, {and} \bibinfo{person}{Eeva
  Raita}.} \bibinfo{year}{2012}\natexlab{}.
\newblock \showarticletitle{Habits make smartphone use more pervasive}.
\newblock \bibinfo{journal}{\emph{Personal and Ubiquitous Computing}}
  \bibinfo{volume}{16}, \bibinfo{number}{1} (\bibinfo{date}{Jan.}
  \bibinfo{year}{2012}), \bibinfo{pages}{105--114}.
\newblock
\showISSN{1617-4917}
\urldef\tempurl%
\url{https://doi.org/10.1007/s00779-011-0412-2}
\showDOI{\tempurl}


\bibitem[Oulasvirta et~al\mbox{.}(2005)]%
        {oulasvirta2005bursts}
\bibfield{author}{\bibinfo{person}{Antti Oulasvirta}, \bibinfo{person}{Sakari
  Tamminen}, \bibinfo{person}{Virpi Roto}, {and} \bibinfo{person}{Jaana
  Kuorelahti}.} \bibinfo{year}{2005}\natexlab{}.
\newblock \showarticletitle{Interaction in 4-second bursts: the fragmented
  nature of attentional resources in mobile HCI}. In
  \bibinfo{booktitle}{\emph{Proceedings of the SIGCHI Conference on Human
  Factors in Computing Systems}} (Portland, Oregon, USA)
  \emph{(\bibinfo{series}{CHI '05})}. \bibinfo{publisher}{Association for
  Computing Machinery}, \bibinfo{address}{New York, NY, USA},
  \bibinfo{pages}{919–928}.
\newblock
\showISBNx{1581139985}
\urldef\tempurl%
\url{https://doi.org/10.1145/1054972.1055101}
\showDOI{\tempurl}


\bibitem[Padmakumar and He(2024)]%
        {padmakumar2024diversity}
\bibfield{author}{\bibinfo{person}{Vishakh Padmakumar} {and}
  \bibinfo{person}{He He}.} \bibinfo{year}{2024}\natexlab{}.
\newblock \bibinfo{title}{Does Writing with Language Models Reduce Content
  Diversity?}
\newblock
\newblock
\showeprint[arxiv]{2309.05196}~[cs.CL]
\urldef\tempurl%
\url{https://arxiv.org/abs/2309.05196}
\showURL{%
\tempurl}


\bibitem[Palin et~al\mbox{.}(2019)]%
        {palin2019mobiletyping}
\bibfield{author}{\bibinfo{person}{Kseniia Palin}, \bibinfo{person}{Anna~Maria
  Feit}, \bibinfo{person}{Sunjun Kim}, \bibinfo{person}{Per~Ola Kristensson},
  {and} \bibinfo{person}{Antti Oulasvirta}.} \bibinfo{year}{2019}\natexlab{}.
\newblock \showarticletitle{How do People Type on Mobile Devices? Observations
  from a Study with 37,000 Volunteers}. In
  \bibinfo{booktitle}{\emph{Proceedings of the 21st International Conference on
  Human-Computer Interaction with Mobile Devices and Services}} (Taipei,
  Taiwan) \emph{(\bibinfo{series}{MobileHCI '19})}.
  \bibinfo{publisher}{Association for Computing Machinery},
  \bibinfo{address}{New York, NY, USA}, Article \bibinfo{articleno}{9},
  \bibinfo{numpages}{12}~pages.
\newblock
\showISBNx{9781450368254}
\urldef\tempurl%
\url{https://doi.org/10.1145/3338286.3340120}
\showDOI{\tempurl}


\bibitem[Park et~al\mbox{.}(2019)]%
        {Park2019inboxneedfinding}
\bibfield{author}{\bibinfo{person}{Soya Park}, \bibinfo{person}{Amy~X. Zhang},
  \bibinfo{person}{Luke~S. Murray}, {and} \bibinfo{person}{David~R. Karger}.}
  \bibinfo{year}{2019}\natexlab{}.
\newblock \showarticletitle{Opportunities for Automating Email Processing: A
  Need-Finding Study}. In \bibinfo{booktitle}{\emph{Proceedings of the 2019 CHI
  Conference on Human Factors in Computing Systems}} (Glasgow, Scotland Uk)
  \emph{(\bibinfo{series}{CHI '19})}. \bibinfo{publisher}{Association for
  Computing Machinery}, \bibinfo{address}{New York, NY, USA},
  \bibinfo{pages}{1–12}.
\newblock
\showISBNx{9781450359702}
\urldef\tempurl%
\url{https://doi.org/10.1145/3290605.3300604}
\showDOI{\tempurl}


\bibitem[Quinn and Zhai(2016)]%
        {Quinn2016}
\bibfield{author}{\bibinfo{person}{Philip Quinn} {and} \bibinfo{person}{Shumin
  Zhai}.} \bibinfo{year}{2016}\natexlab{}.
\newblock \showarticletitle{A Cost-Benefit Study of Text Entry Suggestion
  Interaction}. In \bibinfo{booktitle}{\emph{Proceedings of the 2016 CHI
  Conference on Human Factors in Computing Systems}} (San Jose, California,
  USA) \emph{(\bibinfo{series}{CHI '16})}. \bibinfo{publisher}{Association for
  Computing Machinery}, \bibinfo{address}{New York, NY, USA},
  \bibinfo{pages}{83–88}.
\newblock
\showISBNx{9781450333627}
\urldef\tempurl%
\url{https://doi.org/10.1145/2858036.2858305}
\showDOI{\tempurl}


\bibitem[{R Core Team}(2020)]%
        {R2020}
\bibfield{author}{\bibinfo{person}{{R Core Team}}.}
  \bibinfo{year}{2020}\natexlab{}.
\newblock \bibinfo{booktitle}{\emph{R: A Language and Environment for
  Statistical Computing}}.
\newblock R Foundation for Statistical Computing, Vienna, Austria.
\newblock
\urldef\tempurl%
\url{https://www.R-project.org}
\showURL{%
\tempurl}


\bibitem[Raptis et~al\mbox{.}(2013)]%
        {raptis2013phonesize}
\bibfield{author}{\bibinfo{person}{Dimitrios Raptis}, \bibinfo{person}{Nikolaos
  Tselios}, \bibinfo{person}{Jesper Kjeldskov}, {and}
  \bibinfo{person}{Mikael~B. Skov}.} \bibinfo{year}{2013}\natexlab{}.
\newblock \showarticletitle{Does size matter? investigating the impact of
  mobile phone screen size on users' perceived usability, effectiveness and
  efficiency.}. In \bibinfo{booktitle}{\emph{Proceedings of the 15th
  International Conference on Human-Computer Interaction with Mobile Devices
  and Services}} (Munich, Germany) \emph{(\bibinfo{series}{MobileHCI '13})}.
  \bibinfo{publisher}{Association for Computing Machinery},
  \bibinfo{address}{New York, NY, USA}, \bibinfo{pages}{127–136}.
\newblock
\showISBNx{9781450322737}
\urldef\tempurl%
\url{https://doi.org/10.1145/2493190.2493204}
\showDOI{\tempurl}


\bibitem[Reeves(2008)]%
        {reeves2008emailover50}
\bibfield{author}{\bibinfo{person}{B. Reeves}.}
  \bibinfo{year}{2008}\natexlab{}.
\newblock \bibinfo{booktitle}{\emph{Teach Yourself - The Internet and Email for
  the Over 50s}}.
\newblock \bibinfo{publisher}{McGraw-Hill}.
\newblock
\showISBNx{9780071582834}
\urldef\tempurl%
\url{https://books.google.de/books?id=gz04JwAACAAJ}
\showURL{%
\tempurl}


\bibitem[Reimers and Gurevych(2019)]%
        {reimers2019sbert}
\bibfield{author}{\bibinfo{person}{Nils Reimers} {and} \bibinfo{person}{Iryna
  Gurevych}.} \bibinfo{year}{2019}\natexlab{}.
\newblock \showarticletitle{Sentence-BERT: Sentence Embeddings using Siamese
  BERT-Networks}. In \bibinfo{booktitle}{\emph{Proceedings of the 2019
  Conference on Empirical Methods in Natural Language Processing}}.
  \bibinfo{publisher}{Association for Computational Linguistics}.
\newblock
\urldef\tempurl%
\url{https://arxiv.org/abs/1908.10084}
\showURL{%
\tempurl}


\bibitem[Robertson et~al\mbox{.}(2021)]%
        {Robertson2021cantreply}
\bibfield{author}{\bibinfo{person}{Ronald~E Robertson},
  \bibinfo{person}{Alexandra Olteanu}, \bibinfo{person}{Fernando Diaz},
  \bibinfo{person}{Milad Shokouhi}, {and} \bibinfo{person}{Peter Bailey}.}
  \bibinfo{year}{2021}\natexlab{}.
\newblock \showarticletitle{“I Can’t Reply with That”: Characterizing
  Problematic Email Reply Suggestions}. In
  \bibinfo{booktitle}{\emph{Proceedings of the 2021 CHI Conference on Human
  Factors in Computing Systems}} (Yokohama, Japan) \emph{(\bibinfo{series}{CHI
  '21})}. \bibinfo{publisher}{Association for Computing Machinery},
  \bibinfo{address}{New York, NY, USA}, Article \bibinfo{articleno}{724},
  \bibinfo{numpages}{18}~pages.
\newblock
\showISBNx{9781450380966}
\urldef\tempurl%
\url{https://doi.org/10.1145/3411764.3445557}
\showDOI{\tempurl}


\bibitem[Salehi et~al\mbox{.}(2017)]%
        {salehi2017communicatecontext}
\bibfield{author}{\bibinfo{person}{Niloufar Salehi}, \bibinfo{person}{Jaime
  Teevan}, \bibinfo{person}{Shamsi Iqbal}, {and} \bibinfo{person}{Ece Kamar}.}
  \bibinfo{year}{2017}\natexlab{}.
\newblock \showarticletitle{Communicating Context to the Crowd for Complex
  Writing Tasks}. In \bibinfo{booktitle}{\emph{Proceedings of the 2017 ACM
  Conference on Computer Supported Cooperative Work and Social Computing}}
  (Portland, Oregon, USA) \emph{(\bibinfo{series}{CSCW '17})}.
  \bibinfo{publisher}{Association for Computing Machinery},
  \bibinfo{address}{New York, NY, USA}, \bibinfo{pages}{1890–1901}.
\newblock
\showISBNx{9781450343350}
\urldef\tempurl%
\url{https://doi.org/10.1145/2998181.2998332}
\showDOI{\tempurl}


\bibitem[{Superhuman}(2024)]%
        {superhuman2024video}
\bibfield{author}{\bibinfo{person}{{Superhuman}}.}
  \bibinfo{year}{2024}\natexlab{}.
\newblock \bibinfo{title}{Superhuman {AI} for {iPhone} and {iPad}}.
\newblock
\newblock
\urldef\tempurl%
\url{https://www.youtube.com/watch?v=ijfi16JJguI}
\showURL{%
\tempurl}


\bibitem[Tankelevitch et~al\mbox{.}(2024)]%
        {Tankelevitch2024metacognitive}
\bibfield{author}{\bibinfo{person}{Lev Tankelevitch}, \bibinfo{person}{Viktor
  Kewenig}, \bibinfo{person}{Auste Simkute}, \bibinfo{person}{Ava~Elizabeth
  Scott}, \bibinfo{person}{Advait Sarkar}, \bibinfo{person}{Abigail Sellen},
  {and} \bibinfo{person}{Sean Rintel}.} \bibinfo{year}{2024}\natexlab{}.
\newblock \showarticletitle{The Metacognitive Demands and Opportunities of
  Generative AI}. In \bibinfo{booktitle}{\emph{Proceedings of the CHI
  Conference on Human Factors in Computing Systems}} (Honolulu, HI, USA)
  \emph{(\bibinfo{series}{CHI '24})}. \bibinfo{publisher}{Association for
  Computing Machinery}, \bibinfo{address}{New York, NY, USA}, Article
  \bibinfo{articleno}{680}, \bibinfo{numpages}{24}~pages.
\newblock
\showISBNx{9798400703300}
\urldef\tempurl%
\url{https://doi.org/10.1145/3613904.3642902}
\showDOI{\tempurl}


\bibitem[{The Copilot Connection}(2023)]%
        {copilotconnection2023video}
\bibfield{author}{\bibinfo{person}{{The Copilot Connection}}.}
  \bibinfo{year}{2023}\natexlab{}.
\newblock \bibinfo{title}{M365 {Copilot} demo - {Outlook} on mobile}.
\newblock
\newblock
\urldef\tempurl%
\url{https://www.youtube.com/watch?v=Ru8OUhdKYpI}
\showURL{%
\tempurl}


\bibitem[Wei et~al\mbox{.}(2022)]%
        {wei2022chain}
\bibfield{author}{\bibinfo{person}{Jason Wei}, \bibinfo{person}{Xuezhi Wang},
  \bibinfo{person}{Dale Schuurmans}, \bibinfo{person}{Maarten Bosma},
  \bibinfo{person}{Fei Xia}, \bibinfo{person}{Ed Chi}, \bibinfo{person}{Quoc~V
  Le}, \bibinfo{person}{Denny Zhou}, {et~al\mbox{.}}}
  \bibinfo{year}{2022}\natexlab{}.
\newblock \showarticletitle{Chain-of-thought prompting elicits reasoning in
  large language models}.
\newblock \bibinfo{journal}{\emph{Advances in neural information processing
  systems}}  \bibinfo{volume}{35} (\bibinfo{year}{2022}),
  \bibinfo{pages}{24824--24837}.
\newblock


\bibitem[Wobbrock et~al\mbox{.}(2011)]%
        {wobbrock2011art}
\bibfield{author}{\bibinfo{person}{Jacob~O. Wobbrock}, \bibinfo{person}{Leah
  Findlater}, \bibinfo{person}{Darren Gergle}, {and} \bibinfo{person}{James~J.
  Higgins}.} \bibinfo{year}{2011}\natexlab{}.
\newblock \showarticletitle{The aligned rank transform for nonparametric
  factorial analyses using only anova procedures}. In
  \bibinfo{booktitle}{\emph{Proceedings of the SIGCHI Conference on Human
  Factors in Computing Systems}} (Vancouver, BC, Canada)
  \emph{(\bibinfo{series}{CHI '11})}. \bibinfo{publisher}{Association for
  Computing Machinery}, \bibinfo{address}{New York, NY, USA},
  \bibinfo{pages}{143–146}.
\newblock
\showISBNx{9781450302289}
\urldef\tempurl%
\url{https://doi.org/10.1145/1978942.1978963}
\showDOI{\tempurl}


\bibitem[Yuan et~al\mbox{.}(2022)]%
        {Yuan2022}
\bibfield{author}{\bibinfo{person}{Ann Yuan}, \bibinfo{person}{Andy Coenen},
  \bibinfo{person}{Emily Reif}, {and} \bibinfo{person}{Daphne Ippolito}.}
  \bibinfo{year}{2022}\natexlab{}.
\newblock \showarticletitle{Wordcraft: Story Writing With Large Language
  Models}. In \bibinfo{booktitle}{\emph{Proceedings of the 27th International
  Conference on Intelligent User Interfaces}} (Helsinki, Finland)
  \emph{(\bibinfo{series}{IUI '22})}. \bibinfo{publisher}{Association for
  Computing Machinery}, \bibinfo{address}{New York, NY, USA},
  \bibinfo{pages}{841–852}.
\newblock
\showISBNx{9781450391443}
\urldef\tempurl%
\url{https://doi.org/10.1145/3490099.3511105}
\showDOI{\tempurl}


\bibitem[Zheng and Gao(2024)]%
        {zheng2024waitingtime}
\bibfield{author}{\bibinfo{person}{Jian Zheng} {and} \bibinfo{person}{Ge Gao}.}
  \bibinfo{year}{2024}\natexlab{}.
\newblock \showarticletitle{Fragmented Moments, Balanced Choices: How Do People
  Make Use of Their Waiting Time?}. In \bibinfo{booktitle}{\emph{Proceedings of
  the CHI Conference on Human Factors in Computing Systems}} (Honolulu, HI,
  USA) \emph{(\bibinfo{series}{CHI '24})}. \bibinfo{publisher}{Association for
  Computing Machinery}, \bibinfo{address}{New York, NY, USA}, Article
  \bibinfo{articleno}{713}, \bibinfo{numpages}{14}~pages.
\newblock
\showISBNx{9798400703300}
\urldef\tempurl%
\url{https://doi.org/10.1145/3613904.3642608}
\showDOI{\tempurl}


\end{thebibliography}

\appendix

\section{Appendix}

This appendix lists our prompt templates (\cref{sec:appendix_prompts}), questionnaires (\cref{sec:appendix_questionnaires}) and additional figures (\cref{sec:appendix_extra_figures}).

\subsection{Prompting}\label{sec:appendix_prompts}
Our final prototype used the prompt templates shown below. For a better overview, we shortened them for this appendix by leaving out the concrete few-shot examples and only indicating their position in the templates. See the project repository (link in \cref{sec:conclusion}) for the full prompts including these examples.

Contained variables are defined as follows:

\begin{itemize}
    \item sender: The full name of the sender.
    \item email\_text: Text content of the received email.
    \item existing\_reply: Either `This is the reply you have written so far: ``{existing\_text}'' ' or an empty string if there is no existing text.
    \item attribute: ``accepting'', or ``neutral'', or ``declining'' (two generations for each attribute)
    \item referenced\_text: The sentence that was selected on.
    \item input: Prompt that the user gave.
\end{itemize}

\subsubsection{Sentence-level support, without user input}\label{sec:appendix_sentence_without_input_prompt} %
\begin{lstlisting}
    System: "You are answering an email sentence by sentence. For each given sentence think of a suitable reply. The reply should only answer the selected sentence."
    ... few-shot examples...
    User: "You are Jamie Doe and have received this email from {sender}: '{email_text}'.
    {existing_reply} Formulate a short, {attribute} reply to this selected part of the email: '{referenced_text}'. Only output the short reply in one or two sentences."
\end{lstlisting}

\subsubsection{Sentence-level support, with user input} %
\begin{lstlisting}
    System: "You are answering an email sentence by sentence. For each given sentence think of a suitable reply. The reply should only answer the selected sentence."
    ... few-shot examples...
    User: "You are Jamie Doe and have received this email from {sender}:"{email_text}".
    {existing_reply} Formulate a short reply to this selected part of the email: "{referenced_text}". Incorporate this information into your reply: "{input}".  Only output the short reply in one or two sentences."
\end{lstlisting}

\subsubsection{Improve Email}\label{subsec:appendix_improve_email_prompt} %
\begin{lstlisting}
    System: "You have received an email and have drafted a reply. Now you review your draft and make some final edits to make it sound better. You output the entire improved email at once and nothing else."
    ... few-shot examples...
    User: "You are Jamie Doe and have received this email from {sender}:"{email_text}"
    You have written this reply as an answer:"{existing_reply}"
    You improve this email by fixing any mistakes and adding an email greeting or sign-off if missing. You also make sure to make it sound better but you do not change the content of the email. At last you only output the well formatted email."
\end{lstlisting}

\subsubsection{Message-level reply generation} \mbox{}
\begin{lstlisting}
    System: "You have received an email and are writing a response to it."
    ... few-shot examples...
    You are Jamie Doe and have received this email from {sender}:"{email_text}"
    You answer with a well written email following these instructions: "{input}". You make sure to add a greeting and a sign-off. You do not make anything up that is not mentioned in the instruction. You double check that the email is well formatted."
\end{lstlisting}

\subsection{Questionnaires}\label{sec:appendix_questionnaires}

\subsubsection{Favourite Mode}
\begin{enumerate}
  \item 
    \textbf{Question:} Which mode did you prefer for answering emails? \\
    \textbf{Question Type:} single-choice \\
    \textbf{Answer Options:} 
    \begin{itemize}
      \item Sentence-based suggestions
      \item Single prompt suggestion
      \item Without AI-Support
      \item Depends (please describe) [free-response]
    \end{itemize}
  \item 
    \textbf{Question:} Why did you prefer this mode? \\
    \textbf{Question Type:} free-response

  \item 
    \textbf{Question (Optional):} Did you face any problems, issues or bugs during your participation in this study? \\
    \textbf{Question Type:} free-response
\end{enumerate}

\subsubsection{Demographic Questionnaire}
\begin{enumerate}

\item
  \textbf{Question:} What gender do you identify with? \\
  \textbf{Question Type:} single-choice \\
  \textbf{Answer Options:}
  \begin{itemize}
    \item Woman
    \item Man
    \item Non-Binary
    \item Prefer not to disclose
    \item Prefer to self-describe: [free-response]
  \end{itemize}

\item
  \textbf{Question:} How well do you speak English? \\
  \textbf{Question Type:} single-choice \\
  \textbf{Answer Options:}
  \begin{itemize}
    \item No knowledge of English
    \item Speak poorly (beginner knowledge)
    \item Fairly well (intermediate knowledge)
    \item Well (advanced knowledge)
    \item Very well (proficient in English)
    \item Native speaker
  \end{itemize}

\item
  \textbf{Question:} How old are you? \\
  \textbf{Question Type:} numeric response

\item
  \textbf{Question:} What is your current occupation? \\
  \textbf{Question Type:} free-response

\item
  \textbf{Question:} What is the highest academic level you have achieved? \\
  \textbf{Question Type:} single-choice \\
  \textbf{Answer Options:}
  \begin{itemize}
    \item High School Diploma or equivalent
    \item Bachelor's Degree
    \item Master's Degree
    \item Doctoral Degree
    \item Other: [free-response]
  \end{itemize}

\item
  \textbf{Question:} How often do you reply to emails? \\
  \textbf{Question Type:} single-choice \\
  \textbf{Answer Options:}
  \begin{itemize}
    \item Never
    \item Less than monthly
    \item At least once a month
    \item At least once a week
    \item Daily
    \item More than 10 times a day
    \item Other: [free-response]
  \end{itemize}

\item
  \textbf{Question:} What devices do you use to answer emails on? \\
  \textbf{Question Type:} multiple-select \\
  \textbf{Answer Options:}
  \begin{itemize}
    \item Desktop-PC
    \item Laptop
    \item Tablet
    \item Smartphone
    \item Smartwatch
    \item Other: [free-response]
  \end{itemize}

\item
  \textbf{Question:} Do you have experience with AI writing support (including for email)? \\
  \textbf{Question Type:} multiple-select \\
  \textbf{Answer Options:}
  \begin{itemize}
    \item No, I have no experience writing with AI support
    \item Writing with word- or sentence-suggestions
    \item Writing with auto-correction
    \item Writing with auto-completion
    \item Using ChatGPT (or similar)
    \item Using the smart reply feature
    \item Other: [free-response]
  \end{itemize}

\item
  \textbf{Question:} Where do you usually reply to emails? \\
  \textbf{Question Type:} multiple-select \\
  \textbf{Answer Options:}
  \begin{itemize}
    \item On the go
    \item At home
    \item At the office
    \item Somewhere else: [free-response]
  \end{itemize}

\item
  \textbf{Question:} What context do most of your emails have? \\
  \textbf{Question Type:} multiple-select \\
  \textbf{Answer Options:}
  \begin{itemize}
    \item Business
    \item Private
    \item Other: [free-response]
  \end{itemize}

\end{enumerate}

\subsection{Additional Figures}\label{sec:appendix_extra_figures}
\begin{figure*}[h!]
    \centering
    \includegraphics[width=\linewidth]{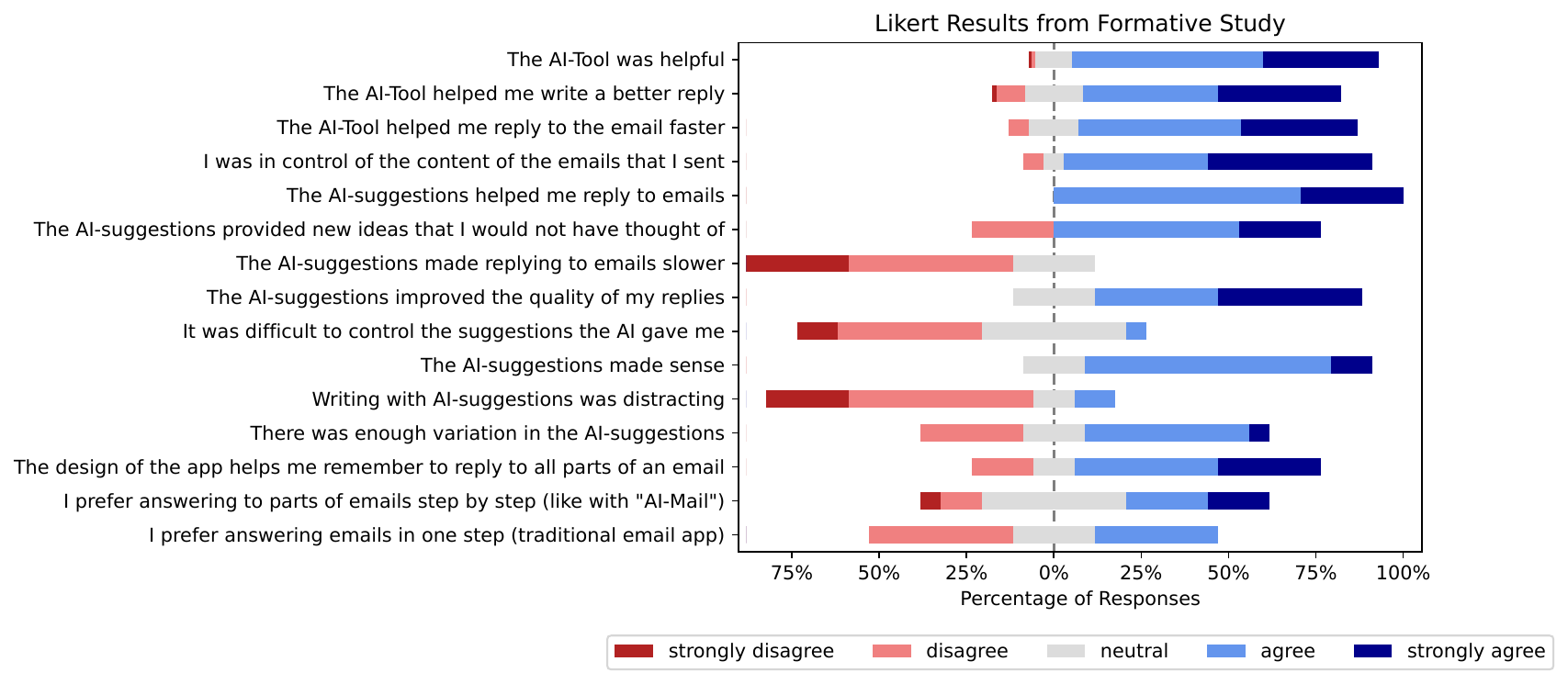}
    \caption{Likert results from the formative study (\cref{sec:formative_study}). The first three items were asked via an in-app feedback screen after each email, while the others were part of the final questionnaire at the end of the study.}
    \Description{This figure presents the Likert scale results from the formative study, which evaluates participant feedback on various aspects of using the AI tool for email writing. The responses are categorised into five levels of agreement: strongly disagree, disagree, neutral, agree, and strongly agree. The figure indicates that the majority of participants had a positive experience with the AI tool, finding it helpful in writing faster, higher-quality replies with enough variation in suggestions, though some reported mixed feelings about control and distractions.}
    \label{fig:likert_items_formative_study}
\end{figure*}

\begin{figure*}[h!]
    \centering
    \includegraphics[width=\linewidth]{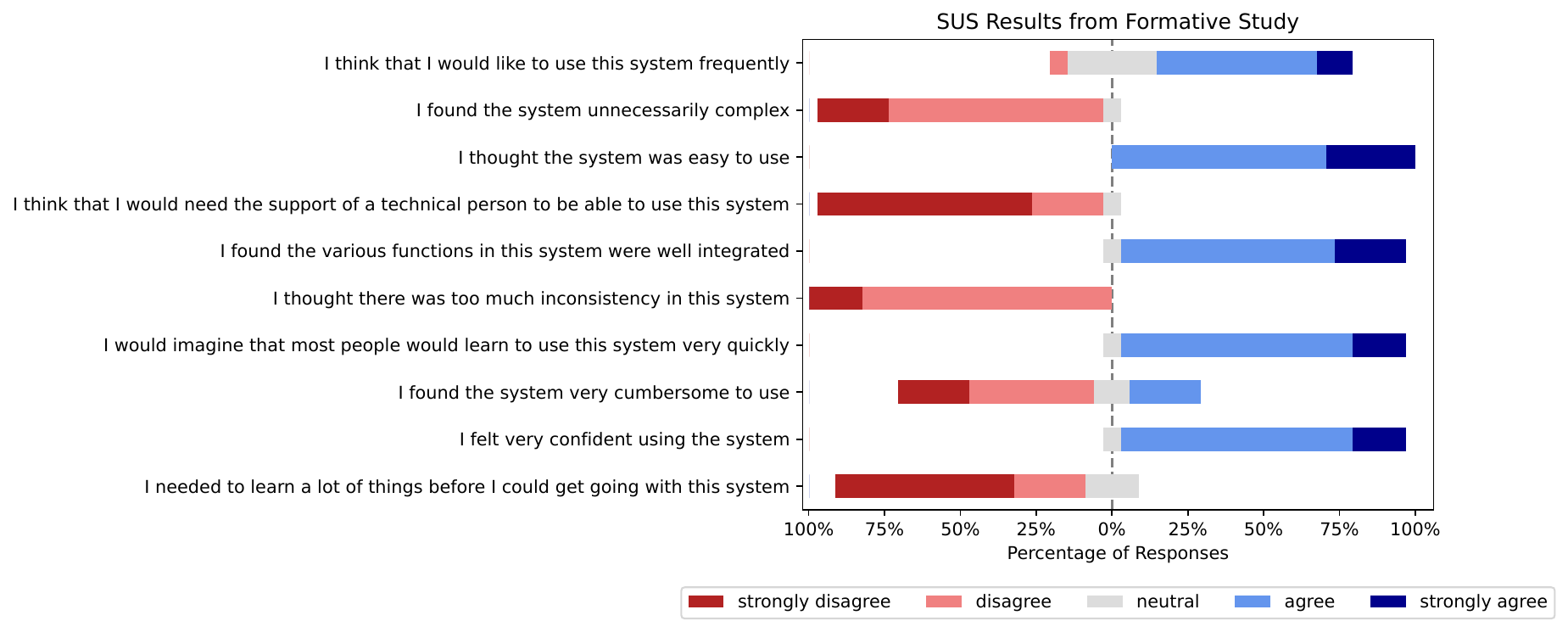}
    \caption{\revision{SUS \cite{brooke1996sus} results from the formative study (\cref{sec:formative_study}).}}
    \Description{This figure presents the SUS Likert scale results from the formative study. The responses are categorised into five levels of agreement: strongly disagree, disagree, neutral, agree, and strongly agree. The figure indicates that the majority of participants had a positive experience with the AI tool.}
    \label{fig:sus_items_formative_study}
\end{figure*}

\begin{figure*}[h!]
    \centering
    \includegraphics[width=0.6\linewidth]{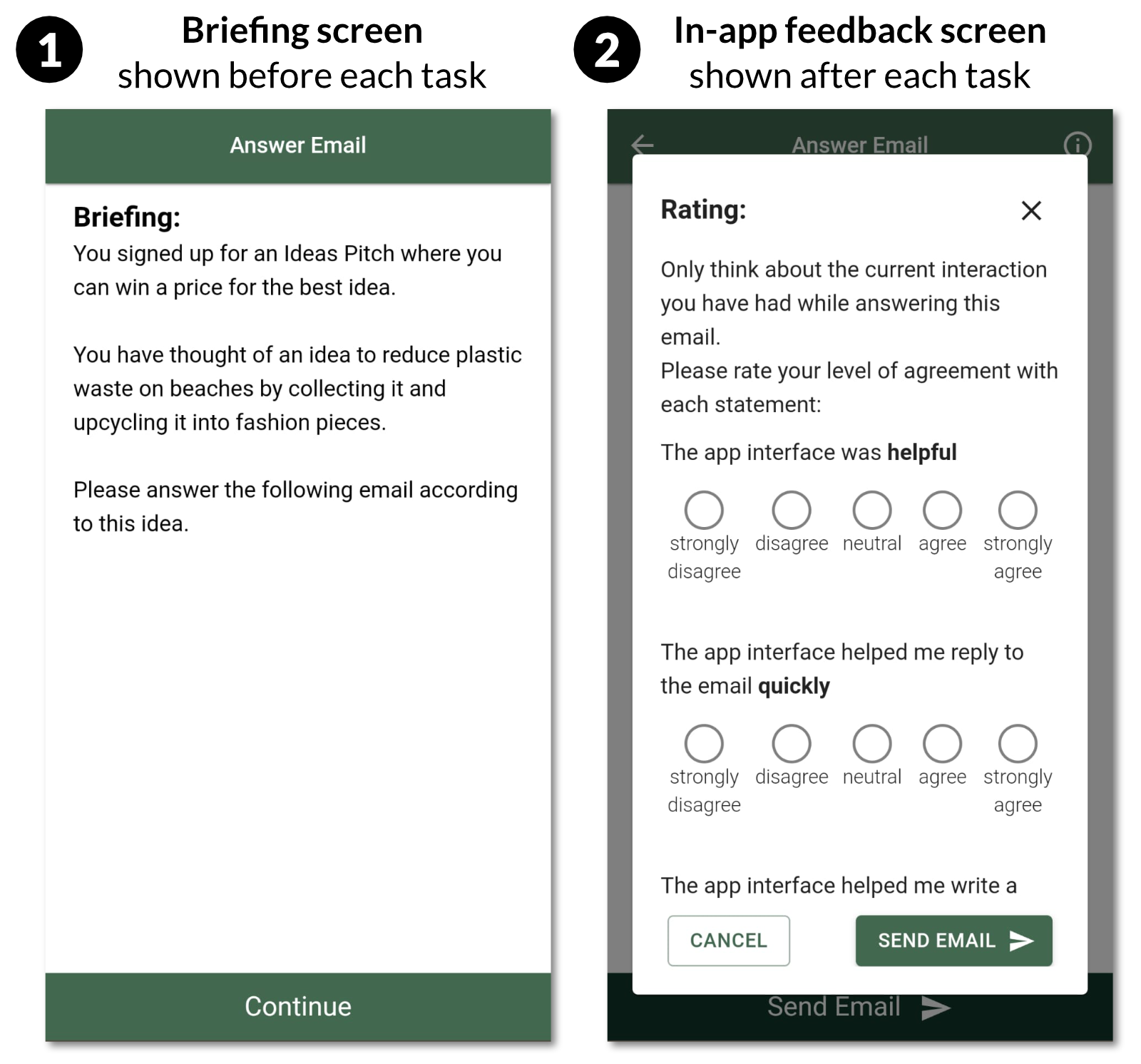}
    \caption{The study-specific UI screens in our prototype: A screen showing the briefing before each email task \textit{(1)}, and a screen asking users to rate four Likert items after each email task \textit{(2)}.}
    \Description{This figure presents two study-specific UI screens from the prototype used in the research:
    Briefing Screen (Left Panel):
    This screen is shown before each email task.
    It provides a brief description of the task, explaining the scenario or context that the participant needs to respond to.
    The screen includes instructions to "Answer the following email according to this idea" and has a "Continue" button to move forward.
    In-App Feedback Screen (Right Panel):
    This screen appears after each email task, asking participants to rate their experience.
    The specific statements in the example ask about the helpfulness of the interface.
    Participants can choose from five options ranging from "strongly disagree" to "strongly agree" for each statement, and submit their feedback by pressing the "Send Email" button.}
    \label{fig:briefing_and_feedback}
\end{figure*}

\begin{figure*}[h!]
    \centering
    \includegraphics[width=0.6\linewidth]{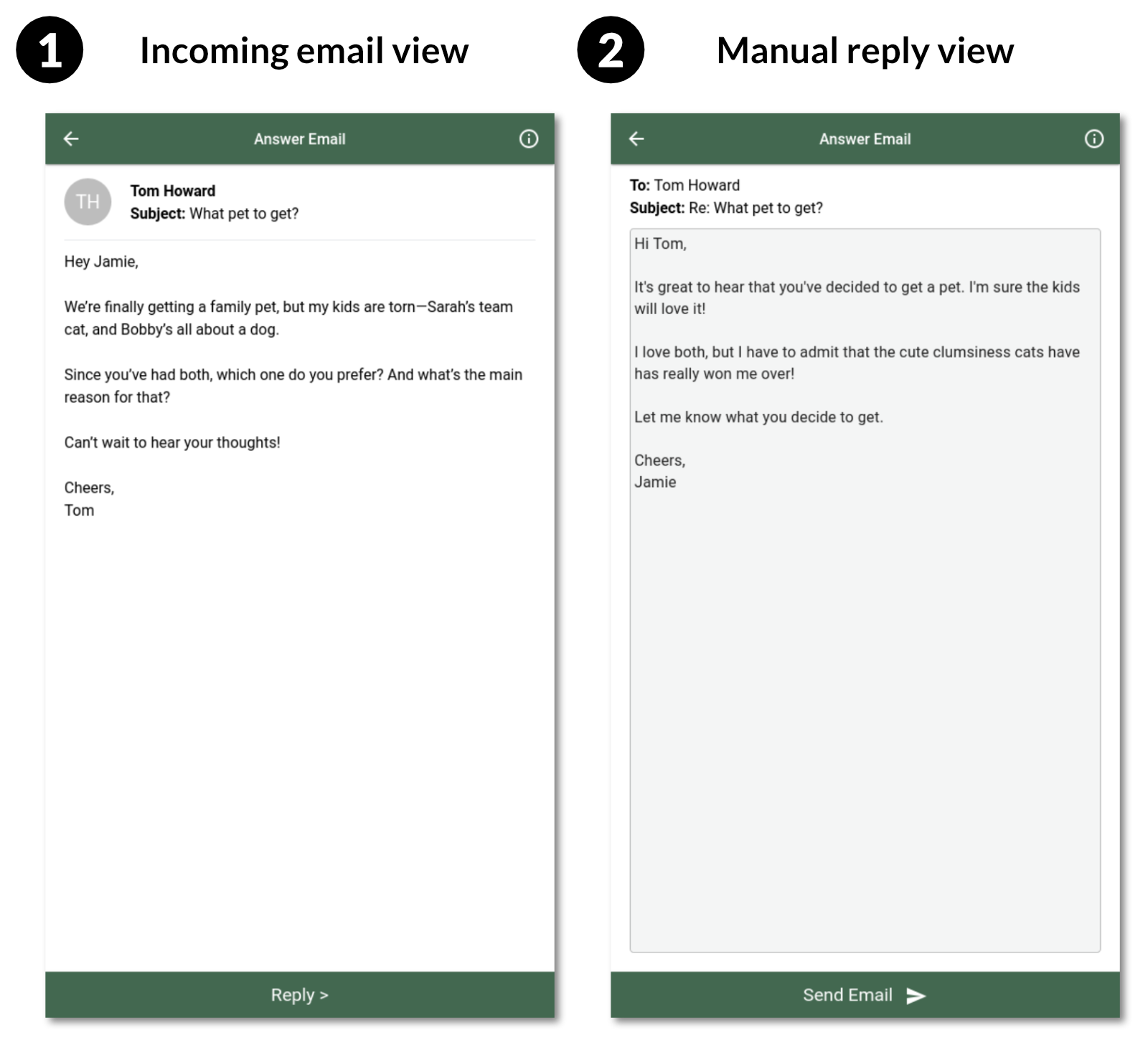}
    \caption{The UI design for manual typing (\modemanual) used in the study. It has one screen to show the incoming email \textit{(1)} and one with a text box to type the reply manually \textit{(2)}. This figure shows the state after typing a reply, as an example. As is usual on mobile devices, the keyboard opened from the bottom when tapping on the text field.}
    \Description{This figure illustrates the UI design for manual typing (NoAI) used in the study, showcasing two views:
    Incoming Email View (Left Panel):
    This screen displays the email received by the user.
    In this example, the email is about the subject "What pet to get?" asking for advice on choosing a family pet between a cat or a dog.
    At the bottom of the screen, there is a "Reply" button, allowing the user to initiate a response.
    Manual Reply View (Right Panel):
    Once the "Reply" button is pressed, the user is taken to the reply screen, where they can type their response manually in a text box.
    The "Send Email" button is at the bottom.}
    \label{fig:baseline_uis_manual}
\end{figure*}

\begin{figure*}[h!]
    \centering
    \includegraphics[width=0.8\linewidth]{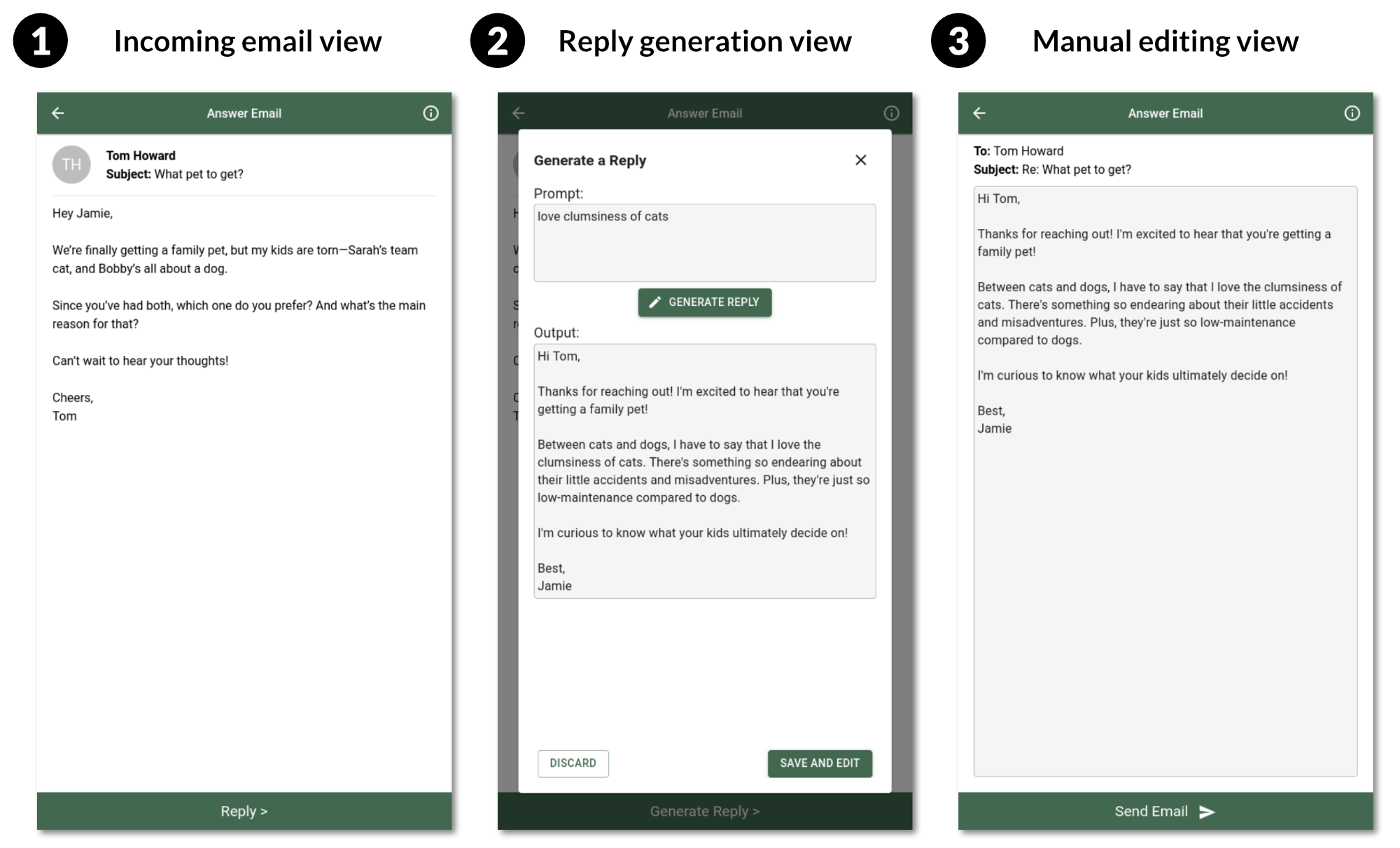}
    \caption{The UI design for \modemailtxt{} (\modemail) used in the study. The first screen shows the incoming email \textit{(1)}. On the reply generation screen \textit{(2)}, users can generate a full message suggestion, optionally guided by entering a prompt in the text box at the top. Finally, the manual screen \textit{(3)} allowed users to freely edit the generated draft.}
    \Description{This figure shows the UI design for message-level reply generation (MSG) used in the study, illustrating three stages of the interaction:
    Incoming Email View (Left Panel):
    Similar to the manual UI, this screen displays the incoming email.
    A "Reply" button is available at the bottom.
    Reply Generation View (Middle Panel):
    In this screen, users can generate a full message reply using AI.
    There is an optional prompt field at the top, where users can enter specific keywords or topics to guide the AI's response.
    Once the prompt is entered, users can press the "Generate Reply" button, and the AI will generate a suggested email, displayed in the output field below.
    The user can either discard the generated reply or choose to save and edit it.
    Manual Editing View (Right Panel):
    After generating the AI-suggested reply, users are taken to this screen to manually edit the draft.
    The "Send Email" button at the bottom allows the user to send the edited reply.}
    \label{fig:baseline_uis_msg}
\end{figure*}

Here we provide additional figures.
\cref{fig:likert_items_formative_study} shows the Likert and \cref{fig:sus_items_formative_study} the SUS \cite{brooke1996sus} results from the formative study (\cref{sec:formative_study}).
The other figures show the UIs used in the study: \cref{fig:briefing_and_feedback} shows the screens for the briefing and in-app feedback. \cref{fig:baseline_uis_manual} shows the UI of the manual mode (\modemanual), and \cref{fig:baseline_uis_msg} shows the UI of the \modemailtxt{} (\modemail). 

\end{document}